\documentclass[preprint,1p,times,authoryear]{elsarticle}
\pdfoutput=1

\usepackage{amssymb}
\usepackage{amsmath}

\journal{Journal of Information Security and Applications (JISA)}

\RequirePackage{adjustbox, array, booktabs, csquotes, forest, graphicx, longtable, multirow, multicol, pifont, scrextend, siunitx, subcaption, url, natbib}

\usepackage{xcolor}
\usepackage{colortbl}
\usepackage{pgfplots}
\usepgfplotslibrary{polar}
\usepgfplotslibrary{colormaps}
\usepackage[utf8]{inputenc}
\usepackage[right]{lineno}
\PassOptionsToPackage{hyphens}{url}
\usepackage[colorlinks=true, linkcolor=blue, citecolor=blue, urlcolor=blue]{hyperref}
\usepackage{cleveref}

\begin{document}

\begin{frontmatter}



\title{On the Validity of Traditional Vulnerability Scoring Systems for Adversarial Attacks against LLMs} 


\author[label1]{Atmane Ayoub MANSOUR BAHAR\corref{corr}} 
\ead{atmane.mansourbahar@gmail.com}
\author[label2]{Ahmad Samer WAZAN}
\ead{Ahmad.Wazan@zu.ac.ae}
\affiliation[label1]{organization={Research Assistant},
            state={Algiers},
            country={Algeria}}

\affiliation[label2]{organization={College of Technological Innovation Zayed University},
            state={Abu Dhabi},
            country={United Arab Emirates}}
            
\cortext[corr]{Corresponding author}

\begin{abstract}
\textbf{Purpose} - This research investigates the effectiveness of established vulnerability metrics, such as the Common Vulnerability Scoring System (CVSS), in evaluating attacks on Large Language Models (LLMs), with a focus on Adversarial Attacks (AAs). The study explores the influence of both general and specific metric factors in determining vulnerability scores, providing new perspectives on potential enhancements to these metrics.

\noindent \textbf{Approach} - This study adopts a quantitative approach, calculating and comparing the coefficient of variation of vulnerability scores across 56 adversarial attacks on LLMs. The attacks, sourced from various research papers, and obtained through online databases, were evaluated using multiple vulnerability metrics. Scores were determined by averaging the values assessed by three distinct LLMs.

\noindent \textbf{Findings} - The results indicate that existing scoring-systems yield vulnerability scores with minimal variation across different attacks, suggesting that many of the metric factors are inadequate for assessing adversarial attacks on LLMs. This is particularly true for context-specific factors or those with predefined value sets, such as those in CVSS. These findings support the hypothesis that current vulnerability metrics, especially those with rigid values, are limited in evaluating AAs on LLMs, highlighting the need for the development of more flexible, generalized metrics tailored to such attacks.

\noindent \textbf{Value} - This research offers a fresh analysis of the effectiveness and applicability of established vulnerability metrics, particularly in the context of adversarial attacks on Large Language Models, both of which have gained significant attention in recent years. Through extensive testing and calculations, the study underscores the limitations of these metrics and opens up new avenues for improving and refining vulnerability assessment frameworks specifically tailored for LLMs.
\end{abstract}


\begin{highlights}
\item \textbf{Evaluation of Vulnerability Metrics:} Assessed the effectiveness of traditional metrics like CVSS, DREAD, OWASP Risk Rating, and SSVC in analyzing adversarial attacks (AAs) against Large Language Models (LLMs).
\item \textbf{Comprehensive Dataset:} Analyzed 56 adversarial attacks, categorized into distinct types such as Jailbreaks, Prompt Injection, and Model Extraction.
\item \textbf{Quantitative Analysis:} Demonstrated minimal variation in traditional vulnerability scores across attack types, highlighting inadequacies in context-specific metrics.
\item \textbf{LLM Integration:} Introduced a novel approach combining assessments from three state-of-the-art LLMs (e.g., GPT-4o) with human-in-the-loop verification for vulnerability scoring.
\item \textbf{Call for New Metrics:} Proposed the development of flexible, LLM-specific vulnerability assessment frameworks to address the unique characteristics of attacks targeting LLMs.
\end{highlights}

\begin{keyword}
Adversarial Attacks \sep Large Language Models \sep Vulnerability Metrics \sep Risk Assessment \sep Descriptive Statistics
\end{keyword}

\end{frontmatter}



\section{Introduction}
\label{sec:introduction}
Large Language Models (LLMs) have recently become a cornerstone in artificial intelligence (AI) research and application, thanks to their remarkable ability to understand and generate human-like text \citep{brown2020language}. LLMs such as GPT \citep{radford2018improving}, BERT \citep{devlin2018bert}, and others have achieved widespread adoption in a variety of fields, including Natural Language Processing (NLP), machine translation, and conversational AI, due to their capacity to generalize across diverse tasks \citep{vaswani2017attention}.  
However, this surge in popularity has also exposed LLMs to a myriad of vulnerabilities, becoming an attractive target for various security threats \citep{goodfellow2014explaining,abdali2024securing, liu2024exploring}.

One of the most significant threats to LLMs is Adversarial Attacks (AAs) \citep{shayegani2023survey, peng2024jailbreaking, zhao2024evaluating}, which are typically designed to fool Machine Learning (ML) models by modifying input data or introducing carefully-crafted inputs that cause the model to behave inappropriately \citep{szegedy2013intriguing, goodfellow2014explaining}. These attacks often remain indistinguishable to humans but significantly impact the model’s decision-making process, posing a significant threat to LLMs, as they can compromise the integrity, reliability, and security of applications that rely on these models \citep{carlini2017towards}. One significant example is the Crescendo attack \citep{russinovich2024greatwritearticlethat}. This sophisticated method manipulates LLMs by gradually escalating a conversation with benign prompts that evolve into more harmful requests, effectively bypassing safety mechanisms. 
Therefore, protecting LLMs has become a critical concern for researchers and practitioners alike \citep{zou2024adversarial, kumar2023certifying}.

To effectively secure LLMs against AAs, it is crucial to assess and rank these threats based on their severity and potential impact on the model. For instance, some attacks, like Prompt Injection \citep{liu2024formalizing}, are easy to execute and widely applicable, making them higher-priority threats. Others, like Backdoor attacks \citep{li2021backdoor}, may require greater sophistication but can cause significant long-term damage \citep{greshake2023more}. This prioritization allows security teams to focus on the most dangerous attacks first for mitigation efforts. Existing vulnerability metrics, such as the Common Vulnerability Scoring System (CVSS) \citep{schiffman2005complete} and OWASP Risk Rating \citep{owaspOWASPRisk}, are commonly used to evaluate the danger level of attacks on traditional systems, taking into account factors such as attack vector, attack complexity, and impact. However, their applicability to LLMs remains questionable.

Most existing vulnerability metrics are tailored for assessing \textbf{technical} vulnerabilities in software or network systems. 
In contrast, AAs on LLMs often target the model’s \textbf{decision-making} capabilities and may not result in traditional technical-impacts, such as data breaches or service outages \citep{zhang2020adversarial}. For example, attacks like Jailbreaks \citep{chu2024comprehensive}, which manipulate the model’s outputs to bypass ethical or safety constraints, cannot easily be classified as technical vulnerabilities. These attacks focus on manipulating the model's behavior rather than exploiting system-level weaknesses. In other terms, the context-specific factors used in existing metrics, such as CVSS, do not adequately account for the unique characteristics of LLMs or the nature of AAs. Consequently, they may be ill-suited for assessing the risk posed by these attacks on LLMs.

In this study, we aim to evaluate the suitability of known vulnerability metrics in assessing Adversarial Attacks against LLMs. We hypothesize that: \textbf{\textit{`the factors used by traditional metrics may not be fully applicable to attacks on LLMs'}}, because many of these factors are not designed to capture the nuances of AAs. 

To test this hypothesis, we evaluated \textbf{56} different AAs across four widely used vulnerability metrics. Each attack was assessed using three distinct LLMs, and the scores were averaged to provide a final assessment. This multi-faceted approach aims to provide a nuanced understanding of how well current metrics can distinguish between varying levels of threat posed by different adversarial strategies, as relying solely on human judgment for security assessments would require domain experts, and human evaluation could introduce biases.

Our findings indicate that average scores across diverse attacks exhibit \textbf{low variability}, suggesting that many of the existing metric factors may not offer fair distinctions among all types of adversarial threats on LLMs. Furthermore, we observe that metrics incorporating more generalized factors tend to yield better differentiation among adversarial attacks, indicating a potential pathway for refining vulnerability assessments tailored for LLMs.

The contributions of this paper are fourfold. 
\begin{itemize}
    \item We provide a taxonomy of the various classification criteria of Adversarial Attacks existing in the literature, showing the logic followed in classifying AAs into multiple types. 
    \item We present a list of 56 AAs specifically targeting LLMs, which serve as our test scenarios.
    \item We provide a comprehensive evaluation of some vulnerability metrics, in the context of AAs targeting LLMs, using differential statistics to analyse the variations of metric scores across different attacks.
    \item We suggest that future work should focus on developing more general and LLM-specific vulnerability metrics that can effectively capture the unique characteristics of AAs targeting these models.
\end{itemize}  

This paper is structured in seven parts. We start in Section \ref{sec:method} by detailing the procedures we employed in this study, especially concerning the data collection, vulnerability assessments through LLMs, and mathematical analysis of the results. After that, we present in Section \ref{sec:adversarial-attacks} an overview of AAs and their existing classifications. In Section \ref{sec:aas-llms}, we present a detailed list of AAs on LLMs, and propose a classification based on the danger level in Section \ref{sec:danger-level}.
Sections \ref{sec:evaluation}, \ref{sec:discussion}, and \ref{sec:suggestions} encompasses respectively, the evaluation of the vulnerability metrics on LLMs, the discussion of the results, and the perspectives for future enhancements.

\section{Methods}
\label{sec:method}
In this section, we outline the methodology adopted to evaluate vulnerabilities in attacks targeting Large Language Models using established metrics such as DREAD \citep{michael2006security}, CVSS \citep{schiffman2005complete}, OWASP Risk Rating \citep{owaspOWASPRisk}, and Stakeholder-Specific Vulnerability Categorization (SSVC) \citep{spring2021prioritizing}.

Our approach involves three key steps depicted below in Figure \ref{fig:framework} : data collection, assessment, and statistical interpretation.

\begin{figure}[h]
\centering 
\includegraphics[width=\textwidth]{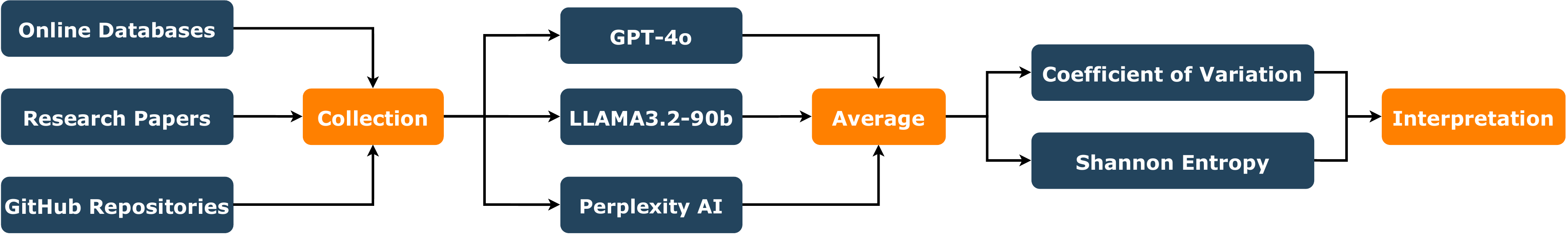}
\caption{Research process}
\label{fig:framework}
\end{figure}

\subsection{Data collection}
The first step in our methodology was to gather a comprehensive dataset of AAs targeting LLMs. To ensure a thorough and systematic approach, we began by reviewing the literature on these attacks, exploring existing types and classifications. This step provided a broad understanding of the main categories of attacks commonly observed in the context of ML and NLP systems.

Following this foundational review, we focused on identifying recent AAs specifically targeting LLMs. These attacks were grouped into \textbf{seven} primary types: Jailbreaks (White-box and Black-box) \citep{xu2024comprehensive}, Prompt Injections \citep{liu2023prompt}, Evasion attacks \citep{wang2023evasion}, Model-Inference (Membership Inference) attacks \citep{hu2022membership}, Model-Extraction attacks \citep{gencc2023taxonomic}, and Poisoning/Trojan/Backdoor attacks \citep{tian2022comprehensive, li2021backdoor, liu2020survey}. For each type, we selected \textbf{eight} representative attacks, prioritizing those published in recent research or demonstrated in practical scenarios. This effort resulted in a list of 56 attacks, covering a diverse range of threat vectors and methodologies.

To enable a systematic ranking of these attacks based on their potential danger, we decided to assess each attack using \textbf{vulnerability metrics}. By applying multiple metrics, we aimed to provide a multi-faceted evaluation of each attack’s severity and to ensure that the dataset would serve as a robust basis for further analysis and interpretation.

\subsection{Score assessments}
To evaluate the severity and danger level of the 56 gathered attacks, we began by identifying widely recognized vulnerability assessment metrics to ensure a comprehensive analysis. After careful consideration, we selected four metrics: DREAD \citep{michael2006security}, CVSS \citep{schiffman2005complete}, OWASP Risk Rating \citep{owaspOWASPRisk}, and SSVC \citep{spring2021prioritizing}. These metrics were chosen for their broad adoption and their focus on different factors, enabling a more nuanced understanding of the vulnerabilities. Since the Adversarial Attacks we collected are recent and not yet assessed in the literature, calculating their scores became essential to address this gap.

Manually assessing 56 attacks across four metrics is a daunting task, requiring extensive \textbf{expertise} from security analysts, system administrators, and other domain experts. The process involves interpreting complex scenarios, considering varying factors for each metric, and ensuring consistency between all evaluations. Completing such an effort manually could take months or even years, which is impractical given the fast-evolving nature of adversarial threats.

To overcome this challenge and accelerate the process, we leveraged the capabilities of LLMs to perform \textbf{semi-automated scoring}. Specifically, we utilized three state-of-the-art models: GPT-4o \citep{openai2024}, LLAMA3.2-90b \citep{dubey2024llama}, and Perplexity AI \citep{perplexity}. Each model operated independently, assessing the attacks and vulnerabilities according to the factors defined by the selected metrics. For each scoring factor, we calculated the average score provided by the three LLMs, rounded to the closest unit.

This approach offers several advantages. First, it enables \textbf{rapid} assessments. Second, using multiple LLMs increases the robustness of the results by \textbf{minimizing biases} or errors from any single model. Furthermore, the models’ advanced text-processing capabilities allow them to analyze the \textbf{contextual details} of each attack and provide scores that align with the logic of the vulnerability metrics.

A recent work of \citet{chopra2024chatnvdadvancingcybersecurityvulnerability} proves that LLMs are able to identify and analyze software vulnerabilities; but that they can lead to misinterpretations or oversights in understanding complex vulnerabilities. To address such potential inconsistencies in the assessments, we incorporated a \textbf{Human-in-the-Loop (HitL)} verification process. We reviewed the logic and reasoning behind each LLM-provided score to ensure its accuracy and reliability. This step was essential to mitigate any errors or misinterpretations that might arise from the LLMs, especially when handling complex scenarios.

To validate this methodology, we tested it on a set of Common Vulnerabilities and Exposures (CVEs) that already have human validated scores with both CVSS and SSVC \citep{spring2021prioritizing} to measure the gap, as shown in Table \ref{tab:llm-justification}. The details of each factor are further explained in Section \ref{subsec:scoring-systems}.

\begin{table}[h]
\centering
\caption{Comparison between some existing and LLM-generated CVSS and SSVC values} 
\vspace{2mm}
\label{tab:llm-justification}
\adjustbox{max width=\textwidth}{
\begin{tabular}{c c c}

    \hline    
    \multirow{2}{*}{\textbf{CVE-ID}}  & \multirow{2}{*}{\textbf{SSVC Values}} &  \multirow{2}{*}{\textbf{CVSS (2.0 or 3.0) Values}} \cr \\
    \hline
    
    \multirow{3}{*}{\textbf{CVE-2014-0751}} & \multirow{2}{*}{\textbf{NVD:} E:N/U:L/T:\textcolor{blue!80}{T}/P:\textcolor{blue!80}{S}} & \multirow{2}{*}{\textbf{NVD:} AV:N/AC:L/Au:N/C:P/I:P/A:P} \cr & \multirow{2}{*}{\textbf{LLMs:} E:N/U:L/T:\textcolor{orange!80}{P}/P:\textcolor{orange!80}{M}} & \multirow{2}{*}{\textbf{LLMs:} AV:N/AC:L/Au:N/C:P/I:P/A:P} \cr \\
    \hline

    \multirow{3}{*}{\textbf{CVE-2015-1014}} & \multirow{2}{*}{\textbf{NVD:} E:N/U:\textcolor{blue!80}{L}/T:T/P:S} & \multirow{2}{*}{\textbf{NVD:} AV:L/AC:L/PR:L/UI:R/S:U/C:H/I:H/A:H} \cr & \multirow{2}{*}{\textbf{LLMs:} E:N/U:\textcolor{orange!80}{E}/T:T/P:S} & \multirow{2}{*}{\textbf{LLMs:} AV:L/AC:L/PR:L/UI:R/S:U/C:H/I:H/A:H} \cr \\
    \hline
 
    \multirow{3}{*}{\textbf{CVE-2015-5374}} & \multirow{2}{*}{\textbf{NVD:} E:\textcolor{blue!80}{A}/U:L/T:P/P:S} & \multirow{2}{*}{\textbf{NVD:} AV:N/AC:L/Au:N/C:N/I:N/A:C} \cr & \multirow{2}{*}{\textbf{LLMs:} E:\textcolor{orange!80}{P}/U:L/T:P/P:S} & \multirow{2}{*}{\textbf{LLMs:} AV:N/AC:L/Au:N/C:N/I:N/A:C} \cr \\
    \hline

    \multirow{3}{*}{\textbf{CVE-2017-3183}} & \multirow{2}{*}{\textbf{NVD:} E:N/U:E/T:\textcolor{blue!80}{T}/P:M} & \multirow{2}{*}{\textbf{NVD:} AV:N/AC:L/PR:L/UI:N/S:U/C:H/I:H/A:\textcolor{blue!80}{H}} \cr & \multirow{2}{*}{\textbf{LLMs:} E:N/U:E/T:\textcolor{orange!80}{P}/P:M} & \multirow{2}{*}{\textbf{LLMs:} AV:N/AC:L/PR:L/UI:N/S:U/C:H/I:H/A:\textcolor{orange!80}{N}} \cr \\
    \hline

    \multirow{3}{*}{\textbf{CVE-2017-5638}} & \multirow{2}{*}{\textbf{NVD:} E:A/U:S/T:T/P:\textcolor{blue!80}{M}} & \multirow{2}{*}{\textbf{NVD:} AV:N/AC:L/PR:N/UI:N/S:U/C:H/I:H/A:H} \cr & \multirow{2}{*}{\textbf{LLMs:} E:A/U:S/T:T/P:\textcolor{orange!80}{S}} & \multirow{2}{*}{\textbf{LLMs:} AV:N/AC:L/PR:N/UI:N/S:U/C:H/I:H/A:H} \cr \\
    \hline

    \multirow{3}{*}{\textbf{CVE-2017-9590}} & \multirow{2}{*}{\textbf{NVD:} E:P/U:E/T:\textcolor{blue!80}{T}/P:M} & \multirow{2}{*}{\textbf{NVD:} AV:N/AC:H/PR:N/UI:N/S:U/C:H/I:N/A:N} \cr & \multirow{2}{*}{\textbf{LLMs:} E:P/U:E/T:\textcolor{orange!80}{P}/P:M} & \multirow{2}{*}{\textbf{LLMs:} AV:N/AC:H/PR:N/UI:N/S:U/C:H/I:N/A:N} \cr \\
    \hline

    \multirow{3}{*}{\textbf{CVE-2018-14781}} & \multirow{2}{*}{\textbf{NVD:} E:P/U:L/T:P/P:M} & \multirow{2}{*}{\textbf{NVD:} AV:A/AC:H/PR:N/UI:N/S:U/C:N/I:H/A:N} \cr & \multirow{2}{*}{\textbf{LLMs:} E:P/U:L/T:P/P:M} & \multirow{2}{*}{\textbf{LLMs:} AV:A/AC:H/PR:N/UI:N/S:U/C:N/I:H/A:N} \cr \\
    \hline

    \multirow{3}{*}{\textbf{CVE-2019-2691}} & \multirow{2}{*}{\textbf{NVD:} E:N/U:\textcolor{blue!80}{E}/T:P/P:M} & \multirow{2}{*}{\textbf{NVD:} AV:N/AC:L/PR:H/UI:N/S:U/C:N/I:N/A:H} \cr & \multirow{2}{*}{\textbf{LLMs:} E:N/U:\textcolor{orange!80}{S}/T:P/P:M} & \multirow{2}{*}{\textbf{LLMs:} AV:N/AC:L/PR:H/UI:N/S:U/C:N/I:N/A:H} \cr \\
    \hline

    \multirow{3}{*}{\textbf{CVE-2019-9042}} & \multirow{2}{*}{\textbf{NVD:} E:\textcolor{blue!80}{A}/U:L/T:T/P:M} & \multirow{2}{*}{\textbf{NVD:} AV:N/AC:L/PR:\textcolor{blue!80}{H}/UI:N/S:U/C:H/I:H/A:H} \cr & \multirow{2}{*}{\textbf{LLMs:} E:\textcolor{orange!80}{P}/U:L/T:T/P:M} & \multirow{2}{*}{\textbf{LLMs:} AV:N/AC:L/PR:\textcolor{orange!80}{N}/UI:N/S:U/C:H/I:H/A:H} \cr \\
    \hline

\end{tabular}
}
\end{table}
The results demonstrate that our approach of aggregating the assessments of three LLMS yields scores closely aligned with existing assessments, with few differences related mainly to the advancements of technologies from the first assessment of those vulnerabilities to today. For instance, vulnerabilities such as \texttt{`CVE-2015-5374'} and \texttt{`CVE-2019-9042'} became less active than before, making their exploitation value with SSVC change from \emph{Active} to \emph{Proof-of-Concept} (refer to Section \ref{subsubsec:ssvc} for more details).

This experiment also shows that combining the computational efficiency of LLMs with human oversight represents a practical solution for scoring new and unassessed attacks in the absence of readily available experts. This innovative approach not only saves time but also ensures a balanced and consistent evaluation process, enabling a deeper understanding of vulnerabilities and their potential impact.

\subsection{Results interpretations}
Our approach provided a multi-dimensional analysis of Adversarial Attacks against LLMs by leveraging four distinct vulnerability assessment metrics: DREAD, CVSS, OWASP Risk Rating, and SSVC. This comprehensive evaluation allowed us to gain a broad perspective on how these metrics reflect the severity and impact of attacks, as well as their usefulness in ranking and understanding vulnerabilities in the LLM context.

To assess the utility and added value of each factor within the metrics, we analyzed their \textbf{variability} across the 56 attacks, grouped by attack type. For the \textbf{quantitative metrics} (DREAD and OWASP Risk Rating), we calculated the \textbf{coefficient of variation (CV)} for each factor to measure the relative dispersion of scores. For the \textbf{qualitative metrics} (CVSS and SSVC), we used \textbf{entropy} \citep{6773024} to quantify the diversity or uniformity of categorical values.

\section{Adversarial Attacks}
\label{sec:adversarial-attacks}

The rise of AAs in the field of Machine Learning has posed significant security challenges, especially for Large Language Models. These attacks exploit the vulnerabilities inherent in AI models by manipulating inputs to achieve unintended or harmful outputs. This section provides a detailed exploration of AAs, beginning with their formal definition and an analysis of why they are considered particularly dangerous to LLMs. Then it introduces various types and classifications of AAs, offering insight into the range of attack strategies used to compromise LLMs. Understanding these elements is crucial for designing more robust defenses and enhancing the security of AI-driven systems.

\subsection{Definition}
Adversarial Attacks are \textbf{intentional manipulations} of input data designed to exploit vulnerabilities in ML models \citep{finlayson2019adversarial}. The concept of adversarial examples was first introduced in the domain of Image Recognition by \citet{szegedy2013intriguing}, and it has since been widely explored across different ML tasks, including NLP \citep{qiu2022adversarial, dong2022adversarial, zhang2020adversarial}. In the context of LLMs, adversarial inputs are carefully crafted to cause the model to produce incorrect, biased, or harmful \textbf{outputs} \citep{kumar2024adversarial}. Unlike traditional errors, AAs are not random; but are \textbf{strategically} designed to exploit the decision boundaries of models by altering inputs in ways imperceptible to humans and effective against ML models \citep{carlini2017adversarial}. These attacks can involve minimal changes, such as swapping words, inserting seemingly harmless phrases, or restructuring sentences, that lead to dramatically different responses from the model, often having severe real-world consequences \citep{ibitoye2019threat, kumar2023impact}, particularly in safety-critical applications such as autonomous driving, healthcare diagnostics, and security systems \citep{papernot2016limitations}. For example, an Adversarial Attack could lead an autonomous vehicle to misinterpret road signs, resulting in catastrophic accidents \citep{eykholt2018robust, zhou2024stealthy}. 


On top of that, AAs can come in various forms, each exploiting different aspects of LLMs. These attacks can be broadly categorized based on the attacker's knowledge, the nature of the perturbations, and the model's vulnerability. The following section will explore the different types and classifications of AAs, showing that each type has distinct strategies and potential impacts on LLMs.

\subsection{Classifications of AAs}
\begin{figure}[h]
\centering 
\includegraphics[width=\textwidth]{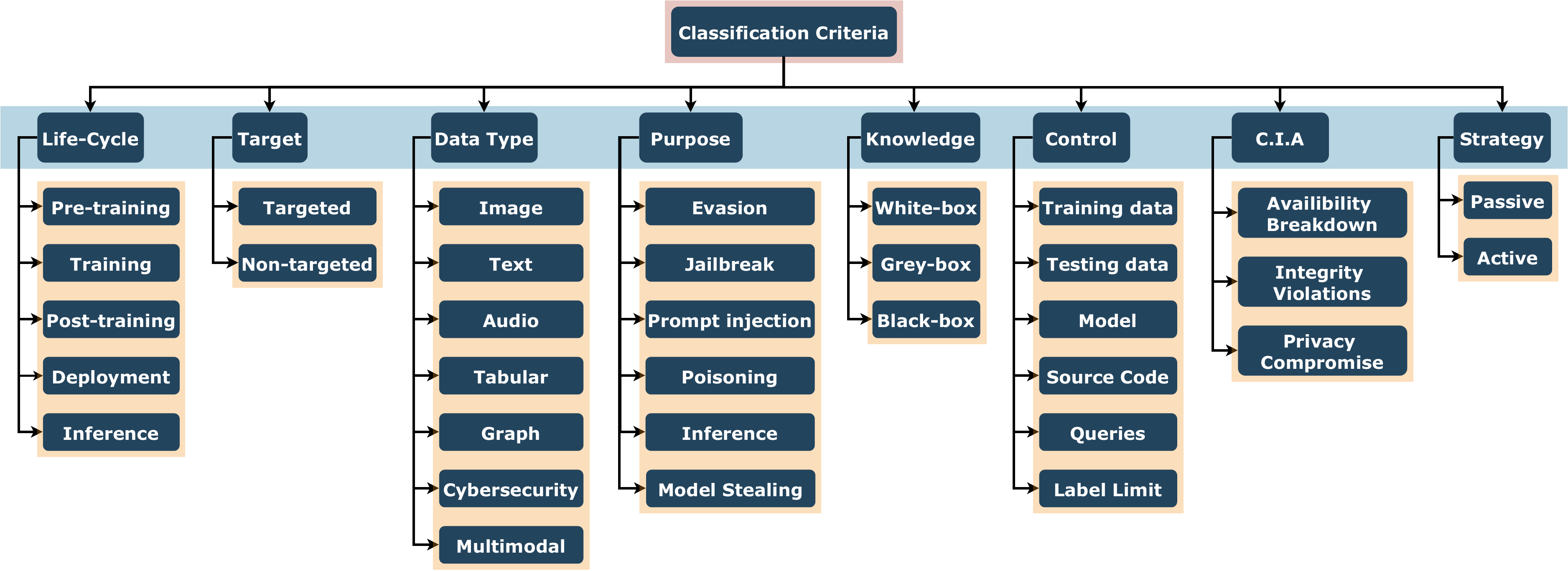}
\caption{Taxonomy of the classification criterias of Adversarial Attacks}
\label{fig:classifications}
\end{figure}

Adversarial Attacks have been classified in various ways in the literature, offering different perspectives on how AAs operate and their potential impact on Machine Learning models. In this section, we have gathered the most common classifications of AAs, based on criterias such as their purpose, target, the attacker's knowledge and strategy, life-cycle stages, CIA\footnote{Confidentiality, Integrity, and Availability} triad, and the type of data and control involved. We depict these classifications in Figure \ref{fig:classifications}, and each one will be discussed in detail in the following subsections.

\subsubsection{Based on the Purpose}
One of the most widely used ways to classify AAs is by analyzing their \textbf{intended purpose} \citep{sidechannelThreatsMachine, visoWhatAdversarial, hdmstuttgartWhenBecomes}. Attacks can be designed either to evade detection by a model or to cause intentional misclassifications, thereby compromising the system’s integrity or exploiting its weaknesses. Based on these overarching objectives, several distinct types of AAs have emerged, including: Evasion attacks \citep{wang2023evasion, li2021backdoor, liu2020survey}, Jailbreak attacks \citep{xu2024comprehensive}, Prompt Injections \citep{liu2023prompt}, Model Inference attacks \citep{hu2022membership}, Model Extraction (Stealing) attacks \citep{gencc2023taxonomic}, and Poisoning/Trojan/Backdoor attacks \citep{tian2022comprehensive, liu2020survey, li2021backdoor}. Each type targets different aspects of an ML system, posing unique challenges to the robustness and security of the models.
\paragraph{\textbf{Evasion attacks:}} In evasion attacks, adversaries craft inputs that evade detection or mislead the model into making incorrect classifications. For instance, small changes in an image may lead a computer vision model to misclassify it, while adversarial inputs in LLMs can bypass content filters \citep{chen2019using, ayub2020model, badr2023novel}.
\paragraph{\textbf{Model jailbreaking:}} In jailbreak attacks, the attacker manipulates the model to bypass restrictions or constraints set by the system such as ethical filters. For example, bypassing content filters in a chatbot by providing a carefully crafted prompt that tricks the model into generating restricted outputs \citep{deng2023jailbreaker, deng2024masterkey}.
\paragraph{\textbf{Prompt injections (PI):}} In prompt injections, the attacker provides maliciously designed prompts that cause the model to follow unintended instructions or generate harmful outputs. Unlike jailbreaking, PI typically involves inserting harmful instructions within regular inputs rather than overriding system-level restrictions \citep{simonwillisonPromptInjection}.
An example is injecting hidden instructions within user input to manipulate a language model's behavior in ways not intended by the developers \citep{liu2023prompt, lee2024prompt, liu2023prompt2}.
\paragraph{\textbf{Model inference:}} In Model (or Membership) Inference attacks, adversaries aim to determine whether specific data was part of the training set. By analyzing model outputs, they can infer sensitive or proprietary information from the training data, posing significant privacy risks \citep{song2019membership, rahman2018membership}.
\paragraph{\textbf{Model Extraction:}} Model Extraction (or Stealing) involves probing a black-box model to reconstruct its functionality or recover sensitive data. For example, an attacker could steal a proprietary financial model by systematically querying it and analyzing its responses \citep{yuan2022attack, shen2022model}.
\paragraph{\textbf{Poisoning/Trojan/Backdoors:}} 
These attacks aims at the integrity of the model during the training phase. The attacker injects malicious data (poisoning) or patterns (trojan, backdoor) into the training set to influence the model's behavior at inference time. For instance, a Poisoning scenario would be introducing mislabeled data to reduce model accuracy \citep{yerlikaya2022data}, and a Trojan scenario would be embedding a hidden trigger in the training data to activate malicious behavior later, such as a trigger that could cause a traffic-light recognition model to classify a red light as a green light in autonomous driving cars \citep{ding2019trojan}.

\subsubsection{Based on the Target}
Adversarial attacks can also be classified based on their target, which refers to whether the attack is aimed at causing a specific or arbitrary misclassification \citep{sidechannelThreatsMachine}.
\paragraph{\textbf{Targeted:}} In targeted attacks, the attacker aims to manipulate the model into misclassifying an input into a specific, incorrect class \citep{carlini2018audio}. For example, an attacker might craft an input to make a stop sign consistently classified as a yield sign.
\paragraph{\textbf{Non-targeted:}} In non-targeted attacks, the goal is to cause the model to misclassify the input, but the specific incorrect class is irrelevant to the attacker \citep{wu2019untargeted}. For instance, an adversarial input could cause a stop sign to be classified as any incorrect traffic sign.

\subsubsection{Based on the Attacker's Knowledge}
Another existing classification is categorizing AAs by the amount of knowledge the attacker has about the target model \citep{oprea2023adversarial, hdmstuttgartWhenBecomes}. These categories typically include white-box, black-box, and sometimes grey-box attacks, although grey-box is not always explicitly classified.
\paragraph{\textbf{White-box:}} In white-box attacks, the attacker has full access to the model's architecture, parameters, and training data, allowing them to exploit the model's gradients for highly effective adversarial examples. For instance, using gradient-based methods, an adversary can precisely manipulate inputs to deceive the model \citep{guo2021gradient, liu2020boosting}.
\paragraph{\textbf{Black-box:}} In black-box attacks, the attacker has no direct access to the model's internals and can only interact with it by sending queries and observing outputs. Despite this limitation, attackers can use techniques like transfer learning, where adversarial examples generated on a surrogate model are used to attack the target model \citep{huang2019black, liu2016delving}.
\paragraph{\textbf{Grey-box:}} In grey-box attacks, the attacker has partial knowledge of the model, such as knowing the architecture but lacking access to the exact parameters or training data. These attacks may combine both white-box and black-box techniques to exploit vulnerabilities effectively \citep{xu2021grey, ma2023grey}.

\subsubsection{Based on the Life-Cycle}
Adversarial attacks can be categorised also by when they occur in the machine learning pipeline, with some references focusing on the training and deployment phases only \citep{oprea2023adversarial}, and others adding phases such as pre-training, post-training, and inference phase \citep{wu2023attacks}.
\paragraph{\textbf{Pre-training:}} Pre-training attacks are conducted before the model training begins, often during the data collection phase. For example, poisoned-data injection into a dataset to compromise the model's integrity once training commences \citep{li2020invisible, liu2022poisonedencoder}.
\paragraph{\textbf{Training:}} Training-phase attacks occur during the actual model training process. A notable example is backdoor injection, where adversaries embed specific triggers in the training data to manipulate the model's behavior later \citep{xie2019dba, du2022ppt}.
\paragraph{\textbf{Post-training:}} Post-training attacks take place immediately after the training process concludes, before the model is deployed. These attacks might involve modifying the model parameters in a way that alters its predictions without detection \citep{qi2021subnet, zhao2019fault}.
\paragraph{\textbf{Deployment:}} Deployment-phase attacks are executed after the model has been deployed on a hardware device, such as a server or mobile device. An example includes modifying model parameters in memory through techniques like bit-flipping, which can lead to unexpected behaviors \citep{Chen_2021_ICCV, bai2021targetedattackdeepneural}.
\paragraph{\textbf{Inference:}} Inference attacks are performed by querying the model with test samples. A specific instance is backdoor activation, where an adversary triggers the model's malicious behaviors by providing inputs that match the previously embedded backdoor conditions \citep{Dong_2018_CVPR, 45818}.

\subsubsection{Based on the CIA Violation}
A fifth classification of AAs is made according to the targeted aspect of the CIA triad, which encompasses confidentiality, integrity, and availability violations \citep{9099439, oprea2023adversarial}.
\paragraph{\textbf{Availability breakdown:}} In availability breakdown attacks, the attacker aims to degrade the model's performance during testing or deployment. This can involve energy-latency attacks that manipulate queries to exhaust system resources, leading to denial of service or reduced responsiveness \citep{9581273, biggio2013poisoningattackssupportvector}.
\paragraph{\textbf{Integrity violations:}} Integrity violation attacks target the accuracy and reliability of the model’s outputs, resulting in incorrect predictions. For instance, poisoning attacks during training can introduce malicious data, causing the model to produce erroneous results when deployed \citep{geiping2021witchesbrewindustrialscale, 8685687}. 
\paragraph{\textbf{Privacy compromise:}} Privacy compromise attacks focus on extracting sensitive information about the model or its training data. Model-extraction attacks exemplify this by allowing an adversary to reconstruct the model’s functionalities or retrieve confidential data used during training \citep{251526, sideChannel}.

\subsubsection{Based on the Type of Control}
Adversarial attacks are classified in other sources based on the type of control the attacker exerts over various elements of the ML model, with some highlighting the control of training and testing data \citep{9099439}, and others \citep{oprea2023adversarial} proposing more aspects of control, such as the control of the model, source code, and queries, as well as a limited control on the data labels.
\paragraph{\textbf{Training data:}} In training data attacks, the attacker manipulates the training dataset by inserting or modifying samples. An example is data poisoning attacks, where malicious inputs are added to influence the model's learning process \citep{wang2018datapoisoningattacksonline}.
\paragraph{\textbf{Testing data:}} Testing data attacks involve altering the input samples during the model's deployment phase. Backdoor poisoning attacks serve as an example, where specific triggers are embedded in the testing data to manipulate the model’s predictions under certain conditions \citep{Saha_Subramanya_Pirsiavash_2020}. 
\paragraph{\textbf{Model:}} Model attacks occur when the attacker gains control over the model's parameters, often by altering the updates applied during training. This can happen in Federated Learning (FL) environments, where malicious model updates are sent to compromise the integrity of the aggregated model \citep{flpoison}.
\paragraph{\textbf{Source code:}} Source code attacks involve modifying the underlying code of the model, which can include changes to third-party libraries, especially those that are open source. This allows attackers to introduce vulnerabilities directly into the model's functionality \citep{codebert}. 
\paragraph{\textbf{Queries:}} Query-based attacks allow the attacker to gather information about the model by submitting various inputs and analyzing the outputs. Black-box evasion attacks exemplify this, as adversaries attempt to craft inputs that evade detection while learning about the model's behavior through its responses \citep{9780921}.
\paragraph{\textbf{Label limit:}} In label limit attacks, the attacker does not have control over the labels associated with the training data. An example is clean-label poisoning attacks, where the adversary influences the model without altering the labels themselves, making detection more difficult \citep{NEURIPS2018_22722a34}. 

\subsubsection{Based on the Type of Data}
An seventh classification of Adversarial attacks is based on the type of data they target, highlighting the diverse methodologies employed across different modalities. Some underline attacks targeting data types as images, text, tabulars, cybersecurity, and even multimodal \citep{oprea2023adversarial}, while other works mention attacks on audio data \citep{8424625}, and graph-based data \citep{pmlr-v80-dai18b}.
\paragraph{\textbf{Image:}} In image-based attacks, the attacker crafts adversarial images designed to cause misclassification. An example includes perturbing images to deceive object detectors or image classifiers, leading to incorrect identification \citep{Subramanya_2019_ICCV}.
\paragraph{\textbf{Text:}} Text attacks involve modifying text inputs to mislead NLP models. For instance, an adversary might introduce typos or antonyms to trick sentiment analysis tools or text classifiers into generating false outputs \citep{Garg_2020}. 
\paragraph{\textbf{Tabular:}} Tabular data attacks target models that operate on structured data, often seen in applications like finance or healthcare. A common example is poisoning attacks, where malicious entries are inserted into tabular datasets to manipulate model behavior \citep{cartella2021adversarialattackstabulardata}.
\paragraph{\textbf{Audio:}} Audio-based attacks involve crafting adversarial noise or altering audio inputs to cause misclassification in systems like voice recognition. For example, specific sound patterns can be designed to mislead voice-activated systems, resulting in incorrect command interpretations \citep{audioattack}.
\paragraph{\textbf{Graphs:}} Graph-based attacks manipulate graph structures and attributes to deceive Graph Neural Networks (GNNs). An attacker might alter edges or node features to induce misclassification or misleading outputs from graph-based models \citep{gnnattack}. 
\paragraph{\textbf{Cybersecurity:}} In the cybersecurity domain, AAs target systems like malware detection or intrusion detection systems. An example is poisoning a spam email classifier, where attackers introduce deceptive emails to degrade the model's performance \citep{spamdetect}.
\paragraph{\textbf{Multimodal:}} Multimodal attacks involve exploiting systems that integrate multiple data types. In these cases, attackers might gain insights by submitting queries that encompass different modalities, such as text and image combinations \citep{wu2024adversarialattacksmultimodalagents}.

\subsubsection{Based on the Strategy}
Last but not least, Adversarial attacks can also be categorized based on the strategy employed by the attacker, distinguishing between passive and active approaches \citep{9099439}.
\paragraph{\textbf{Passive:}} In passive attacks, the attacker seeks to gather information about the application or its users without actively interfering with the system's operation. An example is reverse engineering, where an adversary analyzes a black-box classifier to extract its functionalities and gain insights into its behavior \citep{CHIANG1994107}.
\paragraph{\textbf{Active:}} Active attacks are designed to disrupt the normal functioning of an application. The attacker may implement poisoning attacks that introduce malicious inputs, aiming to trigger misclassifications or degrade the model's performance during operation \citep{NEURIPS2020_8ce6fc70}.

\section{Adversarial Attacks on LLMs}
\label{sec:aas-llms}

In recent years, LLMs have been increasingly targeted by AAs \citep{shayegani2023survey, kumar2024adversarial, YAO2024100211}, posing various threats to their reliability, safety, and security. These attacks can take multiple forms and serve distinct purposes, each exploiting different vulnerabilities within the model or its deployment. In this section, we present a comprehensive taxonomy of 56 recent AAs targeting LLMs, following the purpose-based classification of AA (refer to Section \ref{sec:adversarial-attacks}). We consider 7 types of AAs: White-box Jailbreak attacks, Black-box Jailbreak attack, Prompt Injection, Evasion Attacks, Model Extraction, Model Inference, and Poisoning/Trojan/Backdoor. Each attack type includes 8 prominent examples, which are detailed in the following subsections.

\subsection{Jailbreak Attacks}
The type of AAs that we begin with are model Jailbreaking attacks, which are designed to bypass safety measures. We consider two approaches in jailbreak attacks according to the \textbf{targeted model}: White-box, and Black-box model jailbreaking.

\subsubsection{White-box attacks}
\label{subsec:whitebox-jailbreak}
The first type are White-box Jailbreak attacks, where the attacker has \textbf{complete} access to the model's architecture, parameters, and training data. This level of knowledge allows the attacker to design specific inputs that exploit vulnerabilities in the model, often related to the model gradients, in order to bypass its restrictions or safety measures. 

\subsubsection{Black-box attacks}
\label{subsec:blackbox-jailbreak}
The second type of attacks are Black-box Jailbreak attack, in which, in contrast to white-box attacks, the attacker has \textbf{no access} to the model's internal workings or training data. Instead, the attacker can only interact with the model by providing inputs and observing the outputs, often relying on trial and error to discover effective prompts able to bypass the model's safeguards.

\noindent We present in Table \ref{tab:jailbreaks} a list of recent white-box and black-box jailbreak attacks existing in the literature \citep{githubLlmspMain}, and if they are Open Source (OS) or not, as each has a different strategy and implementation.

\begin{table}[h]
\centering
\caption{Examples of jailbreak attacks against LLMs}
\label{tab:jailbreaks}
\adjustbox{max width = \textwidth}{
\begin{tabular}{c c c c}%

    \hline
    
    \multirow{2}{*}{\textbf{Type}} & \multirow{2}{*}{\textbf{Attack}} & \multirow{2}{*}{\textbf{Concept}} &  \multirow{2}{*}{\textbf{OS?}} \cr \\
     \hline

     \multirow{16}{*}{\textbf{W-box}} & \multirow{2}{*}{\textbf{GCG \citep{zou2023universaltransferableadversarialattacks}}} & \multirow{2}{*}{Adding adversarial suffixes using greedy and gradient-based searches}  & \multirow{2}{*}{\checkmark} \cr \\
    \cline{2-4}

    & \multirow{2}{*}{\textbf{Visual Mod. \citep{niu2024efficientllmjailbreakingintroducingvisual}}} & \multirow{2}{*}{Jailbreaking an LLM using a corresponding Multimodal LLM} & \multirow{2}{*}{\ding{55}} \cr \\
    \cline{2-4}

    & \multirow{2}{*}{\textbf{PGD \citep{geisler2024attackinglargelanguagemodels}}} & \multirow{2}{*}{Jailbreaking attack using Projected Gradient Descent} & \multirow{2}{*}{\ding{55}} \cr  \\
    \cline{2-4}

    & \multirow{2}{*}{\textbf{SCAV \citep{xu2024uncoveringsafetyriskslarge}}} &  \multirow{2}{*}{Guiding Jailbreak attacks against white-box LLMs} & \multirow{2}{*}{\ding{55}} \cr \\
    \cline{2-4}

    & \multirow{2}{*}{\textbf{Soft Prp. \citep{schwinn2024softpromptthreatsattacking}}}  & \multirow{2}{*}{Attacking the continuous embedding representation of input tokens} & \multirow{2}{*}{\checkmark} \cr \\
    \cline{2-4}

    & \multirow{2}{*}{\textbf{DrAttack \citep{li2024drattackpromptdecompositionreconstruction}}} &  \multirow{2}{*}{Decomposition and Reconstruction of prompts for LLM jailbreaking} & \multirow{2}{*}{\checkmark} \cr \\
    \cline{2-4}

    & \multirow{2}{*}{\textbf{RADIAL \citep{du2024analyzinginherentresponsetendency}}} & \multirow{2}{*}{Generating instructions based on LLMs' Inherent Response Tendency} &  \multirow{2}{*}{\ding{55}} \cr \\
    \cline{2-4}

    & \multirow{2}{*}{\textbf{ReNeLLM \citep{ding2024wolfsheepsclothinggeneralized}}}  &  \multirow{2}{*}{Using generalized and nested jailbreak prompts to fool LLMs} & \multirow{2}{*}{\checkmark} \cr \\
    \hline

    \multirow{16}{*}{\textbf{B-box}} & \multirow{2}{*}{\textbf{PAIR \citep{chao2024jailbreakingblackboxlarge}}} & \multirow{2}{*}{Automatic jailbreaking of black box LLMs}  & \multirow{2}{*}{\ding{55}} \cr \\
    \cline{2-4}

    & \multirow{2}{*}{\textbf{Privacy att. \citep{li2023multistepjailbreakingprivacyattacks}}}  &  \multirow{2}{*}{Extracting people-information memorised by GPT-4o} & \multirow{2}{*}{\checkmark} \cr \\
    \cline{2-4}

    & \multirow{2}{*}{\textbf{DAN \citep{shen2024donowcharacterizingevaluating}}} & \multirow{2}{*}{Tricking GPT-4o to break its policies with a role-play} & \multirow{2}{*}{\checkmark} \cr \\
    \cline{2-4}

    & \multirow{2}{*}{\textbf{Ad. Att. \citep{andriushchenko2024jailbreakingleadingsafetyalignedllms}}} &  \multirow{2}{*}{Adding adversarial suffixes using random searches} & \multirow{2}{*}{\checkmark} \cr \\
    \cline{2-4}

    & \multirow{2}{*}{\textbf{GCQ \citep{hayase2024querybasedadversarialpromptgeneration}}}  & \multirow{2}{*}{Enhancing GCG algorithm using best-first search algorithm} & \multirow{2}{*}{\ding{55}} \cr \\
    \cline{2-4}

    & \multirow{2}{*}{\textbf{PAL \citep{sitawarin2024palproxyguidedblackboxattack}}} &  \multirow{2}{*}{Token-level attack using gradients from an open-source proxy} & \multirow{2}{*}{\checkmark} \cr  \\
    \cline{2-4}

    & \multirow{2}{*}{\textbf{IRIS \citep{ramesh2024gpt4jailbreaksnearperfectsuccess}}} & \multirow{2}{*}{Using the same LLM to target itself} &  \multirow{2}{*}{\ding{55}} \cr \\
    \cline{2-4}

    & \multirow{2}{*}{\textbf{Tastle \citep{xiao2024distractlargelanguagemodels}}}  &  \multirow{2}{*}{Framework of black-box jailbreak for automated red-teaming} & \multirow{2}{*}{\ding{55}} \cr \\
    \hline

\end{tabular}
}
\end{table}

\subsection{Prompt Injection}
\label{subsec:prompt-inject}
The third type of attacks that we illustrate are Prompt injections,  where the adversary manipulates the input prompts and queries to deceive the model into producing unintended or harmful outputs. This technique is ranked among the most dangerous attacks against LLMs by \citet{owaspOWASPLarge}. To illustrate the diverse strategies attackers employ to exploit LLMs with PIs, we have gathered eight different attacks, utilizing both direct injections, where the attacker append a malicious input to a prompt, and indirect injection methods, where the attacker append malicious prompts through file or external inputs. These attacks are presented in Table \ref{tab:prompt-injections}
\begin{table}[h]
\centering
\caption{Examples of prompt injection attacks against LLMs}
\label{tab:prompt-injections}
\adjustbox{max width = \textwidth}{
\begin{tabular}{c c c}%

    \hline
    
    \multirow{2}{*}{\textbf{Attack}}  & \multirow{2}{*}{\textbf{Concept}} &  \multirow{2}{*}{\textbf{OS?}} \cr \\
    \hline

    \multirow{2}{*}{\textbf{Ign. Pp. \citep{perez2022ignorepreviouspromptattack}}} & \multirow{2}{*}{A direct prompt injection technique to mislead the LLM in ignoring instructions} & \multirow{2}{*}{\checkmark} \cr \\
    \hline

    \multirow{2}{*}{\textbf{Ind. PI \citep{greshake2023youvesignedforcompromising}}}  &  \multirow{2}{*}{An indirect prompt injection technique through file input to compromise LLMs} & \multirow{2}{*}{\checkmark} \cr  \\
    \hline

    \multirow{2}{*}{\textbf{Frm. PI \citep{liu2024formalizingbenchmarkingpromptinjection}}} & \multirow{2}{*}{General framework for formalizing prompt injection in LLMs} & \multirow{2}{*}{\checkmark} \cr \\
    \hline

    \multirow{2}{*}{\textbf{Mlt. PI \citep{bagdasaryan2023abusingimagessoundsindirect}}} &  \multirow{2}{*}{Using images and sounds for indirect prompt injection in multi-modal LLMs} & \multirow{2}{*}{\checkmark} \cr \\
    \hline

    \multirow{2}{*}{\textbf{Unv. PI \citep{liu2024automaticuniversalpromptinjection}}}  & \multirow{2}{*}{An automatic and indirect prompt injection attack} & \multirow{2}{*}{\checkmark} \cr \\
    \hline

    \multirow{2}{*}{\textbf{Vrt. PI \citep{yan2024backdooringinstructiontunedlargelanguage}}} & \multirow{2}{*}{Backdooring a prompt injection under a triggered scenario} & \multirow{2}{*}{\checkmark} \cr \\
    \hline

    \multirow{2}{*}{\textbf{Chat Tmp. \citep{wei2024hiddenplainsightexploring}}} & \multirow{2}{*}{Creating misleading contexts acceptance elicitation and word anonymization} &  \multirow{2}{*}{\ding{55}} \cr \\
    \hline

    \multirow{2}{*}{\textbf{JudgeDeceiver \citep{shi2024optimizationbasedpromptinjectionattack}}} & \multirow{2}{*}{Deceiving LLM-as-a-Judge to choose a response among multiple choices} & \multirow{2}{*}{\ding{55}} \cr \\
    \hline

\end{tabular}
}
\end{table}

\subsection{Evasion Attacks}
\label{subsec:evasion-attacks}
The forth type of attacks we illustrate are Evasion attacks, in which attackers aim to deceive language models by crafting inputs designed to bypass detection or classification. These attacks often target sentiment analysis and text classification models, seeking to manipulate their outputs through subtle modifications. In Table \ref{tab:evasion-attacks}, we have gathered eight different examples and techniques of evasion attacks presented in the literature, some of which employ text perturbations to alter the original input, while others leverage LLMs to generate sophisticated evasion samples against their counterparts.
\begin{table}[h]
\centering
\caption{Examples of evasion attacks against LLMs}
\label{tab:evasion-attacks}
\adjustbox{max width = \textwidth}{
\begin{tabular}{c c c}%

    \hline
    
    \multirow{2}{*}{\textbf{Attack}}  & \multirow{2}{*}{\textbf{Concept}} & \multirow{2}{*}{\textbf{OS?}} \cr \\
    \hline

    \multirow{2}{*}{\textbf{Hot-Flip \citep{ebrahimi2018hotflipwhiteboxadversarialexamples}}} & \multirow{2}{*}{Flipping letters in a word to mislead the LLM to make incorrect classifications} & \multirow{2}{*}{\ding{55}} \cr \\
    \hline

    \multirow{2}{*}{\textbf{PWWS \citep{ren-etal-2019-generating}}}  &  \multirow{2}{*}{Changing some words with their synonyms to mislead text classification tasks} & \multirow{2}{*}{\checkmark} \cr \\
    \hline

    \multirow{2}{*}{\textbf{Typo-Att. \citep{pruthi2019combatingadversarialmisspellingsrobust}}} & \multirow{2}{*}{Preforming character-level perturbations on a QWERTY keyboard} & \multirow{2}{*}{\checkmark} \cr \\
    \hline

   \multirow{2}{*}{\textbf{VIPER \citep{eger2020textprocessinglikehumans}}} &  \multirow{2}{*}{Changing some letters to symbols in harmful words to avoid detection} & \multirow{2}{*}{\checkmark} \cr \\
    \hline

    \multirow{2}{*}{\textbf{Checklist \citep{ribeiro-etal-2020-beyond}}}  & \multirow{2}{*}{Performing Word-level perturbations using a predifined word checklist} & \multirow{2}{*}{\checkmark} \cr \\
    \hline

    \multirow{2}{*}{\textbf{BERT-Att. \citep{li-etal-2020-bert-attack}}} & \multirow{2}{*}{Using BERT to generate adversarial samples against other LLMs} & \multirow{2}{*}{\checkmark} \cr \\
    \hline

    \multirow{2}{*}{\textbf{GBDA \citep{guo2021gradientbasedadversarialattackstext}}} & \multirow{2}{*}{Gradient-based white box attack using words flipping to mislead text classifiers} &  \multirow{2}{*}{\checkmark} \cr \\
    \hline

    \multirow{2}{*}{\textbf{TF-Att. \citep{li2024tfattacktransferablefastadversarial}}}  &  \multirow{2}{*}{Generating adversarial examples with critical units of sentences using LLMs} & \multirow{2}{*}{\ding{55}} \cr \\
    \hline

\end{tabular}
}
\end{table}

\subsection{Model Extraction}
\label{subsec:model-extraction}
Model extraction attacks are the fifth type we illustrate in this section. These attacks aim to recreate or steal a language model's functionality by querying it and using the responses to reconstruct the model, this poses a significant threat as they allow adversaries to duplicate proprietary models without access to their internal details. We present below in Table \ref{tab:extraction-attacks}, eight examples of Model Extraction attacks, showcasing different methods adversaries use to probe black-box LLMs and either extract training data of the model, or precise personal information of users.
\begin{table}[h]
\centering
\caption{Examples of model extraction attacks against LLMs}
\label{tab:extraction-attacks}
\adjustbox{max width = \textwidth}{
\begin{tabular}{c c c}%

    \hline
    
    \multirow{2}{*}{\textbf{Attack}} & \multirow{2}{*}{\textbf{Concept}} &  \multirow{2}{*}{\textbf{OS?}} \cr \\
    \hline

    \multirow{2}{*}{\textbf{User Extr. \citep{274574}}} & \multirow{2}{*}{Extracting personal data of users memorised by LLMs using model queries} & \multirow{2}{*}{\ding{55}} \cr \\
    \hline

    \multirow{2}{*}{\textbf{LLM Tricks \citep{yu2023bagtrickstrainingdata}}}  &  \multirow{2}{*}{Tricks to enhance data extraction capabilities on LLMs} & \multirow{2}{*}{\checkmark} \cr  \\
    \hline

    \multirow{2}{*}{\textbf{PII Leakage \citep{lukas2023analyzingleakagepersonallyidentifiable}}} & \multirow{2}{*}{Extraction/Inference attacks for analysing personally identifiable information (PII)} & \multirow{2}{*}{\checkmark} \cr \\
    \hline

    \multirow{2}{*}{\textbf{ETHICIST \citep{zhang2023ethicisttargetedtrainingdata}}} & \multirow{2}{*}{Data extraction with Loss Smoothed Soft Prompting} & \multirow{2}{*}{\checkmark} \cr \\
    \hline

    \multirow{2}{*}{\textbf{Scalable Extr. \citep{nasr2023scalableextractiontrainingdata}}}  &  \multirow{2}{*}{Extracting training data from Production LLMs} & \multirow{2}{*}{\ding{55}} \cr \\
    \hline
    
    \multirow{2}{*}{\textbf{Output2Prompt \citep{zhang2024extractingpromptsinvertingllm}}} & \multirow{2}{*}{Extracting user prompts by knowing only their outputs} & \multirow{2}{*}{\checkmark} \cr \\
    \hline

   \multirow{2}{*}{\textbf{PII Compass \citep{nakka2024piicompassguidingllmtraining}}} &  \multirow{2}{*}{Extracting phone numbers from LLM using black-box queries} & \multirow{2}{*}{\ding{55}} \cr \\
    \hline

    \multirow{2}{*}{\textbf{Alpaca-Vicuna \citep{kassem2024alpacavicunausingllms}}}  & \multirow{2}{*}{Using an  LLM to perform data extraction on another LLM} & \multirow{2}{*}{\ding{55}} \cr \\
    \hline

\end{tabular}
}
\end{table}

\subsection{Model Inference}
\label{subsec:model-inference}
Model inference (or Membership Inference) are the sixth type of attacks we focus on in this study. These attacks determine whether specific data samples, especially sensitive information, were part of the training set of an LLM. These attacks can compromise the privacy of users or organizations by revealing training data patterns. We gathered in Table \ref{tab:inference-attacks} eight examples of model inference attacks, which demonstrate how attackers exploit LLMs to infer confidential training data and gain insights into the model's behavior.
\begin{table}[h]
\centering
\caption{Examples of inference attacks against LLMs}
\label{tab:inference-attacks}
\adjustbox{max width = \textwidth}{
\begin{tabular}{c c c}%

    \hline
    
    \multirow{2}{*}{\textbf{Attack}} & \multirow{2}{*}{\textbf{Concept}} &  \multirow{2}{*}{\textbf{OS?}} \cr \\
    \hline

    \multirow{2}{*}{\textbf{LIRA \citep{carlini2022membershipinferenceattacksprinciples}}} & \multirow{2}{*}{Combining difficulty scores and well-Calibrated Gaussian Likelihood Estimate} & \multirow{2}{*}{\checkmark} \cr \\
    \hline

    \multirow{2}{*}{\textbf{Ngb. Comp. \citep{mattern2023membershipinferenceattackslanguage}}}  & \multirow{2}{*}{Detecting training data using neighbor text comparison} & \multirow{2}{*}{\ding{55}} \cr \\
    \hline

    \multirow{2}{*}{\textbf{PII Leakage \citep{lukas2023analyzingleakagepersonallyidentifiable}}} & \multirow{2}{*}{Extraction/Inference attacks for analysing PII leakage} & \multirow{2}{*}{\checkmark} \cr \\
    \hline

    \multirow{2}{*}{\textbf{Data Detect. \citep{shi2024detectingpretrainingdatalarge}}} & \multirow{2}{*}{Detecting pretraining samples of an LLM using minimal probabilities} &  \multirow{2}{*}{\checkmark} \cr \\
    \hline

    \multirow{2}{*}{\textbf{ProPILE \citep{kim2023propileprobingprivacyleakage}}} & \multirow{2}{*}{Probing framework to assess the likelihood of a PII in the training set} & \multirow{2}{*}{\ding{55}} \cr \\
    \hline
    
    \multirow{2}{*}{\textbf{MIA-LLM \citep{fu2024practicalmembershipinferenceattacks}}} & \multirow{2}{*}{Membership Inference based on Self-calibrated Probabilistic Variation} & \multirow{2}{*}{\checkmark} \cr \\
    \hline

   \multirow{2}{*}{\textbf{DeCop \citep{duarte2024decopdetectingcopyrightedcontent}}} & \multirow{2}{*}{Detecting copyrighted content in training sets using multiple-choice questions} & \multirow{2}{*}{\checkmark} \cr \\
    \hline

    \multirow{2}{*}{\textbf{ConRecall \citep{wang2024conrecalldetectingpretrainingdata}}}  & \multirow{2}{*}{Using Contrastive Decoding to detect LLM's pre-training data} & \multirow{2}{*}{\checkmark} \cr \\
    \hline

\end{tabular}
}
\end{table}

\subsection{Poisoning/Trojan/Backdoors}
\label{subsec:poisoning}
The last attacks on LLM we show are Poisoning, Trojan, and Backdoor attacks, which involve injecting malicious data or hidden triggers during the training phase of an LLM. This can lead to incorrect or dangerous behavior at deployment, allowing attackers to manipulate the model's responses. We have compiled in Table \ref{tab:poisoning-attacks} eight examples of these attacks, where adversaries either corrupt the training process with poisoned data, or plant triggers to exploit models during inference, demonstrating the serious risks these methods pose to LLMs.
\begin{table}[h]
\centering
\caption{Examples of poisoning, trojan, and backdoor attacks against LLMs}
\label{tab:poisoning-attacks}
\adjustbox{max width = \textwidth}{
\begin{tabular}{c c c}%

    \hline
    
    \multirow{2}{*}{\textbf{Attack}}  & \multirow{2}{*}{\textbf{Concept}} &  \multirow{2}{*}{\textbf{OS?}} \cr \\
    \hline

    \multirow{2}{*}{\textbf{TrojLLM \citep{NEURIPS2023_cf04d01a}}} & \multirow{2}{*}{Inserting Trojans into text prompts in black-box LLM APIs} & \multirow{2}{*}{\checkmark} \cr \\
    \hline

    \multirow{2}{*}{\textbf{Bst-of-Vnm. \citep{baumgärtner2024bestofvenomattackingrlhfinjecting}}}  & \multirow{2}{*}{Attacking RLHF by injecting Poisoned Preference Data} & \multirow{2}{*}{\ding{55}} \cr \\
    \hline

    \multirow{2}{*}{\textbf{CodeBreaker \citep{yan2024llmassistedeasytotriggerbackdoorattack}}} & \multirow{2}{*}{Instering Backdoors on code-completion LLMs to sugget vulnerable code} & \multirow{2}{*}{\checkmark} \cr \\
    \hline

    \multirow{2}{*}{\textbf{Rtv. Poison. \citep{zhang2024humanimperceptibleretrievalpoisoningattacks}}} & \multirow{2}{*}{Misleading LLMs during the RAG process with malicious documents} & \multirow{2}{*}{\ding{55}} \cr \\
    \hline

    \multirow{2}{*}{\textbf{Clinical LLM \citep{clinicalLLM}}} & \multirow{2}{*}{Editing LLMs to reveal serious implications in clinical settings} & \multirow{2}{*}{\ding{55}} \cr \\
    \hline
    
    \multirow{2}{*}{\textbf{BackdoorLLM \citep{li2024backdoorllmcomprehensivebenchmarkbackdoor}}} & \multirow{2}{*}{Comprehensive benchmark for studying backdoor attacks on LLMs} & \multirow{2}{*}{\checkmark} \cr \\
    \hline

   \multirow{2}{*}{\textbf{CBA \citep{huang2024compositebackdoorattackslarge}}} & \multirow{2}{*}{Composite Backdoor Attacks against LLMs} & \multirow{2}{*}{\checkmark} \cr \\
    \hline

    \multirow{2}{*}{\textbf{TA² \citep{wang2024trojanactivationattackredteaming}}}  & \multirow{2}{*}{Injecting trojan steering vectors into the activation layers of LLMs} & \multirow{2}{*}{\checkmark} \cr \\
    \hline

\end{tabular}
}
\end{table}

\section{Classification of Adversarial attacks on LLMs based on their danger level}
\label{sec:danger-level}
After presenting the existing classifications of AAs and some of the most-recent attacks against LLMs, we propose in this section a new criterion for classifying AAs on LLMs. We present the idea and methodology in the following subsections.

\subsection{Principle}
Seeing the list of AAs on LLMs presented in Section \ref{sec:aas-llms} and how frequent they are, one question that comes across the mind is what attacks should be \textbf{mitigated first} to secure LLMs? In order to answer this question, we need to rank the available attacks based on their \textbf{danger level} against LLMs in order to know what attacks is a model most-vulnerable to. This can be done by calculating the vulnerability score those of attacks using Vulnerability-assessment metrics \citep{Shah2015}.
\subsection{Vulnerability-assessment Metrics}
\label{subsec:scoring-systems}
Vulnerability assessment metrics are critical tools for evaluating and ranking potential security threats based on their severity and likelihood of exploitation. Various methodologies, such as DREAD \citep{michael2006security}, CVSS \citep{schiffman2005complete}, OWASP Risk Rating \citep{owaspOWASPRisk}, and SSVC \citep{spring2021prioritizing}, provide frameworks for assessing vulnerabilities by considering different factors, including technical attributes, potential impacts, and contextual elements. By systematically analyzing attacks based on their danger, these assessment tools facilitate informed decision-making in an ever-evolving threat landscape, allowing organizations to strengthen their security posture and better protect their assets.
\begin{figure}[h]
\centering 
\includegraphics[width=1\textwidth]{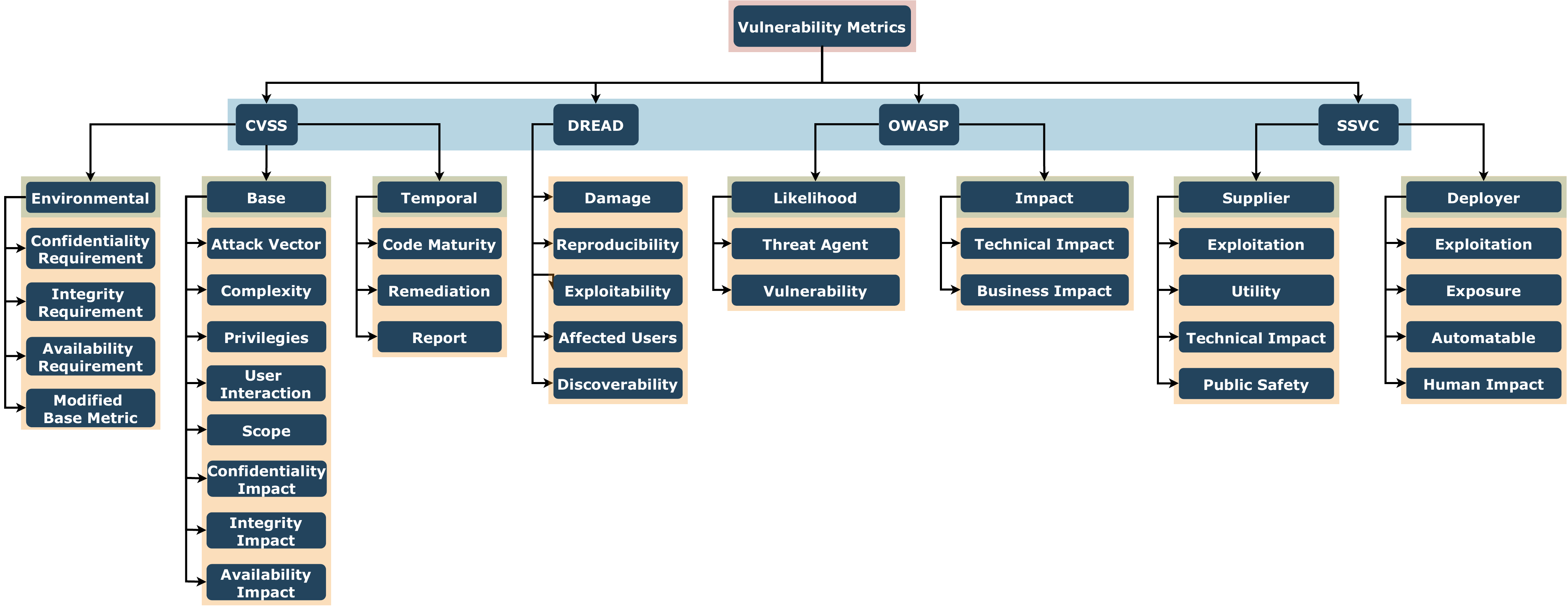}
\caption{Examples of known vulnerability assessment metrics}
\label{fig:metrics}
\end{figure}

\subsubsection{DREAD \citep{michael2006security}} \label{subsubsec:dread}
Originally developed by Microsoft, DREAD is a qualitative risk assessment model that ranks, prioritizes, and evaluates the severity of vulnerabilities and potential threats based on five factors: Damage potential (D), Reproducibility (R), Exploitability (E), Affected users (A), and Discoverability (D) of the attack.

\paragraph{Calculations}
The vulnerability score is calculated with DREAD as an average score of the five factors, each assessed with a value out of 10. The details of each factor and their values are shown in Table \ref{tab:factors-dread} below. 

\begin{table}[h]
\centering
\caption{Metric factors of DREAD \citep{michael2006security}}
\label{tab:factors-dread}
\begin{tabular}{c c c}

    \hline
    
    \multirow{2}{*}{\textbf{Factor}}  & \multirow{2}{*}{\textbf{Definition}} &  \multirow{2}{*}{\textbf{Values}} \cr \\
    \hline

    \multirow{2}{*}{\textbf{Damage Potential (D)}} & \multirow{2}{*}{How much damage can be caused} & \multirow{10}{*}{[1 (Low), 10 (High)]} \cr \\
    \cline{1-2}

    \multirow{2}{*}{\textbf{Reproducibility (R)}} & \multirow{2}{*}{How easy is it to reproduce the attack}   \cr \\
     \cline{1-2}

    \multirow{2}{*}{\textbf{Exploitability (E)}} & \multirow{2}{*}{How easy is it to exploit the vulnerability}   \cr \\
    \cline{1-2}

    \multirow{2}{*}{\textbf{Affected Users (A)}} & \multirow{2}{*}{How many users would be affected}  \cr \\
    \cline{1-2}

    \multirow{2}{*}{\textbf{Discoverability (D)}} & \multirow{2}{*}{How easy is it to discover the vulnerability}  \cr \\
    \hline

\end{tabular}
\end{table}

A value in the range $[1, 4[$ is labeled as \textbf{`Low'} in the level of criticality, a value in the range $[4, 7[$ labeled as \textbf{`Medium'} in criticality, and values over 7 are labeled as \textbf{`High'} in criticality.
The final score is calculated following this equation:
\begin{equation}
    Score  = (D + R + E + A + D) / 5    
\end{equation}

\paragraph{Limitations}
The DREAD model, previously popular for qualitative risk assessment, has several limitations that have reduced its use in favor of more structured frameworks. First of all, its five categories (Damage, Reproducibility, Exploitability, Affected Users, and Discoverability) are highly subjective, leading to inconsistent scoring and prioritization across different assessors and organizations. Moreover, DREAD overlooks contextual factors like the specific environment and business impact, limiting its adaptability for complex needs. Finally, it also fails to account for dynamic threats or mitigation measures, making it less effective for ongoing risk management.

\subsubsection{CVSS (Common Vulnerability Scoring System) \citep{schiffman2005complete}} \label{subsubsec:cvss}
Created by the FIRST\footnote{\url{https://www.first.org/}} (Forum of Incident Response and Security Teams), the CVSS is an industry-standard scoring system for rating the severity of software vulnerabilities out of 10. It is encompasses three main metrics:
\begin{itemize}
    \item \textbf{Base Metrics:} Represent the vulnerabilities that are constant over time. It contains factor related to the exploitability of an attack (how easy it is to exploit the vulnerability) like the Attack Vector (AV), Attack Complexity (AC), Privileges Required (PR), User Interaction (UI), and the Scope (S) of the attack.
    And factors related to the impact of an attack on the CIA triad, such as Confidentiality Impact (C), Integrity Impact (I), and Availability Impact (A).
    \item \textbf{Temporal Metrics (Optional):} Represent the vulnerabilities that might change over time in order to update the base score, it encompasses three factors, Exploit Code Maturity (E), Remediation Level (RL), and Report Confidence (RC).
    \item \textbf{Environmental Metrics (Optional):} Vulnerabilities that are unique to a user environment, such as the Confidentiality Requirements (CR), Integrity Requirement (IR), Availability Requirement (AR), and the modified Base Metrics.
\end{itemize}

\paragraph{Calculations}
The values in CVSS factors are not explicitly numerical; but selected from a specific range of choices, with each qualitative value having a corresponding coefficient. The details of each factor and of the Base Metric and their values according to CVSS version 3.1\footnote{Although version 4.0 is the most recent, version 3.1 is still the most used in vulnerability assessment.} are presented below in Table \ref{tab:factors-cvss}, and their equivalent decimal values are detailed in Table \ref{tab:values-cvss}.

\begin{table}[h]
\centering
\caption{Base metric factors of CVSS 3.1 \citep{firstCVSSV31}}
\label{tab:factors-cvss}
\begin{tabular}{
               >{\centering\arraybackslash} p{0.22\textwidth}%
                >{\centering\arraybackslash}p{0.4\textwidth}%
                >{\centering\arraybackslash}p{0.29\textwidth}}%

    \hline
    
    \multirow{2}{*}{\textbf{Factor}}  & \multirow{2}{*}{\textbf{Definition}} &  \multirow{2}{*}{\textbf{Values}} \cr \\
    \hline

    {\textbf{Attack Vector (AV)}} & \multirow{2}{*}{From where the exploitation is possible} &{Network (N), Adjacent (A), Local (L), Physical (P)}  \\
    \hline

    {\textbf{Attack Complexity (AC)}} & \multirow{2}{*}{How complex is the exploitation} & \multirow{2}{*}{Low (L), High (H)}  \\
    \hline

    {\textbf{Privileges Required (PR)}} & {How much privileges are needed for the exploit} & \multirow{2}{*}{None (N), Low (L), High (H)}  \\
    \hline

    {\textbf{User Interaction (UI)}} & {Is a user interaction required in the compromise} & \multirow{2}{*}{None (N), Required (R)}  \\
    \hline

    \multirow{2}{*}{\textbf{Scope (S)}} & \multirow{2}{*}{Does the scope of the attack change} & \multirow{2}{*}{Unchanged (U), Changed (C)}  \cr \\
    \hline

    {\textbf{Confidentiality Impact (C)}} & \multirow{2}{*}{How much impacted is the confidentiality} & \multirow{5}{*}{None (N), Low (L), High (H)}  \\
    \cline{1-2}

    {\textbf{Integrity Impact (I)}} & {How much impacted is the integrity}  \\
    \cline{1-2}

    {\textbf{Availability Impact (A)}} & \multirow{2}{*}{How much impacted is the availability}  \\
    \hline

\end{tabular}
\end{table}

\begin{table}[h]
\centering
\caption{Numerical values of each CVSS 3.1 factor \citep{firstCVSSV31}}
\label{tab:values-cvss}
\begin{tabular}{c c c}%

    \hline
    
    \multirow{2}{*}{\textbf{Factor}}  & \multirow{2}{*}{\textbf{Value}} &  \multirow{2}{*}{\textbf{Decimal Value}} \cr \\
    \hline

    \multirow{4}{*}{\textbf{AV}} & {Network (N)} & {0.85}  \\
    \cline{2-3}

    & {Adjacent (A)} & {0.62}  \\
     \cline{2-3}

    & {Local (L)} & {0.55}  \\
    \cline{2-3}
    
    & {Physical (P)} & {0.22}  \\
    \hline
    
    \multirow{2}{*}{\textbf{AC}} & {Low (L)} & {0.77}  \\
    \cline{2-3}

    & {High (H)} & {0.44}  \\
    \hline

    \multirow{3}{*}{\textbf{PR}} & {None (N)} & {0.85}  \\
    \cline{2-3}

    & Low (L) & {0.62 (if S = U), 0.68 (if S = C)}  \\
    \cline{2-3}

    & {High (H)} & {0.27 (if S = U), 0.50 (if S = C)}  \\
    \hline

    \multirow{2}{*}{\textbf{UI}} & {None (N)} & {0.85}  \\
    \cline{2-3}

    & {Required (R)} & {0.62}  \\
    \hline

    \multirow{3}{*}{\textbf{C, I ,A}} & {None (N)} & {0.00}  \\
    \cline{2-3}

    & {Low (L)} & {0.22}  \\
    \cline{2-3}

    & {High (H)} & {0.56}  \\
    \hline

\end{tabular}
\end{table}

After assessing a value for each metric, the Base Score of the CVSS is calculated using two different equations depending on the Scope (S), which is either Changed (C) or Unchanged (U). Below are the full details of the equations for both cases: 

\begin{itemize}
    \item [] \textbf{If S = U:}
        \begin{equation}
            \text{Base Score} = \text{roundup}\left(\min\left( \text{Impact}_{U} + \text{Exploitability}, 10 \right)\times 1.08\right)
        \end{equation}
        \begin{equation}
            \text{Impact}_{U} = 6.42 \times \left( 1 - (1 - C) \times (1 - I) \times (1 - A) \right)
        \end{equation}
        \begin{equation}
            \text{Exploitability} = 8.22 \times \text{AV} \times \text{AC} \times \text{PR} \times \text{UI}
        \end{equation}
        
    \item [] \textbf{If S = C:}
        \begin{equation}
            \text{Base Score} = \text{roundup}\left(\min\left( 1.08 \times (\text{Impact}_{C} + \text{Exploitability}), 10 \right)\right)
        \end{equation}
        \begin{equation}
            \text{Impact}_{C} = 7.52 \times \left( I - 0.029 \right) - 3.25 \times \left( I - 0.02 \right)^{15}
        \end{equation}
        \begin{equation}
            I = 1 - (1 - C) \times (1 - I) \times (1 - A)
        \end{equation}
        The exploitability remains the same. 
\end{itemize}
The final Base Score ranges from 0 to 10, with the same criticality assignment as in DREAD, adding to it that a base score of 9 or more is considered a \textbf{`Critical'} vulnerability.

\paragraph{Limitations}
CVSS is a widely used standard for scoring vulnerabilities but has several limitations that affect its real-world effectiveness. Firstly, it tends to oversimplify calculations by focusing on technical aspects like attack complexity and impacts on confidentiality, integrity, and availability, while neglecting business impact and regulatory considerations. Additionally, the Temporal score of CVSS, intended to reflect changing conditions, relies on manual updates rather than real-time adjustments, making it less responsive to evolving threats. Finally, CVSS can be inconsistent, as different organizations may interpret scoring criteria differently, leading to varying assessments for the same vulnerability. 

\subsubsection{OWASP Risk Rating \citep{owaspOWASPRisk}} \label{subsubsec:owasp}
Developed by the Open Web Application Security Project (OWASP)\footnote{\url{https://owasp.org/}}, it is a risk assessment methodology that evaluates vulnerabilities  and categorizes security risks in web applications by assessing likelihood (based on threat agent and vulnerability characteristics) and impact (considering technical and business factors) to produce an overall risk score.
\begin{itemize}
    \item \textbf{Likelihood:} Calculates the probability of the attack to be exploited based on two components:
    \begin{itemize}
        \item \textbf{Threat Agent (TA):} Quantifies the skill level, motivation, opportunity, and size of the threat-agent population
        \item \textbf{Vulnerability (V):} Quantifies the ease of discovery, ease of exploit, awareness, awareness of the system administrators, and the intrusion detection level.
    \end{itemize}
    \item \textbf{Impact:} Calculates the impact or loss produced by the attacks, it encompasses two types of impact:
        \begin{itemize}
        \item \textbf{Technical Impact (TI):} Quantifies the impact on Confidentiality, Integrity, and Availability.
        \item \textbf{Business Impact (BI):} Quantifies the financial damage, reputation damage, non-compliance, and privacy violation
    \end{itemize}
\end{itemize}
\paragraph{Calculations}
The vulnerability score is calculated based on the average score of each component. The values and definitions of each factor of OWASP Risk Rating is presented in Table \ref{tab:factors-owasp}. 

\begin{table}[h]
\centering
\caption{Metric factors of OWASP Risk Rating \citep{owaspOWASPRisk}}
\label{tab:factors-owasp}
\adjustbox{max width=\textwidth}{
\begin{tabular}{c c c}

    \hline
    
    \multirow{2}{*}{\textbf{Factor}}  & \multirow{2}{*}{\textbf{Definition}} &  \multirow{2}{*}{\textbf{Values}} \cr \\
    \hline

    \multirow{2}{*}{\textbf{Skill Level (SL)}} & \multirow{2}{*}{How much expertise is needed} & \multirow{32}{*}{[1 (Low), 10 (High)]} \cr \\
    \cline{1-2}

    \multirow{2}{*}{\textbf{Motivation (M)}} & \multirow{2}{*}{How much motivated is the attacker}   \cr \\
     \cline{1-2}

    \multirow{2}{*}{\textbf{Opportunity (O)}} & \multirow{2}{*}{How easy is it to exploit the vulnerability}   \cr \\
    \cline{1-2}

    \multirow{2}{*}{\textbf{Size of TA (S)}} & \multirow{2}{*}{How many attackers can be there}  \cr \\
    \cline{1-2}
    
    \multirow{2}{*}{\textbf{Ease of Discovery (ED)}} & \multirow{2}{*}{How easy is it to discover the vulnerability}  \cr \\
    \cline{1-2}
    
    \multirow{2}{*}{\textbf{Ease of Exploit (EE)}} & \multirow{2}{*}{How easy is it to exploit the vulnerability}  \cr \\
    \cline{1-2}
    
    \multirow{2}{*}{\textbf{Awareness (A)}} & \multirow{2}{*}{How much aware are the defenders}  \cr \\
    \cline{1-2}
    
    \multirow{2}{*}{\textbf{Intrusion Detect. (ID)}} & \multirow{2}{*}{How difficult is it to detect the attack}  \cr \\
    \cline{1-2}
    
    \multirow{2}{*}{\textbf{Confidentiality (LC)}} & \multirow{2}{*}{How much is confidentiality impacted}  \cr \\
    \cline{1-2}
    
    \multirow{2}{*}{\textbf{Integrity (LI)}} & \multirow{2}{*}{How much is the integrity impacted}  \cr \\
    \cline{1-2}
    
    \multirow{2}{*}{\textbf{Availability (LAV)}} & \multirow{2}{*}{How much is the availability impacted}  \cr \\
    \cline{1-2}
    
    \multirow{2}{*}{\textbf{Financial Dmg. (FD)}} & \multirow{2}{*}{How much financial loss can result}  \cr \\
    \cline{1-2}
    
    \multirow{2}{*}{\textbf{Reputation Dmg. (RD)}} & \multirow{2}{*}{How much the reputation can be harmed}  \cr \\
    \cline{1-2}
    
    \multirow{2}{*}{\textbf{Non-Compliance (NC)}} & \multirow{2}{*}{How much legal violations can happen} \cr \\
    \cline{1-2}
    
    \multirow{2}{*}{\textbf{Privacy Violation (PV)}} & \multirow{2}{*}{How much users' privacy is violated} \cr \\
    \hline

\end{tabular}
}
\end{table}

In this metric, a value in the range $[1, 3[$ is considered an attack of \textbf{`Low'} criticality. The \textbf{`Medium'} criticality range is $[3, 6[$, and the values starting from 6 are labeled as \textbf{`High'} in criticality.
After assessing all the values, the final score is a multiplication between the score of Likelihood and the score of Impact as shown below:
\begin{equation}
    OWASP Score = Likelihood * Impact
\end{equation}
The score of Likelihood is calculated as the mean of the Threat Agent and the Vulnerability scores: 
\begin{equation}
    Likelihood = (Score_{TA} + Score_{V}) / 2
\end{equation}
And the score of Impact is calculated as the mean of the Technical and Business impact scores 
\begin{equation}
    Impact = (Score_{TI} + Score_{BI}) / 2
\end{equation}
Where the score of each component (TA, V, TI, BI) are respectively the average score of their factors:
\begin{equation}
    Score_{TA} = ({Skill Level} + Motivation + Opportunity + Size_{TA}) / 4
\end{equation}
\begin{equation}
    Score_{V} = (Ease of Discovery + Ease of Exploit + Awareness + Intrusion Detection) / 4
\end{equation}
\begin{equation}
    Score_{TI} = (Confidentiality + Integrity + Availability) / 3
\end{equation}
\begin{equation}
    Score_{BI} = (Financial Dmg + Reputation Dmg + NonCompliance + Privacy Violation) / 4
\end{equation}
The rank of the final OWASP severity-score (Low, Medium, High, Critical) is defined based on the combinations shown in the Table \ref{tab:owasp-comb}. For example, if the $Likelihood = 5/10$ (Medium criticality) and the $Impact = 6/10$ (High criticality), the final score according the matrix is \textbf{High}.

\begin{table}[h]
\centering
\caption{Criticality Matrix of OWASP Risk Rating \citep{owaspOWASPRisk}}
\label{tab:owasp-comb}
\renewcommand{\arraystretch}{2}
\begin{tabular}{c | c | c | c | c}

    \hline
    
    \multirow{4}{*}{\textbf{Impact}} & {\textbf{High}} & \cellcolor{yellow!40}{\textbf{Medium}} & \cellcolor{red!40}{\textbf{High}} & \cellcolor{purple!40}{\textbf{Critical}}  \\
    \cline{2-5}

    & {\textbf{Medium}} & \cellcolor{green!40}{\textbf{Low}} & \cellcolor{yellow!40}{\textbf{Medium}} & \cellcolor{red!40}{\textbf{High}}  \\
    \cline{2-5}

    & {\textbf{Low}} & \cellcolor{cyan!40}{\textbf{Note}} & \cellcolor{green!40}{\textbf{Low}} & \cellcolor{yellow!40}{\textbf{Medium}}  \\
    \cline{2-5}

    & & {\textbf{Low}} & {\textbf{Medium}} & {\textbf{High}}  \\
    \hline

    & \multicolumn{4}{c}{{\textbf{Likelihood}}} \\
    \hline

\end{tabular}
\end{table}

\paragraph{Limitations}
The OWASP Risk Rating methodology, though widely used for web application security assessment, has notable limitations. Its reliance on subjective evaluations of factors like threat agent skill and impact severity can result in inconsistent ratings across different assessors and lead to biased outcomes. Additionally, OWASP Risk Rating lacks specificity for environments like cloud or mobile and does not adapt to rapidly changing threat landscapes, making it less responsive in dynamic security contexts. Finally, having many factors increases the complexity of this metric and its reliance on experts knowledge to assess each factor precisely.

\subsubsection{SSVC (Stakeholder-Specific Vulnerability Categorization) \citep{spring2021prioritizing}} \label{subsubsec:ssvc}
The SSVC is a framework that prioritizes vulnerabilities based on qualitative decision trees tailored to specific stakeholder roles, instead of numerical severity scores. The main two stakeholders represented are:
\begin{itemize}
    \item \textbf{Suppliers:} They decide how urgent it is to develop and release patches for their systems based on reports about potential vulnerabilities. Their decision tree is based on factors such as Exploitation, Technical Impact, Utility, and Safety Impact.
    \item \textbf{Deployers:} They decide when and how to deploy the patches developed by the suppliers. Their decision tree is based on similar factors such as Exploitation, System Exposure, Automation, and Human Impact.
\end{itemize}

\paragraph{Calculations} In our case, we consider LLMselves as Suppliers trying to assess the potential vulnerabilities impacting their LLM. Table \ref{tab:factors-ssvc} below show the different factors used in evaluating vulnerabilities using SSVC as a supplier.
\begin{table}[h]
\centering
\caption{Metric factors of SSVC for a supplier \citep{spring2021prioritizing}}
\label{tab:factors-ssvc}
\begin{tabular}{
               >{\centering\arraybackslash} p{0.22\textwidth}%
                >{\centering\arraybackslash}p{0.4\textwidth}%
                >{\centering\arraybackslash}p{0.29\textwidth}}%

    \hline
    
    \multirow{2}{*}{\textbf{Factor}}  & \multirow{2}{*}{\textbf{Definition}} &  \multirow{2}{*}{\textbf{Values}} \cr \\
    \hline

    \multirow{2}{*}{\textbf{Exploitation (E)}} & \multirow{2}{*}{In which state is the exploitation} & None (N), Proof-of-Concept (P), Active (A) \\
    \hline

    \multirow{2}{*}{\textbf{Automatable (A)}} & \multirow{2}{*}{Can the attack be automatable} & \multirow{2}{*}{No (N), Yes (Y)} \cr \\
    \hline

   \multirow{2}{*}{\textbf{Value Density (V)}} & {How valuable is the information accessed by the attacker} & \multirow{2}{*}{Diffuse (D), Concentrated (C)}  \\
    \hline

    \multirow{2}{*}{\textbf{Utility (U)}} & {How much useful is the exploit for the attacker} & Laborious (L), Efficient (E), Super Efficient (S)  \\
    \hline

    {\textbf{Technical Impact (T)}} & {How much impact does the vulnerability do} & \multirow{2}{*}{Partial (P), Total (T)} \\
    \hline

    {\textbf{Public-Safety Impact (S)}} & {How much impact has vulnerability on the public} & \multirow{2}{*}{Minimal (M), Significant (S)} \\
    \hline

\end{tabular}
\end{table}

The value of Utility (U) is calculated based in the values if Automatable (A) and Value Density (V) as follows:

The final decision is taken by following the logic described in Figure \ref{fig:ssvc-tree}. There four main possible outcomes ranked from the lowest priority to the highest one are: Defer, Scheduled, Out-of-cycle, and Immediate. Each one of them represents the emergency level for developing corresponding patches.

\begin{figure}[h]
\centering
\adjustbox{max width=0.9\textwidth}{
\begin{forest}
for tree={
    grow=east,
    draw,
    rounded corners,
    text centered,
    anchor=west,
    minimum height=2em,
    minimum width=3cm,
    edge={->, thick},
    font=\sffamily,
    align=center,
    l sep=5em, 
    edge label={node[midway, above, sloped]{}} 
}
[Start
    [Exploitation
        [Utility, edge label={node[midway, above, sloped]{active}}
            [Technical Impact, edge label={node[midway, above, sloped]{super effective}}
                [Public-Safety Impact, edge label={node[midway, above, sloped]{total}}
                    [Immediate, fill= red!45, edge label={node[midway, above]{significant}}]
                    [Immediate, fill=red!45, edge label={node[midway, above]{minimal}}]
                ]
                [Public-Safety Impact, edge label={node[midway, above, sloped]{partial}}
                    [Immediate, fill= red!45, edge label={node[midway, above]{significant}}]
                    [Immediate, fill=red!45, edge label={node[midway, above]{minimal}}]
                ]
            ]
            [Technical Impact, edge label={node[midway, above, sloped]{efficient}}
                [Public-Safety Impact, edge label={node[midway, above, sloped]{total}}
                    [Immediate, fill= red!45, edge label={node[midway, above]{significant}}]
                    [Out-of-Cycle, fill=orange!45, edge label={node[midway, above]{minimal}}]
                ]
                [Public-Safety Impact, edge label={node[midway, above, sloped]{partial}}
                    [Immediate, fill= red!45, edge label={node[midway, above]{significant}}]
                    [Out-of-Cycle, fill=orange!45, edge label={node[midway, above]{minimal}}]
                ]
            ]
            [Technical Impact, edge label={node[midway, above, sloped]{laborious}}
                [Public-Safety Impact, edge label={node[midway, above, sloped]{total}}
                    [Immediate, fill= red!45, edge label={node[midway, above]{significant}}]
                    [Out-of-Cycle, fill=orange!45, edge label={node[midway, above]{minimal}}]
                ]
                [Public-Safety Impact, edge label={node[midway, above, sloped]{partial}}
                    [Immediate, fill= red!45, edge label={node[midway, above]{significant}}]
                    [Out-of-Cycle, fill=orange!45, edge label={node[midway, above]{minimal}}]
                ]
            ]
        ]
        [Utility, edge label={node[midway, above, sloped]{poc}}
            [Technical Impact, edge label={node[midway, above, sloped]{super effective}}
                [Public-Safety Impact, edge label={node[midway, above, sloped]{total}}
                    [Immediate, fill= red!45, edge label={node[midway, above]{significant}}]
                    [Out-of-Cycle, fill=orange!45, edge label={node[midway, above]{minimal}}]
                ]
                [Public-Safety Impact, edge label={node[midway, above, sloped]{partial}}
                    [Immediate, fill= red!45, edge label={node[midway, above]{significant}}]
                    [Out-of-Cycle, fill=orange!45, edge label={node[midway, above]{minimal}}]
                ]
            ]
            [Technical Impact, edge label={node[midway, above, sloped]{efficient}}
                [Public-Safety Impact, edge label={node[midway, above, sloped]{total}}
                    [Immediate, fill= red!45, edge label={node[midway, above]{significant}}]
                    [Out-of-Cycle, fill=orange!45, edge label={node[midway, above]{minimal}}]
                ]
                [Public-Safety Impact, edge label={node[midway, above, sloped]{partial}}
                    [Immediate, fill= red!45, edge label={node[midway, above]{significant}}]
                    [Scheduled, fill=yellow!45, edge label={node[midway, above]{minimal}}]
                ]
            ]
            [Technical Impact, edge label={node[midway, above, sloped]{laborious}}
                [Public-Safety Impact, edge label={node[midway, above, sloped]{total}}
                    [Immediate, fill= red!45, edge label={node[midway, above]{significant}}]
                    [Scheduled, fill=yellow!45, edge label={node[midway, above]{minimal}}]
                ]
                [Public-Safety Impact, edge label={node[midway, above, sloped]{partial}}
                    [Out-of-Cycle, fill= orange!45, edge label={node[midway, above]{significant}}]
                    [Scheduled, fill=yellow!45, edge label={node[midway, above]{minimal}}]
                ]
            ]
        ]
        [Utility, edge label={node[midway, above, sloped]{none}}
            [Technical Impact, edge label={node[midway, above, sloped]{super effective}}
                [Public-Safety Impact, edge label={node[midway, above, sloped]{total}}
                    [Out-of-Cycle, fill= orange!45, edge label={node[midway, above]{significant}}]
                    [Out-of-Cycle,  fill=orange!45, edge label={node[midway, above]{minimal}}]
                ]
                [Public-Safety Impact, edge label={node[midway, above, sloped]{partial}}
                    [Out-of-Cycle, fill= orange!45, edge label={node[midway, above]{significant}}]
                    [Scheduled, fill=yellow!45, edge label={node[midway, above]{minimal}}]
                ]
            ]
            [Technical Impact, edge label={node[midway, above, sloped]{efficient}}
                [Public-Safety Impact, edge label={node[midway, above, sloped]{total}}
                    [Out-of-Cycle, fill= orange!45, edge label={node[midway, above]{significant}}]
                    [Scheduled,  fill=yellow!45, edge label={node[midway, above]{minimal}}]
                ]
                [Public-Safety Impact, edge label={node[midway, above, sloped]{partial}}
                    [Out-of-Cycle,  fill= orange!45, edge label={node[midway, above]{significant}}]
                    [Scheduled, fill=yellow!45, edge label={node[midway, above]{minimal}}]
                ]
            ]
            [Technical Impact, edge label={node[midway, above, sloped]{laborious}}
                [Public-Safety Impact, edge label={node[midway, above, sloped]{total}}
                    [Out-of-Cycle, fill= orange!45, edge label={node[midway, above]{significant}}]
                    [Scheduled, fill=yellow!45, edge label={node[midway, above]{minimal}}]
                ]
                [Public-Safety Impact, edge label={node[midway, above, sloped]{partial}}
                    [Scheduled, fill= yellow!45, edge label={node[midway, above]{significant}}]
                    [Defer, fill=green!45, edge label={node[midway, above]{minimal}}]
                ]
            ]
        ]
    ]
]
\end{forest}
}
\caption{Decision Tree for Suppliers in SSVC \citep{spring2021prioritizing}}
\label{fig:ssvc-tree}
\end{figure}

\paragraph{Limitations} 
The SSVC metric has several limitations. It relies heavily on qualitative decision points, which may lead to subjective interpretations and inconsistencies across stakeholders. Additionally, the absence of numerical scoring might limit its integration with existing risk management systems that rely on quantitative data, potentially requiring significant adjustments to current workflows. Lastly, SSVC is tailored for specific stakeholder roles, which may make it be less effective in hybrid roles or complex environments where stakeholders overlap.

\section{Assessment of AAs on LLMs with Vulnerability Metrics}
\label{sec:evaluation}

In this section, we present and interpret the results of assessing the criticality of AAs against LLMs, grouped in seven types: White-box Jailbreak, Black-box Jailbreak, Prompt Injection, Evasion attacks, Model Extraction, Model Inference, and Poisoning/Trojan/Backdoor attacks. The detailed scores of these attacks given by the 3 LLMs (GPT-4o, LLAMA, and Perplexity) and their average are presented in \ref{appendix:assessment}. We represent the results in score-vectors and in spider-graph formats for more interpretability. 

Note that for the qualitative factors of CVSS and SSVC, we represent their values \textbf{numerically} in the spider graph following this logic:
\begin{itemize}
    \item For CVSS Factors:
    \begin{itemize}
        \item If they have four values (eg. AV), they are represented with values from 1 to 4.
        \item If have three values (eg. PR, C, I, A), they are represented with values from 1 to 3. 
        \item If they have two values (eg. AC, UI, S), they are represented with the values 2 and 4.
    \end{itemize}
    \item For SSVC Factors:
    \begin{itemize}
        \item If they have three values (eg. E, U), they are represented with values from 1 to 3. 
        \item If they have two values (eg. A, V, T, P), they are represented with the values 1 and 3.
    \end{itemize}
\end{itemize}

\subsection{Assessment of White-box Jailbreak attacks}
We start by evaluating White-box jailbreak attacks, the chosen attacks are the same presented in Section \ref{subsec:whitebox-jailbreak} earlier: (1) GCG \citep{zou2023universaltransferableadversarialattacks}, (2) Visual Modality \citep{niu2024efficientllmjailbreakingintroducingvisual}, (3) PGD \citep{geisler2024attackinglargelanguagemodels}, (4) SCAV \citep{xu2024uncoveringsafetyriskslarge}, (5) Soft Prompt Threats \citep{schwinn2024softpromptthreatsattacking}, (6) DrAttack \citep{li2024drattackpromptdecompositionreconstruction}, (7) RADIAL \citep{du2024analyzinginherentresponsetendency}, (8) ReNeLLM \citep{ding2024wolfsheepsclothinggeneralized}.


\subsubsection{With DREAD}
We start by evaluating the eight White-box jailbreaks attacks using DREAD \citep{michael2006security}. Here are below the attack vectors of each attack:

\begin{itemize}
    \item (1) $\rightarrow$ (D:8/R:9/E:8/A:8/D:6) = \textcolor{red}{7.8 (High)}
    \item (2) $\rightarrow$ (D:6/R:6/E:6/A:6/D:5) = \textcolor{yellow}{5.8 (Medium)}
    \item (3) $\rightarrow$ (D:7/R:7/E:7/A:7/D:5) = \textcolor{yellow}{6.6 (Medium)}
    \item (4) $\rightarrow$ (D:7/R:6/E:5/A:6/D:5) = \textcolor{yellow}{5.8 (Medium)}
    \item (5) $\rightarrow$ (D:8/R:9/E:7/A:7/D:6) = \textcolor{red}{7.4 (High)}
    \item (6) $\rightarrow$ (D:8/R:8/E:7/A:8/D:6) = \textcolor{red}{7.4 (High)}
    \item (7) $\rightarrow$ (D:7/R:6/E:7/A:6/D:5) = \textcolor{yellow}{6.2 (Medium)}
    \item (8) $\rightarrow$ (D:8/R:9/E:8/A:7/D:6) = \textcolor{red}{7.6 (High)}
\end{itemize}

The detailed calculations for these attacks are presented in Table \ref{tab:appendix-whitebox-dread}, with assessments supervised by a \textbf{Human-in-the-Loop (HitL)} to minimize \textbf{misconceptions}. For instance, GPT-4o initially scored 8/10 for the Discoverability factor in DREAD for the first attack \citep{zou2023universaltransferableadversarialattacks}, while LLAMA-3 and Perplexity AI both assigned a score of 6/10. GPT-4o's higher score stemmed from a \textbf{misunderstanding} of the factor’s meaning, interpreting Discoverability as the level of researcher awareness about the threat rather than the ease with which it can be discovered. After \textbf{clarifying} this distinction, GPT-4o revised its score to 5/10, aligning more closely with the intended definition of the metric.

A different issue arose when evaluating the \textbf{impact} of attacks. For example, the Damage factor of the second attack \citep{niu2024efficientllmjailbreakingintroducingvisual} was rated 6/10 by GPT-4o and 5/10 by LLAMA-3, reflecting moderate damage due to situational input requirements, such as specific visual input use cases. However, Perplexity AI assigned a higher score of 8/10, citing potential scenarios where the attack could have a significant impact on the targeted system. In this case, the discrepancy was due to differing \textbf{interpretations} rather than misunderstandings, making it difficult to standardize the scores. To address this, \textbf{averaging} the three scores provided a balanced result, aligning closely with the consensus of GPT-4o and LLAMA-3.

Using this approach, we reduced inconsistencies in the scoring process. The final DREAD scores are illustrated in a spider graph, highlighting that White-box Jailbreak attacks can inflict considerable damage on systems while being relatively easy to reproduce. However, discovering these threats remains a significant challenge.
\begin{center}
\begin{tikzpicture}
\begin{polaraxis}[
    title={Assessment of White-box model jailbreaking attacks with DREAD},
    title style={font=\bfseries\large},
    xtick={0, 72, 144, 216, 288},
    xticklabels={Damage, Reproducibility, Exploitability, Affected users, Discoverability},
    xticklabel style={font=\bfseries, anchor=north}, 
    ymin=0, ymax=10,
    ytick={1,2,3,4,5,6,7,8,9,10},
    yticklabels={1,2,3,4,5,6,7,8,9,10}, 
    yticklabel style={font=\small}, 
    grid=both, 
    major grid style={line width=0.8pt, draw=gray!50},
    minor grid style={line width=0.4pt, draw=gray!20},
    ylabel near ticks,
    legend style={at={(1.3,1)}, anchor=north, legend columns=1}, 
]
    \addplot[line width=1.45pt, mark=*,
        color=blue!80, fill=yellow!20, fill opacity=0.15] 
        coordinates {(0,8) (72,9) (144,8) (216,8) (288,6) (360,8)} 
        \closedcycle;
    \addplot[line width=1.45pt, mark=square*,
        color=red!80, fill=yellow!20, fill opacity=0.1] 
        coordinates {(0,6) (72,6) (144,6) (216,6) (288,5) (360,6)} 
        \closedcycle;
    \addplot[line width=1.45pt, mark=triangle*,
        color=magenta!80, fill=yellow!20, fill opacity=0.1] 
        coordinates {(0,7) (72,7) (144,7) (216,7) (288,5) (360,7)} 
        \closedcycle;
    \addplot[line width=1.45pt, mark=diamond*,
        color=brown!80, fill=yellow!20, fill opacity=0.1] 
        coordinates {(0,7) (72,6) (144,5) (216,6) (288,5) (360,7)} 
        \closedcycle;
    \addplot[line width=1.45pt, mark=pentagon*,
        color=cyan!80, fill=yellow!20, fill opacity=0.1] 
        coordinates {(0,8) (72,9) (144,7) (216,7) (288,6) (360,8)} 
        \closedcycle;
    \addplot[line width=1.45pt, mark=o,
        color=green!80, fill=yellow!20, fill opacity=0.1] 
        coordinates {(0,8) (72,8) (144,7) (216,8) (288,6) (360,8)} 
        \closedcycle;
    \addplot[line width=1.45pt, mark=x,
        color=orange!80, fill=yellow!20, fill opacity=0.1] 
        coordinates {(0,7) (72,6) (144,7) (216,6) (288,5) (360,7)} 
        \closedcycle;
    \addplot[line width=1.45pt, mark=none,
        color=violet!80, fill=yellow!20, fill opacity=0.1] 
        coordinates {(0,8) (72,9) (144,8) (216,7) (288,6) (360,8)} 
        \closedcycle;
    
    \legend{Attack 1, Attack 2, Attack 3, Attack 4, Attack 5, Attack 6, Attack 7, Attack 8}
\end{polaraxis}
\end{tikzpicture}
\end{center}

\subsubsection{With CVSS}
Then, we evaluate the assessment of these attacks using CVSS \citep{schiffman2005complete}. The corresponding CVSS Vectors are shown below:

\begin{itemize}
    \item (1) $\rightarrow$ (AV:N/AC:H/PR:N/UI:N/S:C/C:L/I:H/A:N) = \textcolor{red}{7.5 (High)}
    \item (2) $\rightarrow$ (AV:N/AC:H/PR:N/UI:N/S:C/C:L/I:H/A:N) = \textcolor{red}{7.5 (High)}
    \item (3) $\rightarrow$ (AV:N/AC:H/PR:N/UI:N/S:C/C:L/I:H/A:N) = \textcolor{red}{7.5 (High)}
    \item (4) $\rightarrow$ (AV:N/AC:H/PR:N/UI:N/S:C/C:L/I:H/A:N) = \textcolor{red}{7.1 (High)}
    \item (5) $\rightarrow$ (AV:N/AC:L/PR:L/UI:N/S:C/C:L/I:H/A:N) = \textcolor{red}{8.5 (High)}
    \item (6) $\rightarrow$ (AV:N/AC:L/PR:L/UI:N/S:C/C:L/I:H/A:N) = \textcolor{red}{8.5 (High)}
    \item (7) $\rightarrow$ (AV:N/AC:H/PR:N/UI:R/S:C/C:L/I:H/A:N) = \textcolor{yellow}{6.9 (Medium)}
    \item (8) $\rightarrow$ (AV:N/AC:L/PR:N/UI:R/S:C/C:L/I:H/A:N) = \textcolor{red}{8.2 (High)}
\end{itemize}

The detailed scores are presented in Table \ref{tab:appendix-whitebox-cvss}. During the analysis, some LLMs encountered challenges in correctly interpreting the characteristics of each attack. For instance, GPT-4o initially concluded that White-box jailbreak attacks only impact the Confidentiality of data, with no effect on Integrity—a conclusion that was refuted by the other two LLMs. It was crucial to identify such \textbf{misunderstandings} and \textbf{guide} the models to recognize their errors. Rather than providing direct corrections, we prompted GPT-4o with \textbf{questions} such as: \textit{Do these attacks target Integrity given that they involve manipulation of gradients and embeddings?} This approach enabled the model to identify and rectify its own mistake while fostering greater \textbf{caution} in subsequent assessments. 

This process highlights another key advantage of using multiple LLMs: \textbf{they provide diverse perspectives and explanations}, which help identify and address unconventional or erroneous analyses. Moreover, there was also other slight divergence in scoring factors such as the Scope and User Interaction; but using an averaging method helps align the final scores to the majority.

After averaging the final values, we visualized the scores using a spider chart for clarity. The CVSS scores reveal that White-box attacks are typically executed through the network, requiring low-to-medium privileges and primarily targeting the Integrity of systems.

\begin{center}
\begin{tikzpicture}
\begin{polaraxis}[
    title={Assessment of White-box model jailbreaking attacks with CVSS},
    title style={font=\bfseries\large},
    xtick={0, 45, 90, 135, 180, 225, 270, 315},
    xticklabels={Vector, Complexity, Privileges, User Interaction, Scope, Confidentiality, Integrity, Availability},
    xticklabel style={font=\bfseries, anchor=north}, 
    ymin=0, ymax=4,
    ytick={0,1,2,3,4},
    yticklabels={0,1,2,3,4}, 
    yticklabel style={font=\small}, 
    grid=both, 
    major grid style={line width=0.8pt, draw=gray!50},
    minor grid style={line width=0.4pt, draw=gray!20},
    ylabel near ticks,
    legend style={at={(1.3,1)}, anchor=north, legend columns=1}, 
]
    \addplot[line width=1.45pt, mark=x,
        color=orange!80, fill=cyan!20, fill opacity=0.1] 
        coordinates {(0,1) (45,4) (90,1) (135,2) (180,4) (225,2) (270,3) (315,1) (360,1)} 
        \closedcycle;
    \addplot[line width=1.45pt, mark=pentagon,
        color=orange!80, fill=cyan!20, fill opacity=0.1] 
        coordinates {(0,1) (45,4) (90,1) (135,2) (180,4) (225,2) (270,3) (315,1) (360,1)}
        \closedcycle;
    \addplot[line width=1.45pt, mark=star,
        color=orange!80, fill=cyan!20, fill opacity=0.1] 
        coordinates {(0,1) (45,4) (90,1) (135,2) (180,4) (225,2) (270,3) (315,1) (360,1)}
        \closedcycle;
    \addplot[line width=1.45pt, mark=square,
        color=blue!80, fill=cyan!20, fill opacity=0.1] 
        coordinates {(0,1) (45,4) (90,1) (135,2) (180,4) (225,2) (270,3) (315,1) (360,1)} 
        \closedcycle;
    \addplot[line width=1.45pt, mark=pentagon,
        color=red!80, fill=cyan!20, fill opacity=0.1] 
        coordinates {(0,1) (45,2) (90,2) (135,2) (180,4) (225,2) (270,3) (315,1) (360,1)} 
        \closedcycle;
    \addplot[line width=1.45pt, mark=o,
        color=red!80, fill=cyan!20, fill opacity=0.1] 
        coordinates {(0,1) (45,2) (90,2) (135,2) (180,4) (225,2) (270,3) (315,1) (360,1)} 
        \closedcycle;
    \addplot[line width=1.45pt, mark=pentagon,
        color=magenta!80, fill=cyan!20, fill opacity=0.1] 
        coordinates {(0,1) (45,4) (90,2) (135,4) (180,4) (225,2) (270,3) (315,1) (360,1)} 
        \closedcycle;
    \addplot[line width=1.45pt, mark=none,
        color=green!80, fill=cyan!20, fill opacity=0.1] 
        coordinates {(0,1) (45,2) (90,1) (135,4) (180,4) (225,2) (270,3) (315,1) (360,1)} 
        \closedcycle;
    
    \legend{Attack 1, Attack 2, Attack 3, Attack 4, Attack 5, Attack 6, Attack 7, Attack 8}
\end{polaraxis}
\end{tikzpicture}
\end{center}

\subsubsection{With OWASP Risk Rating}
A third evaluation of white-box jailbreak attacks is done using OWASP RR \citep{owaspOWASPRisk}. The corresponding vulnerability vectors of each attack is:

\begin{itemize}
    \item (1) $\rightarrow$ (SL:7/M:6/O:6/S:6/ED:7/EE:8/A:5/ID:5/LC:5/LI:7/LA:4/FD:7/RD:8/NC:4/PV:4) = \textcolor{red}{3.6 (High)}
    \item (2) $\rightarrow$ (SL:7/M:6/O:5/S:5/ED:5/EE:7/A:5/ID:5/LC:5/LI:6/LA:3/FD:6/RD:7/NC:4/PV:4) = \textcolor{yellow}{2.8 (Medium)}
    \item (3) $\rightarrow$ (SL:7/M:7/O:5/S:6/ED:6/EE:7/A:5/ID:5/LC:5/LI:7/LA:3/FD:6/RD:7/NC:6/PV:5) = \textcolor{yellow}{3.2 (Medium)}
    \item (4) $\rightarrow$ (SL:5/M:6/O:4/S:3/ED:4/EE:6/A:5/ID:4/LC:4/LI:6/LA:1/FD:5/RD:6/NC:3/PV:4) = \textcolor{yellow}{1.9 (Medium)}
    \item (5) $\rightarrow$ (SL:6/M:7/O:6/S:5/ED:6/EE:7/A:6/ID:5/LC:5/LI:7/LA:1/FD:6/RD:7/NC:4/PV:4) = \textcolor{red}{2.8 (High)}
    \item (6) $\rightarrow$ (SL:7/M:7/O:6/S:6/ED:6/EE:7/A:5/ID:5/LC:5/LI:7/LA:2/FD:6/RD:7/NC:4/PV:5) = \textcolor{red}{3.2 (High)}
    \item (7) $\rightarrow$ (SL:6/M:6/O:5/S:4/ED:5/EE:6/A:5/ID:5/LC:4/LI:6/LA:1/FD:5/RD:6/NC:3/PV:3) = \textcolor{yellow}{2.1 (Medium)}
    \item (8) $\rightarrow$ (SL:7/M:7/O:7/S:5/ED:7/EE:7/A:6/ID:5/LC:4/LI:7/LA:1/FD:6/RD:7/NC:4/PV:4) = \textcolor{red}{3.1 (High)}
\end{itemize}

The scores assigned by each LLM are detailed in Table \ref{tab:appendix-whitebox-owasp}. With its multiple factors, the OWASP Risk Rating provided a more comprehensive analysis of each attack. However, we encountered some \textbf{interpretation discrepancies}, particularly with Perplexity AI. This model argued that these attacks have a Medium-to-High impact on Confidentiality—an assessment that differed from its CVSS evaluation of the same attacks. This highlights the inherent \textbf{subjectivity} in scoring, as analyzing identical attacks in separate conversations can yield inconsistent results. In contrast, the other two LLMs provided scores consistent with the CVSS evaluation for Confidentiality and Integrity, along with a Low-to-None impact on Availability. Averaging the scores \textbf{mitigated} such discrepancies while preserving the unique perspectives offered by each LLM, especially in factors like Non-Compliance and Privacy Violation. Notably, LLAMA-3 failed to detect any impact in these areas, whereas GPT-4o and Perplexity AI highlighted their significance.

Another challenge we observed was the tendency of LLMs to rely on \textbf{memorized values} when evaluating attacks across multiple factors. For example, GPT-4o initially assigned identical scores to the first two attacks \citep{zou2023universaltransferableadversarialattacks, niu2024efficientllmjailbreakingintroducingvisual}. Upon prompting it to provide objective and distinct evaluations, GPT-4o revised its scores, adjusting the ED value from 6/10 to 5/10, the Availability impact from 6/10 to 5/10, and the NC value from 6/10 to 7/10. It justified these changes by acknowledging \textbf{similarities} between the attacks while ensuring the scores reflected nuanced differences.

The averaged scores are visualized below for clarity. These results align with the CVSS evaluation in terms of technical impact and ease of exploitation, while also shedding light on the reputational damage that could arise if such attacks are exploited. Moreover, they emphasize that White-box attacks have a medium impact on Non-Compliance and Privacy Violation.

\begin{center}
\begin{tikzpicture}
\begin{polaraxis}[
    title={Assessment of White-box model jailbreaking attacks with OWASP RR},
    title style={font=\bfseries\large},
    xtick={0, 24, 48, 72, 96, 120, 144, 168, 192, 216, 240, 264, 288, 312, 336},
    xticklabels={Skill, Motivation, Opportunity, Size, Discovery, Exploit, Awareness, Intrusion Dtc, Confidentiality, Integrity, Availability, Fnc Dmg, Rpt Dmg, Non-comp, Privacy Vlt},
    xticklabel style={font=\bfseries, anchor=north}, 
    ymin=0, ymax=10,
    ytick={0,1,2,3,4,5,6,7,8,9,10},
    yticklabels={0,1,2,3,4,5,6,7,8,9,10}, 
    yticklabel style={font=\small}, 
    grid=both, 
    major grid style={line width=0.8pt, draw=gray!50},
    minor grid style={line width=0.4pt, draw=gray!20},
    ylabel near ticks,
    legend style={at={(1.3,1)}, anchor=north, legend columns=1}, 
]
    \addplot[line width=1.45pt, mark=square,
        color=blue!80, fill=green!20, fill opacity=0.1] 
        coordinates {(0,7) (24,6) (48,6) (72,6) (96,7) (120,8) (144,5) (168,5) (192,5) (216,7) (240,4) (264,7) (288,8) (312,4) (336,4) (360,7)} 
        \closedcycle;
    \addplot[line width=1.45pt, mark=triangle,
        color=red!80, fill=green!20, fill opacity=0.1] 
        coordinates {(0,7) (24,6) (48,5) (72,5) (96,5) (120,7) (144,5) (168,5) (192,5) (216,6) (240,3) (264,6) (288,7) (312,4) (336,4) (360,7)}
        \closedcycle;
    \addplot[line width=1.45pt, mark=square,
        color=violet!80, fill=green!20, fill opacity=0.1] 
        coordinates {(0,7) (24,7) (48,5) (72,6) (96,6) (120,7) (144,5) (168,5) (192,5) (216,7) (240,3) (264,6) (288,7) (312,6) (336,5) (360,7)}
        \closedcycle;
    \addplot[line width=1.45pt, mark=pentagon*,
        color=green!80, fill=green!20, fill opacity=0.1] 
        coordinates {(0,5) (24,6) (48,4) (72,3) (96,4) (120,6) (144,5) (168,4) (192,4) (216,6) (240,1) (264,5) (288,6) (312,3) (336,4) (360,5)} 
        \closedcycle;
    \addplot[line width=1.45pt, mark=square,
        color=orange!80, fill=green!20, fill opacity=0.1] 
        coordinates {(0,6) (24,7) (48,6) (72,5) (96,6) (120,7) (144,6) (168,5) (192,5) (216,7) (240,1) (264,6) (288,7) (312,4) (336,4) (360,6)} 
        \closedcycle;
    \addplot[line width=1.45pt, mark=diamond,
        color=magenta!80, fill=green!20, fill opacity=0.1] 
        coordinates {(0,7) (24,7) (48,6) (72,6) (96,6) (120,7) (144,5) (168,5) (192,5) (216,7) (240,2) (264,6) (288,7) (312,4) (336,5) (360,7)} 
        \closedcycle;
    \addplot[line width=1.45pt, mark=pentagon,
        color=yellow!80, fill=green!20, fill opacity=0.1] 
        coordinates {(0,6) (24,6) (48,5) (72,4) (96,5) (120,6) (144,5) (168,5) (192,4) (216,6) (240,1) (264,5) (288,6) (312,3) (336,3) (360,6)} 
        \closedcycle;
    \addplot[line width=1.45pt, mark=none,
        color=cyan!80, fill=green!20, fill opacity=0.1] 
        coordinates {(0,7) (24,7) (48,7) (72,5) (96,7) (120,7) (144,6) (168,5) (192,4) (216,7) (240,1) (264,6) (288,7) (312,4) (336,4) (360,7)} 
        \closedcycle;
    
    \legend{Attack 1, Attack 2, Attack 3, Attack 4, Attack 5, Attack 6, Attack 7, Attack 8}
\end{polaraxis}
\end{tikzpicture}
\end{center}

\subsubsection{With SSVC}
Finally, we evaluate these attacks using SSVC \citep{spring2021prioritizing}. The corresponding vectors, as a supplier, are shown below:

\begin{itemize}
    \item (1) $\rightarrow$ (E:P/A:Y/V:C/U:S/T:T/P:S) = \textcolor{red}{Immediate (Very High)}
    \item (2) $\rightarrow$ (E:P/A:N/V:C/U:E/T:T/P:S) = \textcolor{red}{Immediate (Very High)}
    \item (3) $\rightarrow$ (E:P/A:N/V:C/U:E/T:T/P:S) = \textcolor{red}{Immediate (Very High)}
    \item (4) $\rightarrow$ (E:P/A:N/V:C/U:E/T:P/P:M) = \textcolor{yellow}{Scheduled (Medium)}
    \item (5) $\rightarrow$ (E:P/A:Y/V:C/U:S/T:T/P:S) = \textcolor{red}{Immediate (Very High)}
    \item (6) $\rightarrow$ (E:P/A:Y/V:C/U:S/T:T/P:S) = \textcolor{red}{Immediate (Very High)}
    \item (7) $\rightarrow$ (E:P/A:N/V:C/U:E/T:P/P:M) = \textcolor{yellow}{Scheduled (Medium)}
    \item (8) $\rightarrow$ (E:P/A:Y/V:C/U:S/T:T/P:S) = \textcolor{red}{Immediate (Very High)}
\end{itemize}

Table \ref{tab:appendix-whitebox-ssvc} presents the detailed SSVC assessment scores provided by each LLM. As SSVC is relatively \textbf{straightforward} to apply, the LLMs performed the evaluations without significant issues. The primary role of the HitL in this context was to \textbf{interpret} the rationale behind the values assigned by the LLMs, particularly for the Exploitation factor. For instance, when evaluating the second White-box jailbreak attack \citep{niu2024efficientllmjailbreakingintroducingvisual}, GPT-4o determined there was no PoC for the attack, as its implementation was not publicly available, and accordingly assigned it a `None' value. In contrast, LLAMA-3 and Perplexity AI offered a different perspective. Both argued that the paper provided sufficient detail about the attack, making it possible to reproduce with some effort. Consequently, they concluded that a PoC exists. With the majority of models agreeing, the average score reflected their viewpoint, recognizing the presence of a PoC.

The final SSVC scores are visualized below in a spider chart. These results indicate that White-box jailbreak attacks can be automated and highly rewarding, underscoring their significant risks to both technical systems and public safety.

\begin{center}
\begin{tikzpicture}
\begin{polaraxis}[
    title={Assessment of White-box model jailbreaking attacks with SSVC},
    title style={font=\bfseries\large},
    xtick={0, 60, 120, 180, 240, 300},
    xticklabels={Exploitation, Automatable, Value Density, Utility, Technical Imp, Public-Safety Imp},
    xticklabel style={font=\bfseries, anchor=north}, 
    ymin=0, ymax=3,
    ytick={0,1,2,3},
    yticklabels={0,1,2,3}, 
    yticklabel style={font=\small}, 
    grid=both, 
    major grid style={line width=0.8pt, draw=gray!50},
    minor grid style={line width=0.4pt, draw=gray!20},
    ylabel near ticks,
    legend style={at={(1.35,1)}, anchor=north, legend columns=1}, 
]
    \addplot[line width=1.45pt, mark=x,
        color=red!80, fill=red!20, fill opacity=0.1] 
        coordinates {(0,2) (60,3) (120,3) (180,3) (240,3) (300,3) (360,2)} 
        \closedcycle;
    \addplot[line width=1.45pt, mark=pentagon,
        color=blue!80, fill=red!20, fill opacity=0.1] 
        coordinates {(0,2) (60,1) (120,3) (180,2) (240,3) (300,3) (360,2)}
        \closedcycle;
    \addplot[line width=1.45pt, mark=square,
        color=blue!80, fill=red!20, fill opacity=0.1] 
        coordinates {(0,2) (60,1) (120,3) (180,2) (240,3) (300,3) (360,2)}
        \closedcycle;
    \addplot[line width=1.45pt, mark=pentagon,
        color=green!80, fill=red!20, fill opacity=0.1] 
        coordinates {(0,2) (60,1) (120,3) (180,2) (240,1) (300,1) (360,2)} 
        \closedcycle;
    \addplot[line width=1.45pt, mark=pentagon,
        color=red!80, fill=red!20, fill opacity=0.1] 
        coordinates {(0,2) (60,3) (120,3) (180,3) (240,3) (300,3) (360,2)} 
        \closedcycle;
    \addplot[line width=1.45pt, mark=o,
        color=red!80, fill=red!20, fill opacity=0.1] 
        coordinates {(0,2) (60,3) (120,3) (180,3) (240,3) (300,3) (360,2)} 
        \closedcycle;
    \addplot[line width=1.45pt, mark=o,
        color=green!80, fill=red!20, fill opacity=0.1] 
        coordinates {(0,2) (60,1) (120,3) (180,2) (240,1) (300,1) (360,2)} 
        \closedcycle;
    \addplot[line width=1.45pt, mark=none,
        color=red!80, fill=red!20, fill opacity=0.1] 
        coordinates {(0,2) (60,3) (120,3) (180,3) (240,3) (300,3) (360,2)} 
        \closedcycle;
    
    \legend{Attack 1, Attack 2, Attack 3, Attack 4, Attack 5, Attack 6, Attack 7, Attack 8}
\end{polaraxis}
\end{tikzpicture}
\end{center}

\subsection{Assessment of Black-box Jailbreak attacks}
We evaluate now Black-box jailbreak attacks, the eight attacks are the same presented in Section \ref{subsec:blackbox-jailbreak} earlier: (1) Privacy attack on GPT-4o \citep{li2023multistepjailbreakingprivacyattacks}, (2) PAIR \citep{chao2024jailbreakingblackboxlarge}, (3) DAN \citep{shen2024donowcharacterizingevaluating}, (4) Simple Adaptive Attack \citep{andriushchenko2024jailbreakingleadingsafetyalignedllms}, (5) PAL \citep{sitawarin2024palproxyguidedblackboxattack}, (6) GCQ \citep{hayase2024querybasedadversarialpromptgeneration}, (7) IRIS \citep{ramesh2024gpt4jailbreaksnearperfectsuccess}, (8) Tastle \citep{xiao2024distractlargelanguagemodels}.


\subsubsection{With DREAD}
We start with the evaluation using DREAD \citep{michael2006security}. Below are the DREAD vectors of each of the eight Black-box Jailbreak attacks:

\begin{itemize}
    \item (1) $\rightarrow$ (D:8/R:7/E:7/A:8/D:5) = \textcolor{red}{7 (High)}
    \item (2) $\rightarrow$ (D:8/R:8/E:8/A:7/D:5) = \textcolor{red}{7.2 (High)}
    \item (3) $\rightarrow$ (D:8/R:8/E:7/A:7/D:6) = \textcolor{red}{7.2 (High)}
    \item (4) $\rightarrow$ (D:9/R:8/E:8/A:8/D:6) = \textcolor{red}{7.8 (High)}
    \item (5) $\rightarrow$ (D:8/R:9/E:8/A:7/D:5) = \textcolor{red}{7.4 (High)}
    \item (6) $\rightarrow$ (D:8/R:6/E:7/A:7/D:5) = \textcolor{yellow}{6.6 (Medium)}
    \item (7) $\rightarrow$ (D:9/R:8/E:9/A:8/D:5) = \textcolor{red}{7.8 (High)}
    \item (8) $\rightarrow$ (D:8/R:7/E:7/A:7/D:5) = \textcolor{yellow}{6.8 (Medium)}
\end{itemize}

The details are presented in Table \ref{tab:appendix-blackbox-dread}. This time, no misunderstandings occurred, as the corrections made during the DREAD assessment of White-box attacks were already in place. However, some divergences in attack analysis still arose. For instance, the Exploitability factor of the first attack \citep{li2023multistepjailbreakingprivacyattacks} was rated 6/10 by GPT-4o, which noted that the attack requires specific query patterns but is still manageable to execute. In contrast, LLAMA-3 and Perplexity AI assigned a score of 8/10, arguing that the implementation details provided in the paper make the attack easily exploitable.

Another challenge was the potential \textbf{memorization} of values. For example, LLAMA-3 gave identical scores for the fourth, fifth, and seventh attacks \citep{andriushchenko2024jailbreakingleadingsafetyalignedllms, sitawarin2024palproxyguidedblackboxattack, ramesh2024gpt4jailbreaksnearperfectsuccess}, justifying this by highlighting the \textbf{similar characteristics} of these attacks. While this explanation is plausible, as the scores were consistent with those of the other LLMs, averaging the scores across all models helped mitigate these analytical inconsistencies by favoring the majority consensus.

After averaging the scores, we visualized the results in a spider chart for clarity. The DREAD scores indicate that Black-box Jailbreak attacks, like their White-box counterparts, can inflict significant damage while being highly reproducible and exploitable, yet challenging to detect.

\begin{center}
\begin{tikzpicture}
\begin{polaraxis}[
    title={Assessment of Black-box model jailbreaking attacks with DREAD},
    title style={font=\bfseries\large},
    xtick={0, 72, 144, 216, 288},
    xticklabels={Damage, Reproducibility, Exploitability, Affected users, Discoverability},
    xticklabel style={font=\bfseries, anchor=north}, 
    ymin=0, ymax=10,
    ytick={1,2,3,4,5,6,7,8,9,10},
    yticklabels={1,2,3,4,5,6,7,8,9,10}, 
    yticklabel style={font=\small}, 
    grid=both, 
    major grid style={line width=0.8pt, draw=gray!50},
    minor grid style={line width=0.4pt, draw=gray!20},
    ylabel near ticks,
    legend style={at={(1.3,1)}, anchor=north, legend columns=1}, 
]
    \addplot[line width=1.45pt, mark=*,
        color=magenta!80, fill=yellow!20, fill opacity=0.1] 
        coordinates {(0,8) (72,7) (144,7) (216,8) (288,5) (360,8)} 
        \closedcycle;
    \addplot[line width=1.45pt, mark=square*,
        color=violet!80, fill=yellow!20, fill opacity=0.1] 
        coordinates {(0,8) (72,8) (144,8) (216,7) (288,5) (360,8)} 
        \closedcycle;
    \addplot[line width=1.45pt, mark=triangle*,
        color=green!80, fill=yellow!20, fill opacity=0.1] 
        coordinates {(0,8) (72,8) (144,7) (216,7) (288,6) (360,8)} 
        \closedcycle;
    \addplot[line width=1.45pt, mark=diamond*,
        color=yellow!80, fill=yellow!20, fill opacity=0.1] 
        coordinates {(0,9) (72,8) (144,8) (216,8) (288,6) (360,9)} 
        \closedcycle;
    \addplot[line width=1.45pt, mark=pentagon*,
        color=orange!80, fill=yellow!20, fill opacity=0.1] 
        coordinates {(0,8) (72,9) (144,8) (216,7) (288,5) (360,8)} 
        \closedcycle;
    \addplot[line width=1.45pt, mark=o,
        color=red!80, fill=yellow!20, fill opacity=0.1] 
        coordinates {(0,8) (72,6) (144,7) (216,7) (288,5) (360,8)} 
        \closedcycle;
    \addplot[line width=1.45pt, mark=x,
        color=cyan!80, fill=yellow!20, fill opacity=0.1] 
        coordinates {(0,9) (72,8) (144,9) (216,8) (288,5) (360,9)} 
        \closedcycle;
    \addplot[line width=1.45pt, mark=none,
        color=blue!80, fill=yellow!20, fill opacity=0.1] 
        coordinates {(0,8) (72,7) (144,7) (216,7) (288,5) (360,8)} 
        \closedcycle;
    
    \legend{Attack 1, Attack 2, Attack 3, Attack 4, Attack 5, Attack 6, Attack 7, Attack 8}
\end{polaraxis}
\end{tikzpicture}
\end{center}

\subsubsection{With CVSS}
In this second assessment of Black-box Jailbreak, we evaluate the attacks using CVSS \citep{schiffman2005complete}. The corresponding CVSS Vectors are shown below:

\begin{itemize}
    \item (1) $\rightarrow$ (AV:N/AC:L/PR:N/UI:N/S:U/C:L/I:L/A:N) = \textcolor{yellow}{6.5 (Medium)}
    \item (2) $\rightarrow$ (AV:N/AC:L/PR:N/UI:N/S:U/C:L/I:H/A:N) = \textcolor{red}{8.2 (High)}
    \item (3) $\rightarrow$ (AV:N/AC:L/PR:N/UI:N/S:U/C:L/I:L/A:N) = \textcolor{yellow}{6.5 (Medium)}
    \item (4) $\rightarrow$ (AV:N/AC:L/PR:N/UI:N/S:C/C:L/I:N/A:N) = \textcolor{red}{7.2 (High)}
    \item (5) $\rightarrow$ (AV:N/AC:L/PR:N/UI:N/S:C/C:L/I:N/A:N) = \textcolor{red}{7.2 (High)}
    \item (6) $\rightarrow$ (AV:N/AC:H/PR:N/UI:N/S:U/C:L/I:N/A:N) = \textcolor{yellow}{5.4 (Medium)}
    \item (7) $\rightarrow$ (AV:N/AC:L/PR:L/UI:N/S:U/C:L/I:H/A:N) = \textcolor{red}{7.1 (High)}
    \item (8) $\rightarrow$ (AV:N/AC:L/PR:N/UI:N/S:U/C:L/I:H/A:N) = \textcolor{red}{8.2 (High)}
\end{itemize}

Table \ref{tab:appendix-blackbox-cvss} outlines the detailed CVSS scores for the Black-box Jailbreak attacks. As observed with previous assessments, the three LLMs displayed some divergence in evaluating the technical impact of each attack. However, averaging the scores allowed us to establish a balanced consensus that moderated the variations in their evaluations.

One notable issue arose with LLAMA-3 in \textbf{interpreting} the User Interaction factor, which assesses whether a user other than the attacker must interact with the system for the attack to succeed. In the case of Black-box jailbreaks, where most attacks are executed remotely, no additional user interaction is required—a point accurately identified by GPT-4o and Perplexity AI. However, LLAMA-3 initially marked the UI factor as `Required', justifying this based on the attacker’s interaction with the system. The HitL \textbf{clarified} through prompts that the UI factor refers specifically to interactions by users other than the attacker. Following this explanation, LLAMA-3 adjusted its evaluation, aligning with the `None' rating given by the other LLMs.

After averaging the scores, the final results are visualized below in a spider chart. The CVSS scores highlight that Black-box jailbreak attacks are easier to reproduce compared to White-box jailbreaks, require no privileges, and have a low-to-moderate impact on both integrity and confidentiality.

\begin{center}
\begin{tikzpicture}
\begin{polaraxis}[
    title={Assessment of black-box model jailbreaking attacks with CVSS},
    title style={font=\bfseries\large},
    xtick={0, 45, 90, 135, 180, 225, 270, 315},
    xticklabels={Vector, Complexity, Privileges, User Interaction, Scope, Confidentiality, Integrity, Availability},
    xticklabel style={font=\bfseries, anchor=north}, 
    ymin=0, ymax=4,
    ytick={0,1,2,3,4},
    yticklabels={0,1,2,3,4}, 
    yticklabel style={font=\small}, 
    grid=both, 
    major grid style={line width=0.8pt, draw=gray!50},
    minor grid style={line width=0.4pt, draw=gray!20},
    ylabel near ticks,
    legend style={at={(1.3,1)}, anchor=north, legend columns=1}, 
]
    \addplot[line width=1.45pt, mark=x,
        color=red!80, fill=cyan!20, fill opacity=0.1] 
        coordinates {(0,1) (45,2) (90,1) (135,2) (180,2) (225,2) (270,2) (315,1) (360,1)} 
        \closedcycle;
    \addplot[line width=1.45pt, mark=o,
        color=green!80, fill=cyan!20, fill opacity=0.1] 
        coordinates {(0,1) (45,2) (90,1) (135,2) (180,2) (225,2) (270,3) (315,1) (360,1)}
        \closedcycle;
    \addplot[line width=1.45pt, mark=star,
        color=red!80, fill=cyan!20, fill opacity=0.1] 
        coordinates {(0,1) (45,2) (90,1) (135,2) (180,2) (225,2) (270,2) (315,1) (360,1)}
        \closedcycle;
    \addplot[line width=1.45pt, mark=x,
        color=blue!80, fill=cyan!20, fill opacity=0.1] 
        coordinates {(0,1) (45,2) (90,1) (135,2) (180,4) (225,2) (270,1) (315,1) (360,1)} 
        \closedcycle;
    \addplot[line width=1.45pt, mark=pentagon,
        color=blue!80, fill=cyan!20, fill opacity=0.1] 
        coordinates {(0,1) (45,2) (90,1) (135,2) (180,4) (225,2) (270,1) (315,1) (360,1)} 
        \closedcycle;
    \addplot[line width=1.45pt, mark=o,
        color=orange!80, fill=cyan!20, fill opacity=0.1] 
        coordinates {(0,1) (45,4) (90,1) (135,2) (180,2) (225,2) (270,1) (315,1) (360,1)} 
        \closedcycle;
    \addplot[line width=1.45pt, mark=square,
        color=violet!80, fill=cyan!20, fill opacity=0.1] 
        coordinates {(0,1) (45,2) (90,2) (135,2) (180,2) (225,2) (270,3) (315,1) (360,1)} 
        \closedcycle;
    \addplot[line width=1.45pt, mark=none,
        color=green!80, fill=cyan!20, fill opacity=0.1] 
        coordinates {(0,1) (45,2) (90,1) (135,2) (180,2) (225,2) (270,3) (315,1) (360,1)} 
        \closedcycle;
    
    \legend{Attack 1, Attack 2, Attack 3, Attack 4, Attack 5, Attack 6, Attack 7, Attack 8}
\end{polaraxis}
\end{tikzpicture}
\end{center}

\subsubsection{With OWASP Risk Rating}
A third evaluation of is done with OWASP Risk Rating \citep{owaspOWASPRisk}. The corresponding vulnerability vectors of each attack is:

\begin{itemize}
    \item (1) $\rightarrow$ (SL:6/M:8/O:8/S:6/ED:6/EE:7/A:5/ID:6/LC:8/LI:2/LA:1/FD:6/RD:8/NC:5/PV:7) = \textcolor{red}{3.3 (High)}
    \item (2) $\rightarrow$ (SL:5/M:8/O:8/S:6/ED:6/EE:7/A:6/ID:6/LC:8/LI:1/LA:1/FD:6/RD:8/NC:4/PV:6) = \textcolor{red}{3 (High)}
    \item (3) $\rightarrow$ (SL:4/M:8/O:8/S:6/ED:7/EE:8/A:7/ID:7/LC:9/LI:1/LA:1/FD:7/RD:8/NC:5/PV:9) = \textcolor{red}{3.8 (High)}
    \item (4) $\rightarrow$ (SL:7/M:8/O:7/S:5/ED:6/EE:8/A:6/ID:8/LC:8/LI:1/LA:1/FD:6/RD:8/NC:5/PV:8) = \textcolor{red}{3.5 (High)}
    \item (5) $\rightarrow$ (SL:6/M:8/O:8/S:5/ED:6/EE:8/A:5/ID:7/LC:7/LI:1/LA:1/FD:5/RD:7/NC:5/PV:7) = \textcolor{red}{3 (High)}
    \item (6) $\rightarrow$ (SL:6/M:8/O:8/S:6/ED:6/EE:8/A:6/ID:7/LC:7/LI:2/LA:1/FD:6/RD:7/NC:5/PV:7) = \textcolor{red}{3 (High)}
    \item (7) $\rightarrow$ (SL:6/M:9/O:8/S:5/ED:6/EE:8/A:5/ID:7/LC:8/LI:3/LA:1/FD:6/RD:9/NC:7/PV:8) = \textcolor{red}{3.8 (High)}
    \item (8) $\rightarrow$ (SL:6/M:8/O:7/S:5/ED:6/EE:8/A:5/ID:7/LC:8/LI:3/LA:1/FD:6/RD:8/NC:5/PV:8) = \textcolor{red}{3.4 (High)}
\end{itemize}

Table \ref{tab:appendix-blackbox-owasp} presents the detailed OWASP RR assessments conducted using three LLMs. Unlike previous evaluations, no significant errors were observed in the scoring provided by the models. However, some divergence was noted in specific factors. For example, when assessing the Opportunity factor for the sixth attack \citep{hayase2024querybasedadversarialpromptgeneration}, GPT-4o and LLAMA-3 scored it 8/10 and 9/10, respectively, arguing that these attacks target online LLMs, thereby increasing the availability of opportunities for exploitation. In contrast, Perplexity AI assigned a score of 6/10, reasoning that the attacks are not immediately apparent or straightforward to execute, resulting in a medium-to-high Opportunity rating. To maintain \textbf{neutrality} and \textbf{objectivity}, we chose not to modify or influence these values, allowing the models’ perspectives to remain intact. Averaging the scores enabled a \textbf{balanced} consideration of all three points of view.

The final scores are visualized below in the spider chart. The results indicate that Black-box jailbreak attacks have a significant impact on the confidentiality of data, as they can extract sensitive information from the models. In contrast, their impact on integrity is minimal, and they have no impact on availability. The OWASP RR metric further highlights the severe implications these attacks have on privacy violations and the reputation of the targeted organization.

\begin{center}
\begin{tikzpicture}
\begin{polaraxis}[
    title={Assessment of Black-box model jailbreaking attacks with OWASP RR},
    title style={font=\bfseries\large},
    xtick={0, 24, 48, 72, 96, 120, 144, 168, 192, 216, 240, 264, 288, 312, 336},
    xticklabels={Skill, Motivation, Opportunity, Size, Discovery, Exploit, Awareness, Intrusion Dtc, Confidentiality, Integrity, Availability, Fnc Dmg, Rpt Dmg, Non-comp, Privacy Vlt},
    xticklabel style={font=\bfseries, anchor=north}, 
    ymin=0, ymax=10,
    ytick={0,1,2,3,4,5,6,7,8,9,10},
    yticklabels={0,1,2,3,4,5,6,7,8,9,10}, 
    yticklabel style={font=\small}, 
    grid=both, 
    major grid style={line width=0.8pt, draw=gray!50},
    minor grid style={line width=0.4pt, draw=gray!20},
    ylabel near ticks,
    legend style={at={(1.3,1)}, anchor=north, legend columns=1}, 
]
    \addplot[line width=1.45pt, mark=square,
        color=blue!80, fill=green!20, fill opacity=0.1] 
        coordinates {(0,6) (24,8) (48,8) (72,6) (96,6) (120,7) (144,5) (168,6) (192,8) (216,2) (240,1) (264,6) (288,8) (312,5) (336,7) (360,6)} 
        \closedcycle;
    \addplot[line width=1.45pt, mark=triangle,
        color=red!80, fill=green!20, fill opacity=0.1] 
        coordinates {(0,5) (24,8) (48,8) (72,6) (96,6) (120,7) (144,6) (168,6) (192,8) (216,1) (240,1) (264,6) (288,8) (312,4) (336,6) (360,5)}
        \closedcycle;
    \addplot[line width=1.45pt, mark=diamond,
        color=violet!80, fill=green!20, fill opacity=0.1] 
        coordinates {(0,4) (24,8) (48,8) (72,6) (96,7) (120,8) (144,7) (168,7) (192,9) (216,1) (240,1) (264,7) (288,8) (312,5) (336,9) (360,4)}
        \closedcycle;
    \addplot[line width=1.45pt, mark=pentagon*,
        color=cyan!80, fill=green!20, fill opacity=0.1] 
        coordinates {(0,7) (24,8) (48,7) (72,5) (96,6) (120,8) (144,6) (168,8) (192,8) (216,1) (240,1) (264,6) (288,8) (312,5) (336,8) (360,7)} 
        \closedcycle;
    \addplot[line width=1.45pt, mark=pentagon,
        color=orange!80, fill=green!20, fill opacity=0.1] 
        coordinates {(0,6) (24,8) (48,8) (72,5) (96,6) (120,8) (144,5) (168,7) (192,7) (216,1) (240,1) (264,5) (288,7) (312,5) (336,7) (360,6)} 
        \closedcycle;
    \addplot[line width=1.45pt, mark=o,
        color=yellow!80, fill=green!20, fill opacity=0.1] 
        coordinates {(0,6) (24,8) (48,8) (72,6) (96,6) (120,8) (144,6) (168,7) (192,7) (216,2) (240,1) (264,6) (288,7) (312,5) (336,7) (360,6)} 
        \closedcycle;
    \addplot[line width=1.45pt, mark=o,
        color=magenta!80, fill=green!20, fill opacity=0.1] 
        coordinates {(0,6) (24,9) (48,8) (72,5) (96,6) (120,8) (144,5) (168,7) (192,8) (216,3) (240,1) (264,6) (288,9) (312,7) (336,8) (360,6)} 
        \closedcycle;
    \addplot[line width=1.45pt, mark=none,
        color=green!80, fill=green!20, fill opacity=0.1] 
        coordinates {(0,6) (24,8) (48,7) (72,5) (96,6) (120,8) (144,5) (168,7) (192,8) (216,3) (240,1) (264,6) (288,8) (312,5) (336,8) (360,6)} 
        \closedcycle;
    
    \legend{Attack 1, Attack 2, Attack 3, Attack 4, Attack 5, Attack 6, Attack 7, Attack 8}
\end{polaraxis}
\end{tikzpicture}
\end{center}

\subsubsection{With SSVC}
The forth evaluation is performed using SSVC \citep{spring2021prioritizing} in a supplier role. The corresponding vulnerability vectors are detailed below:

\begin{itemize}
    \item (1) $\rightarrow$ (E:P/A:Y/V:C/U:S/T:T/P:S) = \textcolor{red}{Immediate (Very High)}
    \item (2) $\rightarrow$ (E:P/A:Y/V:D/U:E/T:P/P:M) = \textcolor{yellow}{Scheduled (Medium)}
    \item (3) $\rightarrow$ (E:A/A:Y/V:C/U:S/T:T/P:S) = \textcolor{red}{Immediate (Very High)}
    \item (4) $\rightarrow$ (E:P/A:Y/V:C/U:S/T:T/P:S) = \textcolor{red}{Immediate (Very High)}
    \item (5) $\rightarrow$ (E:P/A:Y/V:C/U:S/T:T/P:S) = \textcolor{red}{Immediate (Very High)}
    \item (6) $\rightarrow$ (E:P/A:Y/V:D/U:E/T:P/P:M) = \textcolor{yellow}{Scheduled (Medium)}
    \item (7) $\rightarrow$ (E:A/A:Y/V:C/U:S/T:T/P:S) = \textcolor{red}{Immediate (Very High)}
    \item (8) $\rightarrow$ (E:P/A:Y/V:C/U:S/T:T/P:S) = \textcolor{red}{Immediate (Very High)}
\end{itemize}

Table \ref{tab:appendix-blackbox-ssvc} presents the SSVC scores assigned by the three LLMs. The primary challenge encountered during this assessment was the ability of the LLMs to remain \textbf{up-to-date}. Specifically, some attacks might have been actively exploited in the past but are now less prevalent. For example, in the case of the third and fourth attacks \citep{shen2024donowcharacterizingevaluating, andriushchenko2024jailbreakingleadingsafetyalignedllms}, some LLMs classified these as `Active', while others evaluated them at the `Proof-of-Concept' stage. Determining which LLM is correct in such scenarios is challenging. To address this, we prompted the LLMs to confirm their assessments by asking clarifying questions such as: \textit{`Are there proofs of recent active exploitations of these attacks?'} This approach led to adjustments in certain scores. For instance, LLAMA-3 revised its assessment for the third attack from `Active' to `Proof-of-Concept', explaining that while the attack was previously active, there is no current evidence of active exploitation.

The final scores are visualized below in the spider chart. The results indicate that the SSVC scores align closely with those of DREAD, demonstrating that these Black-box jailbreak attacks are highly dangerous and straightforward to exploit, regardless of whether they target the CIA triad or financial aspects.

\begin{center}
\begin{tikzpicture}
\begin{polaraxis}[
    title={Assessment of Black-box model jailbreaking attacks with SSVC},
    title style={font=\bfseries\large},
    xtick={0, 60, 120, 180, 240, 300},
    xticklabels={Exploitation, Automatable, Value Density, Utility, Technical Imp, Public-Safety Imp},
    xticklabel style={font=\bfseries, anchor=north}, 
    ymin=0, ymax=3,
    ytick={0,1,2,3},
    yticklabels={0,1,2,3}, 
    yticklabel style={font=\small}, 
    grid=both, 
    major grid style={line width=0.8pt, draw=gray!50},
    minor grid style={line width=0.4pt, draw=gray!20},
    ylabel near ticks,
    legend style={at={(1.35,1)}, anchor=north, legend columns=1}, 
]
    \addplot[line width=1.45pt, mark=x,
        color=red!80, fill=red!20, fill opacity=0.1] 
        coordinates {(0,2) (60,3) (120,3) (180,3) (240,3) (300,3) (360,2)} 
        \closedcycle;
    \addplot[line width=1.45pt, mark=pentagon,
        color=blue!80, fill=red!20, fill opacity=0.1] 
        coordinates {(0,2) (60,3) (120,1) (180,2) (240,1) (300,1) (360,2)}
        \closedcycle;
    \addplot[line width=1.45pt, mark=square,
        color=red!80, fill=red!20, fill opacity=0.1] 
        coordinates {(0,3) (60,3) (120,3) (180,3) (240,3) (300,3) (360,2)}
        \closedcycle;
    \addplot[line width=1.45pt, mark=pentagon,
        color=red!80, fill=red!20, fill opacity=0.1] 
        coordinates {(0,2) (60,3) (120,3) (180,3) (240,3) (300,3) (360,2)} 
        \closedcycle;
    \addplot[line width=1.45pt, mark=pentagon,
        color=red!80, fill=red!20, fill opacity=0.1] 
        coordinates {(0,2) (60,3) (120,3) (180,3) (240,3) (300,3) (360,2)} 
        \closedcycle;
    \addplot[line width=1.45pt, mark=o,
        color=blue!80, fill=red!20, fill opacity=0.1] 
        coordinates {(0,2) (60,3) (120,1) (180,2) (240,1) (300,1) (360,2)} 
        \closedcycle;
    \addplot[line width=1.45pt, mark=square,
        color=green!80, fill=red!20, fill opacity=0.1] 
        coordinates {(0,3) (60,3) (120,3) (180,3) (240,3) (300,3) (360,3)} 
        \closedcycle;
    \addplot[line width=1.45pt, mark=none,
        color=red!80, fill=red!20, fill opacity=0.1] 
        coordinates {(0,2) (60,3) (120,3) (180,3) (240,3) (300,3) (360,2)} 
        \closedcycle;
    
    \legend{Attack 1, Attack 2, Attack 3, Attack 4, Attack 5, Attack 6, Attack 7, Attack 8}
\end{polaraxis}
\end{tikzpicture}
\end{center}

For the subsequent assessments, we will present only the results, as the justifications follow the same reasoning outlined for the White-box and Black-box Jailbreak attacks.

\subsection{Assessment of Prompt Injection attacks}
The third assessment is that of PI attacks, we evaluate the attacks described earlier in Section \ref{subsec:prompt-inject}: (1) Ignore Previous Prompt \citep{perez2022ignorepreviouspromptattack}, (2) Indirect Instruction Injection \citep{greshake2023youvesignedforcompromising} (3) Formalised Prompt Injection \citep{liu2024formalizingbenchmarkingpromptinjection}, (4) Injection through file input \citep{bagdasaryan2023abusingimagessoundsindirect}, (5) Universal Prompt Injection \citep{liu2024automaticuniversalpromptinjection}, (6) Virtual Prompt Injection \citep{yan2024backdooringinstructiontunedlargelanguage}, (7) Chat History Tampering \citep{wei2024hiddenplainsightexploring}, (8) JudgeDeceiverAttack \citep{shi2024optimizationbasedpromptinjectionattack}.


\subsubsection{With DREAD}
As done before, we start by evaluating the eight prompt injection attacks using DREAD \citep{michael2006security}, and we find the corresponding vulnerability vectors as follows:

\begin{itemize}
    \item (1) $\rightarrow$ (D:8/R:9/E:8/A:7/D:6) = \textcolor{red}{7.6 (High)}
    \item (2) $\rightarrow$ (D:8/R:8/E:8/A:7/D:6) = \textcolor{red}{7.4 (High)}
    \item (3) $\rightarrow$ (D:7/R:9/E:7/A:6/D:7) = \textcolor{red}{7.2 (High)}
    \item (4) $\rightarrow$ (D:7/R:8/E:8/A:8/D:5) = \textcolor{red}{7.2 (High)}
    \item (5) $\rightarrow$ (D:8/R:9/E:9/A:8/D:6) = \textcolor{red}{8 (High)}
    \item (6) $\rightarrow$ (D:8/R:8/E:7/A:8/D:5) = \textcolor{red}{7.2 (High)}
    \item (7) $\rightarrow$ (D:7/R:6/E:6/A:7/D:5) = \textcolor{yellow}{6.2 (Medium)}
    \item (8) $\rightarrow$ (D:7/R:6/E:7/A:6/D:5) = \textcolor{yellow}{6.2 (Medium)}
\end{itemize}

The detailed scores are shown in Table \ref{tab:appendix-promptinject-dread}, with the final results visualized in the Spider-chart below. The DREAD analysis reveals that Prompt-Injection attacks cause significant damage to systems and impact a wide range of users, but they are comparatively harder to exploit and reproduce than Jailbreak attacks.

\begin{center}
\begin{tikzpicture}
\begin{polaraxis}[
    title={Assessment of Prompt-injection attacks with DREAD},
    title style={font=\bfseries\large},
    xtick={0, 72, 144, 216, 288},
    xticklabels={Damage, Reproducibility, Exploitability, Affected users, Discoverability},
    xticklabel style={font=\bfseries, anchor=north}, 
    ymin=0, ymax=10,
    ytick={1,2,3,4,5,6,7,8,9,10},
    yticklabels={1,2,3,4,5,6,7,8,9,10}, 
    yticklabel style={font=\small}, 
    grid=both, 
    major grid style={line width=0.8pt, draw=gray!50},
    minor grid style={line width=0.4pt, draw=gray!20},
    ylabel near ticks,
    legend style={at={(1.3,1)}, anchor=north, legend columns=1}, 
]
    \addplot[line width=1.45pt, mark=*,
        color=magenta!80, fill=yellow!20, fill opacity=0.1] 
        coordinates {(0,8) (72,9) (144,8) (216,7) (288,6) (360,8)} 
        \closedcycle;
    \addplot[line width=1.45pt, mark=square*,
        color=orange!80, fill=yellow!20, fill opacity=0.1] 
        coordinates {(0,8) (72,8) (144,8) (216,7) (288,6) (360,8)} 
        \closedcycle;
    \addplot[line width=1.45pt, mark=triangle*,
        color=green!80, fill=yellow!20, fill opacity=0.1] 
        coordinates {(0,7) (72,9) (144,7) (216,6) (288,7) (360,7)} 
        \closedcycle;
    \addplot[line width=1.45pt, mark=diamond*,
        color=blue!80, fill=yellow!20, fill opacity=0.1] 
        coordinates {(0,7) (72,8) (144,8) (216,8) (288,5) (360,7)} 
        \closedcycle;
    \addplot[line width=1.45pt, mark=pentagon*,
        color=violet!80, fill=yellow!20, fill opacity=0.5] 
        coordinates {(0,8) (72,9) (144,9) (216,8) (288,6) (360,8)} 
        \closedcycle;
    \addplot[line width=1.45pt, mark=o,
        color=red!80, fill=yellow!20, fill opacity=0.1] 
        coordinates {(0,8) (72,8) (144,7) (216,8) (288,5) (360,8)} 
        \closedcycle;
    \addplot[line width=1.45pt, mark=x,
        color=cyan!80, fill=yellow!20, fill opacity=0.1] 
        coordinates {(0,7) (72,6) (144,6) (216,7) (288,5) (360,7)} 
        \closedcycle;
    \addplot[line width=1.45pt, mark=none,
        color=yellow!80, fill=yellow!20, fill opacity=0.1] 
        coordinates {(0,7) (72,6) (144,7) (216,6) (288,5) (360,7)} 
        \closedcycle;
    
    \legend{Attack 1, Attack 2, Attack 3, Attack 4, Attack 5, Attack 6, Attack 7, Attack 8}
\end{polaraxis}
\end{tikzpicture}
\end{center}

\subsubsection{With CVSS}
The second assessment of PI attacks is done with CVSS \citep{schiffman2005complete}. The corresponding CVSS Vectors are shown below:

\begin{itemize}
    \item (1) $\rightarrow$ (AV:N/AC:L/PR:N/UI:N/S:U/C:N/I:H/A:N) = \textcolor{red}{7.5 (High)}
    \item (2) $\rightarrow$ (AV:N/AC:H/PR:N/UI:R/S:U/C:L/I:H/A:N) = \textcolor{yellow}{5.9 (Medium)}
    \item (3) $\rightarrow$ (AV:N/AC:L/PR:N/UI:N/S:U/C:L/I:H/A:N) = \textcolor{red}{8.2 (High)}
    \item (4) $\rightarrow$ (AV:N/AC:H/PR:N/UI:R/S:C/C:L/I:H/A:N) = \textcolor{yellow}{6.9 (Medium)}
    \item (5) $\rightarrow$ (AV:N/AC:L/PR:N/UI:N/S:C/C:L/I:H/A:N) = \textcolor{purple}{9.3 (Critical)}
    \item (6) $\rightarrow$ (AV:N/AC:H/PR:N/UI:R/S:C/C:L/I:H/A:N) = \textcolor{yellow}{6.9 (Medium)}
    \item (7) $\rightarrow$ (AV:N/AC:H/PR:N/UI:R/S:U/C:L/I:H/A:N) = \textcolor{yellow}{5.9 (Medium)}
    \item (8) $\rightarrow$ (AV:N/AC:H/PR:N/UI:N/S:U/C:L/I:H/A:N) = \textcolor{yellow}{6.5 (Medium)}
\end{itemize}

The detailed CVSS results are presented in Table \ref{tab:appendix-promptinject-cvss}, with the final scores visualized in the Spider-chart below. The analysis indicates that Prompt-Injection attacks share similarities with Jailbreak attacks, as they are primarily executed remotely through the network. However, they are slightly more complex to perform than Jailbreak attacks. These attacks predominantly target system integrity, have a lesser impact on confidentiality, and do not affect availability.
 
\begin{center}
\begin{tikzpicture}
\begin{polaraxis}[
    title={Assessment of Prompt-injection attacks with CVSS},
    title style={font=\bfseries\large},
    xtick={0, 45, 90, 135, 180, 225, 270, 315},
    xticklabels={Vector, Complexity, Privileges, User Interaction, Scope, Confidentiality, Integrity, Availability},
    xticklabel style={font=\bfseries, anchor=north}, 
    ymin=0, ymax=4,
    ytick={0,1,2,3,4},
    yticklabels={0,1,2,3,4}, 
    yticklabel style={font=\small}, 
    grid=both, 
    major grid style={line width=0.8pt, draw=gray!50},
    minor grid style={line width=0.4pt, draw=gray!20},
    ylabel near ticks,
    legend style={at={(1.3,1)}, anchor=north, legend columns=1}, 
]
    \addplot[line width=1.45pt, mark=square*,
        color=blue!80, fill=cyan!20, fill opacity=0.1] 
        coordinates {(0,1) (45,2) (90,1) (135,2) (180,2) (225,1) (270,3) (315,1) (360,1)} 
        \closedcycle;
    \addplot[line width=1.45pt, mark=*,
        color=violet!80, fill=cyan!20, fill opacity=0.1] 
        coordinates {(0,1) (45,4) (90,1) (135,4) (180,2) (225,2) (270,3) (315,1) (360,1)}
        \closedcycle;
    \addplot[line width=1.45pt, mark=square,
        color=red!80, fill=cyan!20, fill opacity=0.1] 
        coordinates {(0,1) (45,2) (90,1) (135,2) (180,2) (225,2) (270,3) (315,1) (360,1)}
        \closedcycle;
    \addplot[line width=1.45pt, mark=pentagon,
        color=orange!80, fill=cyan!20, fill opacity=0.1] 
        coordinates {(0,1) (45,4) (90,1) (135,4) (180,4) (225,2) (270,3) (315,1) (360,1)} 
        \closedcycle;
    \addplot[line width=1.45pt, mark=o,
        color=green!80, fill=cyan!20, fill opacity=0.1] 
        coordinates {(0,1) (45,2) (90,1) (135,2) (180,4) (225,2) (270,3) (315,1) (360,1)} 
        \closedcycle;
    \addplot[line width=1.45pt, mark=o,
        color=orange!80, fill=cyan!20, fill opacity=0.1] 
        coordinates {(0,1) (45,4) (90,1) (135,4) (180,4) (225,2) (270,3) (315,1) (360,1)} 
        \closedcycle;
    \addplot[line width=1.45pt, mark=pentagon,
        color=violet!80, fill=cyan!20, fill opacity=0.1] 
        coordinates {(0,1) (45,4) (90,1) (135,4) (180,2) (225,2) (270,3) (315,1) (360,1)} 
        \closedcycle;
    \addplot[line width=1.45pt, mark=none,
        color=yellow!80, fill=cyan!20, fill opacity=0.1] 
        coordinates {(0,1) (45,4) (90,1) (135,2) (180,2) (225,2) (270,3) (315,1) (360,1)} 
        \closedcycle;
    
    \legend{Attack 1, Attack 2, Attack 3, Attack 4, Attack 5, Attack 6, Attack 7, Attack 8}
\end{polaraxis}
\end{tikzpicture}
\end{center}

\subsubsection{With OWASP Risk Rating}
Another evaluation of Prompt Injection attacks is done with OWASP Risk Rating \citep{owaspOWASPRisk}. The corresponding vulnerability vectors of each attack is:

\begin{itemize}
    \item (1) $\rightarrow$ (SL:6/M:8/O:7/S:5/ED:6/EE:8/A:6/ID:6/LC:3/LI:8/LA:3/FD:6/RD:8/NC:4/PV:4) = \textcolor{red}{3.3 (High)}
    \item (2) $\rightarrow$ (SL:6/M:8/O:7/S:6/ED:6/EE:7/A:5/ID:7/LC:5/LI:8/LA:3/FD:7/RD:8/NC:4/PV:6) = \textcolor{red}{3.8 (High)}
    \item (3) $\rightarrow$ (SL:6/M:8/O:7/S:5/ED:6/EE:8/A:6/ID:6/LC:5/LI:8/LA:3/FD:7/RD:8/NC:4/PV:5) = \textcolor{red}{3.7 (High)}
    \item (4) $\rightarrow$ (SL:6/M:8/O:7/S:6/ED:6/EE:8/A:6/ID:7/LC:6/LI:7/LA:3/FD:7/RD:8/NC:5/PV:7) = \textcolor{purple}{4.1 (Critical)}
    \item (5) $\rightarrow$ (SL:5/M:8/O:7/S:6/ED:6/EE:8/A:6/ID:7/LC:6/LI:7/LA:3/FD:7/RD:8/NC:5/PV:6) = \textcolor{red}{3.8 (High)}
    \item (6) $\rightarrow$ (SL:7/M:8/O:8/S:6/ED:5/EE:8/A:5/ID:8/LC:7/LI:9/LA:3/FD:7/RD:8/NC:5/PV:7) = \textcolor{purple}{4.4 (Critical)}
    \item (7) $\rightarrow$ (SL:6/M:8/O:7/S:6/ED:5/EE:7/A:5/ID:6/LC:4/LI:8/LA:3/FD:6/RD:7/NC:3/PV:5) = \textcolor{yellow}{2.6 (Medium)}
    \item (8) $\rightarrow$ (SL:6/M:7/O:6/S:6/ED:5/EE:6/A:4/ID:6/LC:3/LI:8/LA:1/FD:5/RD:7/NC:2/PV:3) = \textcolor{yellow}{1.9 (Medium)}
\end{itemize}

The assessments conducted using three LLMs are detailed in Table \ref{tab:appendix-prompt-owasp}, and the final scores are depicted in the Spider-chart below for enhanced visualization. The OWASP RR results corroborate that Prompt-Injection attacks exert a greater impact on integrity than on confidentiality and availability. Additionally, they highlight the significant influence these attacks have on privacy violations and reputation damage, which are critical factors beyond the technical scope. Notably, the assessments also reveal a general lack of awareness among public users regarding these specific threats.

\begin{center}
\begin{tikzpicture}
\begin{polaraxis}[
    title={Assessment of Prompt-injection attacks with OWASP RR},
    title style={font=\bfseries\large},
    xtick={0, 24, 48, 72, 96, 120, 144, 168, 192, 216, 240, 264, 288, 312, 336},
    xticklabels={Skill, Motivation, Opportunity, Size, Discovery, Exploit, Awareness, Intrusion Dtc, Confidentiality, Integrity, Availability, Fnc Dmg, Rpt Dmg, Non-comp, Privacy Vlt},
    xticklabel style={font=\bfseries, anchor=north}, 
    ymin=0, ymax=10,
    ytick={0,1,2,3,4,5,6,7,8,9,10},
    yticklabels={0,1,2,3,4,5,6,7,8,9,10}, 
    yticklabel style={font=\small}, 
    grid=both, 
    major grid style={line width=0.8pt, draw=gray!50},
    minor grid style={line width=0.4pt, draw=gray!20},
    ylabel near ticks,
    legend style={at={(1.3,1)}, anchor=north, legend columns=1}, 
]
    \addplot[line width=1.45pt, mark=square,
        color=blue!80, fill=green!20, fill opacity=0.1] 
        coordinates {(0,6) (24,8) (48,7) (72,5) (96,6) (120,8) (144,6) (168,6) (192,3) (216,8) (240,3) (264,6) (288,8) (312,4) (336,4) (360,6)} 
        \closedcycle;
    \addplot[line width=1.45pt, mark=triangle,
        color=red!80, fill=green!20, fill opacity=0.1] 
        coordinates {(0,6) (24,8) (48,7) (72,6) (96,6) (120,7) (144,5) (168,7) (192,5) (216,8) (240,3) (264,7) (288,8) (312,4) (336,6) (360,6)}
        \closedcycle;
    \addplot[line width=1.45pt, mark=square,
        color=violet!80, fill=green!20, fill opacity=0.1] 
        coordinates {(0,6) (24,8) (48,7) (72,5) (96,6) (120,8) (144,6) (168,6) (192,5) (216,8) (240,3) (264,7) (288,7) (312,4) (336,5) (360,6)}
        \closedcycle;
    \addplot[line width=1.45pt, mark=pentagon*,
        color=green!80, fill=green!20, fill opacity=0.1] 
        coordinates {(0,6) (24,8) (48,7) (72,6) (96,6) (120,8) (144,6) (168,7) (192,6) (216,7) (240,3) (264,7) (288,6) (312,5) (336,7) (360,6)} 
        \closedcycle;
    \addplot[line width=1.45pt, mark=square,
        color=orange!80, fill=green!20, fill opacity=0.1] 
        coordinates {(0,5) (24,8) (48,7) (72,6) (96,6) (120,8) (144,6) (168,7) (192,6) (216,7) (240,3) (264,7) (288,7) (312,5) (336,6) (360,5)} 
        \closedcycle;
    \addplot[line width=1.45pt, mark=diamond,
        color=magenta!80, fill=green!20, fill opacity=0.1] 
        coordinates {(0,7) (24,8) (48,8) (72,6) (96,5) (120,8) (144,5) (168,8) (192,7) (216,9) (240,3) (264,7) (288,7) (312,5) (336,7) (360,7)} 
        \closedcycle;
    \addplot[line width=1.45pt, mark=pentagon,
        color=yellow!80, fill=green!20, fill opacity=0.1] 
        coordinates {(0,6) (24,8) (48,7) (72,6) (96,5) (120,7) (144,5) (168,6) (192,4) (216,8) (240,3) (264,6) (288,6) (312,3) (336,5) (360,6)} 
        \closedcycle;
    \addplot[line width=1.45pt, mark=none,
        color=cyan!80, fill=green!20, fill opacity=0.1] 
        coordinates {(0,6) (24,8) (48,6) (72,6) (96,5) (120,6) (144,4) (168,6) (192,3) (216,8) (240,1) (264,5) (288,7) (312,2) (336,3) (360,6)} 
        \closedcycle;
    
    \legend{Attack 1, Attack 2, Attack 3, Attack 4, Attack 5, Attack 6, Attack 7, Attack 8}
\end{polaraxis}
\end{tikzpicture}
\end{center}

\subsubsection{With SSVC}
A last evaluation is performed using SSVC \citep{spring2021prioritizing} as done before, the results of the assessments are presented below:

\begin{itemize}
    \item (1) $\rightarrow$ (E:P/A:Y/V:C/U:S/T:T/P:S) = \textcolor{red}{Immediate (Very High)}
    \item (2) $\rightarrow$ (E:P/A:Y/V:C/U:S/T:T/P:S) = \textcolor{red}{Immediate (Very High)}
    \item (3) $\rightarrow$ (E:P/A:Y/V:C/U:S/T:T/P:S) = \textcolor{red}{Immediate (Very High)}
    \item (4) $\rightarrow$ (E:P/A:Y/V:C/U:S/T:T/P:S) = \textcolor{red}{Immediate (Very High)}
    \item (5) $\rightarrow$ (E:P/A:Y/V:C/U:S/T:T/P:S) = \textcolor{red}{Immediate (Very High)}
    \item (6) $\rightarrow$ (E:P/A:Y/V:C/U:S/T:T/P:S) = \textcolor{red}{Immediate (Very High)}
    \item (7) $\rightarrow$ (E:N/A:N/V:D/U:L/T:P/P:M) = \textcolor{green}{Defer (Low)}
    \item (8) $\rightarrow$ (E:P/A:Y/V:C/U:S/T:T/P:S) = \textcolor{red}{Immediate (Very High)}
\end{itemize}

Table \ref{tab:appendix-promptinject-ssvc} provides the detailed SSVC assessments conducted with the three LLMs. These scores offer additional insights beyond those captured by other metrics, emphasizing that Prompt Injection attacks are highly automatable, posing significant risks to both technical systems and public safety. The averaged results are visualized in the Spider-chart below for enhanced clarity.

\begin{center}
\begin{tikzpicture}
\begin{polaraxis}[
    title={Assessment of Prompt Injection attacks with SSVC},
    title style={font=\bfseries\large},
    xtick={0, 60, 120, 180, 240, 300},
    xticklabels={Exploitation, Automatable, Value Density, Utility, Technical Imp, Public-Safety Imp},
    xticklabel style={font=\bfseries, anchor=north}, 
    ymin=0, ymax=3,
    ytick={0,1,2,3},
    yticklabels={0,1,2,3}, 
    yticklabel style={font=\small}, 
    grid=both, 
    major grid style={line width=0.8pt, draw=gray!50},
    minor grid style={line width=0.4pt, draw=gray!20},
    ylabel near ticks,
    legend style={at={(1.35,1)}, anchor=north, legend columns=1}, 
]
    \addplot[line width=1.45pt, mark=x,
        color=red!80, fill=red!20, fill opacity=0.1] 
        coordinates {(0,2) (60,3) (120,3) (180,3) (240,3) (300,3) (360,2)} 
        \closedcycle;
    \addplot[line width=1.45pt, mark=pentagon,
        color=red!80, fill=red!20, fill opacity=0.1] 
        coordinates {(0,2) (60,3) (120,3) (180,3) (240,3) (300,3) (360,2)} 
        \closedcycle;
    \addplot[line width=1.45pt, mark=square,
        color=red!80, fill=red!20, fill opacity=0.1] 
        coordinates {(0,2) (60,3) (120,3) (180,3) (240,3) (300,3) (360,2)} 
        \closedcycle;
    \addplot[line width=1.45pt, mark=pentagon,
        color=red!80, fill=red!20, fill opacity=0.1] 
        coordinates {(0,2) (60,3) (120,3) (180,3) (240,3) (300,3) (360,2)} 
        \closedcycle;
    \addplot[line width=1.45pt, mark=pentagon,
        color=red!80, fill=red!20, fill opacity=0.1] 
        coordinates {(0,2) (60,3) (120,3) (180,3) (240,3) (300,3) (360,2)} 
        \closedcycle;
    \addplot[line width=1.45pt, mark=o,
        color=red!80, fill=red!20, fill opacity=0.1] 
        coordinates {(0,2) (60,3) (120,3) (180,3) (240,3) (300,3) (360,2)} 
        \closedcycle;
    \addplot[line width=1.45pt, mark=square,
        color=blue!80, fill=red!20, fill opacity=0.1] 
        coordinates {(0,1) (60,1) (120,1) (180,1) (240,1) (300,1) (360,1)} 
        \closedcycle;
    \addplot[line width=1.45pt, mark=none,
        color=red!80, fill=red!20, fill opacity=0.1] 
        coordinates {(0,2) (60,3) (120,3) (180,3) (240,3) (300,3) (360,2)} 
        \closedcycle;
    
    \legend{Attack 1, Attack 2, Attack 3, Attack 4, Attack 5, Attack 6, Attack 7, Attack 8}
\end{polaraxis}
\end{tikzpicture}
\end{center}

\subsection{Assessment of Evasion attacks}
The forth experiment is evaluating eight Evasion attacks described in Section \ref{subsec:evasion-attacks}: (1) Hot Flip \citep{ebrahimi2018hotflipwhiteboxadversarialexamples}, (2) PWWS \citep{ren-etal-2019-generating}, (3) TypoAttack \citep{pruthi2019combatingadversarialmisspellingsrobust}, (4) VIPER \citep{eger2020textprocessinglikehumans}, (5) CheckList \citep{ribeiro-etal-2020-beyond}, (6) BertAttack \citep{li-etal-2020-bert-attack}, (7) GBDA \citep{guo2021gradientbasedadversarialattackstext}, (8) TF-Attack \citep{li2024tfattacktransferablefastadversarial}. 


\subsubsection{With DREAD}
The first evaluation is done using DREAD \citep{michael2006security}. The corresponding vulnerability vectors as follows:

\begin{itemize}
    \item (1) $\rightarrow$ (D:7/R:7/E:6/A:7/D:5) = \textcolor{yellow}{6.4 (Medium)}
    \item (2) $\rightarrow$ (D:7/R:9/E:8/A:7/D:5) = \textcolor{red}{7.2 (High)}
    \item (3) $\rightarrow$ (D:6/R:8/E:6/A:6/D:5) = \textcolor{yellow}{6.2 (Medium)}
    \item (4) $\rightarrow$ (D:8/R:8/E:7/A:7/D:5) = \textcolor{red}{7 (High)}
    \item (5) $\rightarrow$ (D:6/R:9/E:7/A:7/D:6) = \textcolor{red}{7 (High)}
    \item (6) $\rightarrow$ (D:8/R:8/E:8/A:8/D:5) = \textcolor{red}{7.4 (High)}
    \item (7) $\rightarrow$ (D:9/R:8/E:8/A:8/D:5) = \textcolor{red}{7.6 (High)}
    \item (8) $\rightarrow$ (D:8/R:8/E:8/A:8/D:5) = \textcolor{red}{7.4 (High)}
\end{itemize}

Table \ref{tab:appendix-evasion-attack} presents the detailed assessments conducted with the three LLMs, with the final scores visualized in the Spider-chart below. The DREAD evaluation reveals that evasion attacks generally cause medium-to-high damage and are highly reproducible, easily exploitable, and difficult to detect, while having the potential to impact a wide range of users. This underscores the critical need to mitigate such attacks.

\begin{center}
\begin{tikzpicture}
\begin{polaraxis}[
    title={Assessment of Evasion attacks with DREAD},
    title style={font=\bfseries\large},
    xtick={0, 72, 144, 216, 288},
    xticklabels={Damage, Reproducibility, Exploitability, Affected users, Discoverability},
    xticklabel style={font=\bfseries, anchor=north}, 
    ymin=0, ymax=10,
    ytick={1,2,3,4,5,6,7,8,9,10},
    yticklabels={1,2,3,4,5,6,7,8,9,10}, 
    yticklabel style={font=\small}, 
    grid=both, 
    major grid style={line width=0.8pt, draw=gray!50},
    minor grid style={line width=0.4pt, draw=gray!20},
    ylabel near ticks,
    legend style={at={(1.3,1)}, anchor=north, legend columns=1}, 
]
    \addplot[line width=1.45pt, mark=*,
        color=cyan!80, fill=yellow!20, fill opacity=0.1] 
        coordinates {(0,7) (72,7) (144,6) (216,7) (288,5) (360,7)} 
        \closedcycle;
    \addplot[line width=1.45pt, mark=square*,
        color=yellow!80, fill=yellow!20, fill opacity=0.1] 
        coordinates {(0,7) (72,9) (144,8) (216,7) (288,5) (360,7)} 
        \closedcycle;
    \addplot[line width=1.45pt, mark=triangle*,
        color=green!80, fill=yellow!20, fill opacity=0.1] 
        coordinates {(0,6) (72,8) (144,6) (216,6) (288,5) (360,6)} 
        \closedcycle;
    \addplot[line width=1.45pt, mark=square,
        color=magenta!80, fill=yellow!20, fill opacity=0.1] 
        coordinates {(0,8) (72,8) (144,7) (216,7) (288,5) (360,8)} 
        \closedcycle;
    \addplot[line width=1.45pt, mark=pentagon*,
        color=violet!80, fill=yellow!20, fill opacity=0.1] 
        coordinates {(0,6) (72,9) (144,7) (216,7) (288,6) (360,6)} 
        \closedcycle;
    \addplot[line width=1.45pt, mark=pentagon*,
        color=red!80, fill=yellow!20, fill opacity=0.1] 
        coordinates {(0,8) (72,8) (144,8) (216,8) (288,5) (360,8)} 
        \closedcycle;
    \addplot[line width=1.45pt, mark=x,
        color=blue!80, fill=yellow!20, fill opacity=0.1] 
        coordinates {(0,9) (72,8) (144,8) (216,8) (288,5) (360,9)} 
        \closedcycle;
    \addplot[line width=1.45pt, mark=none,
        color=orange!80, fill=yellow!20, fill opacity=0.1] 
        coordinates {(0,8) (72,8) (144,8) (216,8) (288,5) (360,8)} 
        \closedcycle;
    
    \legend{Attack 1, Attack 2, Attack 3, Attack 4, Attack 5, Attack 6, Attack 7, Attack 8}
\end{polaraxis}
\end{tikzpicture}
\end{center}

\subsubsection{With CVSS}
The second assessment of Evasion attacks is done with CVSS \citep{schiffman2005complete}. The corresponding CVSS Vectors are shown below:

\begin{itemize}
    \item (1) $\rightarrow$ (AV:N/AC:L/PR:N/UI:N/S:U/C:N/I:H/A:N) = \textcolor{red}{7.5 (High)}
    \item (2) $\rightarrow$ (AV:N/AC:L/PR:N/UI:N/S:U/C:N/I:H/A:N) = \textcolor{red}{7.5 (High)}
    \item (3) $\rightarrow$ (AV:N/AC:H/PR:N/UI:N/S:U/C:N/I:H/A:N) = \textcolor{yellow}{5.9 (Medium)}
    \item (4) $\rightarrow$ (AV:N/AC:H/PR:N/UI:N/S:U/C:N/I:H/A:N) = \textcolor{yellow}{5.9 (Medium)}
    \item (5) $\rightarrow$ (AV:N/AC:L/PR:N/UI:N/S:U/C:N/I:H/A:N) = \textcolor{red}{7.5 (High)}
    \item (6) $\rightarrow$ (AV:N/AC:L/PR:N/UI:N/S:U/C:N/I:H/A:N) = \textcolor{red}{7.5 (High)}
    \item (7) $\rightarrow$ (AV:N/AC:H/PR:N/UI:N/S:U/C:N/I:H/A:N) = \textcolor{yellow}{5.9 (Medium)}
    \item (8) $\rightarrow$ (AV:N/AC:L/PR:N/UI:N/S:U/C:N/I:H/A:N) = \textcolor{red}{7.5 (High)}
\end{itemize}

Table \ref{tab:appendix-evasion-cvss} displays the scores provided by the three LLMs along with their average. For enhanced clarity and ease of interpretation, the score vectors are visualized in the Spider-chart below.

The CVSS evaluations reveal a consistent scoring pattern for evasion attacks, emphasizing their typical characteristics. These attacks are often performed over a network, require minimal complexity, and do not necessitate privileges or user interaction. While they have no impact on data Confidentiality or Availability, they can significantly affect Integrity.

\begin{center}
\begin{tikzpicture}
\begin{polaraxis}[
    title={Assessment of Evasion attacks with CVSS},
    title style={font=\bfseries\large},
    xtick={0, 45, 90, 135, 180, 225, 270, 315},
    xticklabels={Vector, Complexity, Privileges, User Interaction, Scope, Confidentiality, Integrity, Availability},
    xticklabel style={font=\bfseries, anchor=north}, 
    ymin=0, ymax=4,
    ytick={0,1,2,3,4},
    yticklabels={0,1,2,3,4}, 
    yticklabel style={font=\small}, 
    grid=both, 
    major grid style={line width=0.8pt, draw=gray!50},
    minor grid style={line width=0.4pt, draw=gray!20},
    ylabel near ticks,
    legend style={at={(1.3,1)}, anchor=north, legend columns=1}, 
]
    \addplot[line width=1.45pt, mark=square*,
        color=blue!80, fill=cyan!20, fill opacity=0.1] 
        coordinates {(0,1) (45,2) (90,1) (135,2) (180,2) (225,1) (270,3) (315,1) (360,1)} 
        \closedcycle;
    \addplot[line width=1.45pt, mark=pentagon,
        color=blue!80, fill=cyan!20, fill opacity=0.1] 
        coordinates {(0,1) (45,2) (90,1) (135,2) (180,2) (225,1) (270,3) (315,1) (360,1)}
        \closedcycle;
    \addplot[line width=1.45pt, mark=square,
        color=violet!80, fill=cyan!20, fill opacity=0.1] 
        coordinates {(0,1) (45,4) (90,1) (135,2) (180,2) (225,1) (270,3) (315,1) (360,1)}
        \closedcycle;
    \addplot[line width=1.45pt, mark=pentagon,
        color=violet!80, fill=cyan!20, fill opacity=0.1] 
        coordinates {(0,1) (45,4) (90,1) (135,2) (180,2) (225,1) (270,3) (315,1) (360,1)} 
        \closedcycle;
    \addplot[line width=1.45pt, mark=x,
        color=blue!80, fill=cyan!20, fill opacity=0.1] 
        coordinates {(0,1) (45,2) (90,1) (135,2) (180,2) (225,1) (270,3) (315,1) (360,1)} 
        \closedcycle;
    \addplot[line width=1.45pt, mark=o,
        color=blue!80, fill=cyan!20, fill opacity=0.1] 
        coordinates {(0,1) (45,2) (90,1) (135,2) (180,2) (225,1) (270,3) (315,1) (360,1)} 
        \closedcycle;
    \addplot[line width=1.45pt, mark=triangle,
        color=violet!80, fill=cyan!20, fill opacity=0.1] 
        coordinates {(0,1) (45,4) (90,1) (135,2) (180,2) (225,1) (270,3) (315,1) (360,1)} 
        \closedcycle;
    \addplot[line width=1.45pt, mark=none,
        color=blue!80, fill=cyan!20, fill opacity=0.1] 
        coordinates {(0,1) (45,2) (90,1) (135,2) (180,2) (225,1) (270,3) (315,1) (360,1)} 
        \closedcycle;
    
    \legend{Attack 1, Attack 2, Attack 3, Attack 4, Attack 5, Attack 6, Attack 7, Attack 8}
\end{polaraxis}
\end{tikzpicture}
\end{center}

\subsubsection{With OWASP Risk Rating}
Another evaluation of Evasion attacks is done with OWASP Risk Rating \citep{owaspOWASPRisk}. The corresponding vulnerability vectors of each attack is:

\begin{itemize}
    \item (1) $\rightarrow$ (SL:7/M:7/O:5/S:4/ED:5/EE:7/A:5/ID:6/LC:1/LI:8/LA:0/FD:5/RD:7/NC:4/PV:3) = \textcolor{yellow}{2.2 (Medium)}
    \item (2) $\rightarrow$ (SL:6/M:7/O:6/S:5/ED:5/EE:7/A:5/ID:6/LC:1/LI:8/LA:0/FD:6/RD:7/NC:5/PV:3) = \textcolor{yellow}{2.4 (Medium)}
    \item (3) $\rightarrow$ (SL:6/M:7/O:5/S:5/ED:5/EE:6/A:4/ID:5/LC:1/LI:7/LA:1/FD:5/RD:6/NC:4/PV:3) = \textcolor{yellow}{2 (Medium)}
    \item (4) $\rightarrow$ (SL:7/M:8/O:6/S:5/ED:5/EE:7/A:5/ID:6/LC:2/LI:8/LA:1/FD:7/RD:8/NC:5/PV:4) = \textcolor{red}{3 (High)}
    \item (5) $\rightarrow$ (SL:6/M:7/O:6/S:5/ED:5/EE:7/A:5/ID:6/LC:1/LI:7/LA:1/FD:6/RD:7/NC:5/PV:4) = \textcolor{yellow}{2.5 (Medium)}
    \item (6) $\rightarrow$ (SL:7/M:8/O:7/S:5/ED:6/EE:7/A:6/ID:7/LC:2/LI:9/LA:1/FD:7/RD:8/NC:6/PV:5) = \textcolor{red}{3.5 (High)}
    \item (7) $\rightarrow$ (SL:7/M:8/O:7/S:5/ED:6/EE:7/A:6/ID:7/LC:2/LI:9/LA:1/FD:7/RD:8/NC:6/PV:5) = \textcolor{red}{3.5 (High)}
    \item (8) $\rightarrow$ (SL:7/M:8/O:7/S:6/ED:6/EE:8/A:6/ID:7/LC:1/LI:8/LA:1/FD:7/RD:8/NC:5/PV:4) = \textcolor{red}{3.2 (High)}
\end{itemize}

The detailed OWASP RR scoring is outlined in Table \ref{tab:appendix-evasion-owasp}, offering insights consistent with those from the CVSS assessments. It highlights that evasion attacks demand only a moderate level of skill and motivation to be executed, are easily exploitable, and are relatively unknown to defenders, making them challenging to detect and mitigate. These attacks pose a significant threat to data integrity while remaining harmless to Confidentiality and Availability. Additionally, OWASP RR sheds light on the substantial financial and reputational impact these attacks can impose on targeted organizations.

\begin{center}
\begin{tikzpicture}
\begin{polaraxis}[
    title={Assessment of Evasion attacks with OWASP RR},
    title style={font=\bfseries\large},
    xtick={0, 24, 48, 72, 96, 120, 144, 168, 192, 216, 240, 264, 288, 312, 336},
    xticklabels={Skill, Motivation, Opportunity, Size, Discovery, Exploit, Awareness, Intrusion Dtc, Confidentiality, Integrity, Availability, Fnc Dmg, Rpt Dmg, Non-comp, Privacy Vlt},
    xticklabel style={font=\bfseries, anchor=north}, 
    ymin=0, ymax=10,
    ytick={0,1,2,3,4,5,6,7,8,9,10},
    yticklabels={0,1,2,3,4,5,6,7,8,9,10}, 
    yticklabel style={font=\small}, 
    grid=both, 
    major grid style={line width=0.8pt, draw=gray!50},
    minor grid style={line width=0.4pt, draw=gray!20},
    ylabel near ticks,
    legend style={at={(1.3,1)}, anchor=north, legend columns=1}, 
]
    \addplot[line width=1.45pt, mark=square,
        color=blue!80, fill=green!20, fill opacity=0.1] 
        coordinates {(0,7) (24,7) (48,5) (72,4) (96,5) (120,7) (144,5) (168,6) (192,1) (216,8) (240,0) (264,5) (288,7) (312,4) (336,3) (360,7)} 
        \closedcycle;
    \addplot[line width=1.45pt, mark=triangle,
        color=red!80, fill=green!20, fill opacity=0.1] 
        coordinates {(0,6) (24,7) (48,6) (72,5) (96,5) (120,7) (144,5) (168,6) (192,1) (216,8) (240,0) (264,6) (288,7) (312,5) (336,3) (360,6)}
        \closedcycle;
    \addplot[line width=1.45pt, mark=square,
        color=violet!80, fill=green!20, fill opacity=0.1] 
        coordinates {(0,6) (24,7) (48,5) (72,5) (96,5) (120,6) (144,4) (168,5) (192,1) (216,7) (240,1) (264,5) (288,6) (312,4) (336,3) (360,6)}
        \closedcycle;
    \addplot[line width=1.45pt, mark=pentagon,
        color=green!80, fill=green!20, fill opacity=0.1] 
        coordinates {(0,7) (24,8) (48,6) (72,5) (96,5) (120,7) (144,5) (168,6) (192,2) (216,8) (240,1) (264,7) (288,8) (312,5) (336,4) (360,7)} 
        \closedcycle;
    \addplot[line width=1.45pt, mark=*,
        color=orange!80, fill=green!20, fill opacity=0.1] 
        coordinates {(0,6) (24,7) (48,6) (72,5) (96,5) (120,7) (144,5) (168,6) (192,1) (216,7) (240,1) (264,6) (288,7) (312,5) (336,4) (360,6)} 
        \closedcycle;
    \addplot[line width=1.45pt, mark=o,
        color=magenta!80, fill=green!20, fill opacity=0.1] 
        coordinates {(0,7) (24,8) (48,7) (72,5) (96,6) (120,7) (144,6) (168,7) (192,2) (216,9) (240,1) (264,7) (288,8) (312,6) (336,5) (360,7)} 
        \closedcycle;
    \addplot[line width=1.45pt, mark=pentagon,
        color=magenta!80, fill=green!20, fill opacity=0.1] 
        coordinates {(0,7) (24,8) (48,7) (72,5) (96,6) (120,7) (144,6) (168,7) (192,2) (216,9) (240,1) (264,7) (288,8) (312,6) (336,5) (360,7)} 
        \closedcycle;
    \addplot[line width=1.45pt, mark=none,
        color=cyan!80, fill=green!20, fill opacity=0.1] 
        coordinates {(0,7) (24,8) (48,7) (72,6) (96,6) (120,8) (144,6) (168,7) (192,1) (216,8) (240,1) (264,7) (288,8) (312,5) (336,4) (360,7)} 
        \closedcycle;
    
    \legend{Attack 1, Attack 2, Attack 3, Attack 4, Attack 5, Attack 6, Attack 7, Attack 8}
\end{polaraxis}
\end{tikzpicture}
\end{center}

\subsubsection{With SSVC}
We continue the evaluation of Evasion attacks with SSVC \citep{spring2021prioritizing} as a last metric, the results of the assessments are shown and detailed below:

\begin{itemize}
    \item (1) $\rightarrow$ (E:P/A:Y/V:C/U:S/T:P/P:S) = \textcolor{red}{Immediate (Very High)}
    \item (2) $\rightarrow$ (E:P/A:Y/V:C/U:S/T:P/P:S) = \textcolor{red}{Immediate (Very High)}
    \item (3) $\rightarrow$ (E:A/A:Y/V:C/U:S/T:P/P:S) = \textcolor{red}{Immediate (Very High)}
    \item (4) $\rightarrow$ (E:P/A:Y/V:C/U:S/T:P/P:S) = \textcolor{red}{Immediate (Very High)}
    \item (5) $\rightarrow$ (E:P/A:N/V:D/U:L/T:P/P:M) = \textcolor{yellow}{Scheduled (Medium)}
    \item (6) $\rightarrow$ (E:P/A:Y/V:C/U:S/T:P/P:S) = \textcolor{red}{Immediate (Very High)}
    \item (7) $\rightarrow$ (E:P/A:Y/V:C/U:S/T:P/P:S) = \textcolor{red}{Immediate (Very High)}
    \item (8) $\rightarrow$ (E:P/A:Y/V:C/U:S/T:P/P:S) = \textcolor{red}{Immediate (Very High)}
\end{itemize}

The scores are visualized below in a Spider-chart, with the detailed assessments provided in Table \ref{tab:appendix-evasion-ssvc}. Notably, the SSVC results align with those of DREAD and OWASP RR, emphasizing that evasion attacks are frequently exploited by attackers. Additionally, SSVC highlights that these attacks are highly automatable and rewarding, making them particularly valuable to adversaries. However, it suggests that while evasion attacks pose minimal technical threats to organizations, their primary danger lies in their significant potential to compromise public safety, especially in scenarios involving object detection and classification.

\begin{center}
\begin{tikzpicture}
\begin{polaraxis}[
    title={Assessment of Evasion attacks with SSVC},
    title style={font=\bfseries\large},
    xtick={0, 60, 120, 180, 240, 300},
    xticklabels={Exploitation, Automatable, Value Density, Utility, Technical Imp, Public-Safety Imp},
    xticklabel style={font=\bfseries, anchor=north}, 
    ymin=0, ymax=3,
    ytick={0,1,2,3},
    yticklabels={0,1,2,3}, 
    yticklabel style={font=\small}, 
    grid=both, 
    major grid style={line width=0.8pt, draw=gray!50},
    minor grid style={line width=0.4pt, draw=gray!20},
    ylabel near ticks,
    legend style={at={(1.35,1)}, anchor=north, legend columns=1}, 
]
    \addplot[line width=1.45pt, mark=x,
        color=red!80, fill=red!20, fill opacity=0.1] 
        coordinates {(0,2) (60,3) (120,3) (180,3) (240,1) (300,3) (360,2)} 
        \closedcycle;
    \addplot[line width=1.45pt, mark=pentagon,
        color=red!80, fill=red!20, fill opacity=0.1] 
        coordinates {(0,2) (60,3) (120,3) (180,3) (240,1) (300,3) (360,2)} 
        \closedcycle;
    \addplot[line width=1.45pt, mark=square,
        color=blue!80, fill=red!20, fill opacity=0.1] 
        coordinates {(0,3) (60,3) (120,3) (180,3) (240,1) (300,3) (360,3)} 
        \closedcycle;
    \addplot[line width=1.45pt, mark=pentagon,
        color=red!80, fill=red!20, fill opacity=0.1] 
        coordinates {(0,2) (60,3) (120,3) (180,3) (240,1) (300,3) (360,2)} 
        \closedcycle;
    \addplot[line width=1.45pt, mark=pentagon,
        color=green!80, fill=red!20, fill opacity=0.1] 
        coordinates {(0,2) (60,1) (120,1) (180,1) (240,1) (300,1) (360,2)} 
        \closedcycle;
    \addplot[line width=1.45pt, mark=o,
        color=red!80, fill=red!20, fill opacity=0.1] 
        coordinates {(0,2) (60,3) (120,3) (180,3) (240,1) (300,3) (360,2)} 
        \closedcycle;
    \addplot[line width=1.45pt, mark=square,
        color=red!80, fill=red!20, fill opacity=0.1] 
        coordinates {(0,2) (60,3) (120,3) (180,3) (240,1) (300,3) (360,2)} 
        \closedcycle;
    \addplot[line width=1.45pt, mark=none,
        color=red!80, fill=red!20, fill opacity=0.1] 
        coordinates {(0,2) (60,3) (120,3) (180,3) (240,1) (300,3) (360,2)} 
        \closedcycle;
    
    \legend{Attack 1, Attack 2, Attack 3, Attack 4, Attack 5, Attack 6, Attack 7, Attack 8}
\end{polaraxis}
\end{tikzpicture}
\end{center}

\subsection{Assessment of Model Extraction attacks}
Model Extraction are the fifth attacks we evaluate with the five vulnerability metrics. The attacks were presented earlier in Section \ref{subsec:model-extraction} and are respectively: (1) User Data Extraction \citep{274574}, (2) LLM Tricks \citep{yu2023bagtrickstrainingdata}, (3) Analysing PII Leakage \citep{lukas2023analyzingleakagepersonallyidentifiable}, (4) ETHICIST \citep{zhang2023ethicisttargetedtrainingdata}, (5) Scalable Extraction \citep{nasr2023scalableextractiontrainingdata}, (6) Output2Prompt \citep{zhang2024extractingpromptsinvertingllm}, (7) PII-Compass \citep{nakka2024piicompassguidingllmtraining}, (8) Alpaca VS Vicuna \citep{kassem2024alpacavicunausingllms}.


\subsubsection{With DREAD}
We start evaluating Extraction attacks using DREAD \citep{michael2006security}. The corresponding vulnerability vectors as follows:

\begin{itemize}
    \item (1) $\rightarrow$ (D:9/R:8/E:8/A:8/D:5) = \textcolor{red}{7.6 (High)}
    \item (2) $\rightarrow$ (D:8/R:9/E:8/A:7/D:5) = \textcolor{red}{7.4 (High)}
    \item (3) $\rightarrow$ (D:9/R:8/E:8/A:9/D:6) = \textcolor{red}{8 (High)}
    \item (4) $\rightarrow$ (D:8/R:8/E:7/A:7/D:5) = \textcolor{red}{7 (High)}
    \item (5) $\rightarrow$ (D:8/R:5/E:6/A:8/D:4) = \textcolor{yellow}{6.2 (Medium)}
    \item (6) $\rightarrow$ (D:7/R:7/E:7/A:7/D:6) = \textcolor{yellow}{6.8 (Medium)}
    \item (7) $\rightarrow$ (D:8/R:6/E:7/A:8/D:4) = \textcolor{yellow}{6.8 (Medium)}
    \item (8) $\rightarrow$ (D:7/R:7/E:7/A:6/D:5) = \textcolor{yellow}{6.4 (Medium)}
\end{itemize}

The detailed scores for this fifth type of attack are shown in Table \ref{tab:appendix-model-extraction-dread}, with the score vectors visualized in a Spider-chart below for clarity. The DREAD assessment reveals that the exploitability and discoverability of model extraction attacks vary depending on the specific implementation. However, all these attacks share a high level of danger to systems due to their potential to cause significant damage. Additionally, the analysis highlights that such attacks can directly or indirectly affect multiple users, while remaining relatively challenging to detect.

\begin{center}
\begin{tikzpicture}
\begin{polaraxis}[
    title={Assessment of model extraction attacks with DREAD},
    title style={font=\bfseries\large},
    xtick={0, 72, 144, 216, 288},
    xticklabels={Damage, Reproducibility, Exploitability, Affected users, Discoverability},
    xticklabel style={font=\bfseries, anchor=north}, 
    ymin=0, ymax=10,
    ytick={1,2,3,4,5,6,7,8,9,10},
    yticklabels={1,2,3,4,5,6,7,8,9,10}, 
    yticklabel style={font=\small}, 
    grid=both, 
    major grid style={line width=0.8pt, draw=gray!50},
    minor grid style={line width=0.4pt, draw=gray!20},
    ylabel near ticks,
    legend style={at={(1.3,1)}, anchor=north, legend columns=1}, 
]
    \addplot[line width=1.45pt, mark=*,
        color=magenta!80, fill=yellow!20, fill opacity=0.1] 
        coordinates {(0,9) (72,8) (144,8) (216,8) (288,5) (360,9)} 
        \closedcycle;
    \addplot[line width=1.45pt, mark=square*,
        color=orange!80, fill=yellow!20, fill opacity=0.1] 
        coordinates {(0,8) (72,9) (144,8) (216,7) (288,5) (360,8)} 
        \closedcycle;
    \addplot[line width=1.45pt, mark=triangle*,
        color=green!80, fill=yellow!20, fill opacity=0.1] 
        coordinates {(0,7) (72,9) (144,7) (216,6) (288,7) (360,7)} 
        \closedcycle;
    \addplot[line width=1.45pt, mark=diamond*,
        color=blue!80, fill=yellow!20, fill opacity=0.1] 
        coordinates {(0,9) (72,8) (144,8) (216,9) (288,6) (360,9)} 
        \closedcycle;
    \addplot[line width=1.45pt, mark=pentagon*,
        color=violet!80, fill=yellow!20, fill opacity=0.1] 
        coordinates {(0,8) (72,5) (144,6) (216,8) (288,4) (360,8)} 
        \closedcycle;
    \addplot[line width=1.45pt, mark=o,
        color=red!80, fill=yellow!20, fill opacity=0.1] 
        coordinates {(0,7) (72,7) (144,7) (216,7) (288,6) (360,7)} 
        \closedcycle;
    \addplot[line width=1.45pt, mark=x,
        color=cyan!80, fill=yellow!20, fill opacity=0.1] 
        coordinates {(0,8) (72,6) (144,7) (216,8) (288,4) (360,8)} 
        \closedcycle;
    \addplot[line width=1.45pt, mark=none,
        color=yellow!80, fill=yellow!20, fill opacity=0.1] 
        coordinates {(0,7) (72,7) (144,7) (216,6) (288,5) (360,7)} 
        \closedcycle;
    
    \legend{Attack 1, Attack 2, Attack 3, Attack 4, Attack 5, Attack 6, Attack 7, Attack 8}
\end{polaraxis}
\end{tikzpicture}
\end{center}

\subsubsection{With CVSS}
The second assessment of Model Extraction attacks is done with CVSS \citep{schiffman2005complete}. The corresponding Vectors are:

\begin{itemize}
    \item (1) $\rightarrow$ (AV:N/AC:H/PR:N/UI:N/S:C/C:H/I:N/A:N) = \textcolor{yellow}{6.8 (Medium)}
    \item (2) $\rightarrow$ (AV:N/AC:L/PR:N/UI:N/S:C/C:H/I:N/A:N) = \textcolor{red}{8.6 (High)}
    \item (3) $\rightarrow$ (AV:N/AC:L/PR:N/UI:N/S:C/C:H/I:N/A:N) = \textcolor{red}{8.6 (High)}
    \item (4) $\rightarrow$ (AV:N/AC:H/PR:N/UI:N/S:C/C:H/I:N/A:N) = \textcolor{yellow}{6.8 (Medium)}
    \item (5) $\rightarrow$ (AV:N/AC:H/PR:N/UI:N/S:C/C:H/I:N/A:N) = \textcolor{yellow}{6.8 (Medium)}
    \item (6) $\rightarrow$ (AV:N/AC:H/PR:N/UI:N/S:C/C:H/I:N/A:N) = \textcolor{yellow}{6.8 (Medium)}
    \item (7) $\rightarrow$ (AV:N/AC:H/PR:N/UI:N/S:C/C:H/I:N/A:N) = \textcolor{yellow}{6.8 (Medium)}
    \item (8) $\rightarrow$ (AV:N/AC:L/PR:N/UI:N/S:C/C:H/I:N/A:N) = \textcolor{red}{8.6 (High)}
\end{itemize}

The CVSS scores assigned by each LLM are detailed in Table \ref{tab:appendix-model-extraction-cvss}, with the final vectors visualized in the Spider-chart below. 
Compared to DREAD, CVSS provides more granular insights into the nature of the damage caused by these attacks, particularly their impact on confidentiality. 
Additionally, the assessment highlights that these attacks typically do not require specific privileges or user interaction for execution. 
However, their scope can vary depending on the type of data extracted, making them broader in target range than previous attack types.

\begin{center}
\begin{tikzpicture}
\begin{polaraxis}[
    title={Assessment of model extraction attacks with CVSS},
    title style={font=\bfseries\large},
    xtick={0, 45, 90, 135, 180, 225, 270, 315},
    xticklabels={Vector, Complexity, Privileges, User Interaction, Scope, Confidentiality, Integrity, Availability},
    xticklabel style={font=\bfseries, anchor=north}, 
    ymin=0, ymax=4,
    ytick={0,1,2,3,4},
    yticklabels={0,1,2,3,4}, 
    yticklabel style={font=\small}, 
    grid=both, 
    major grid style={line width=0.8pt, draw=gray!50},
    minor grid style={line width=0.4pt, draw=gray!20},
    ylabel near ticks,
    legend style={at={(1.3,1)}, anchor=north, legend columns=1}, 
]
    \addplot[line width=1.45pt, mark=square*,
        color=blue!80, fill=cyan!20, fill opacity=0.1] 
        coordinates {(0,1) (45,4) (90,1) (135,2) (180,4) (225,3) (270,1) (315,1) (360,1)} 
        \closedcycle;
    \addplot[line width=1.45pt, mark=*,
        color=violet!80, fill=cyan!20, fill opacity=0.1] 
        coordinates {(0,1) (45,2) (90,1) (135,2) (180,4) (225,3) (270,1) (315,1) (360,1)}
        \closedcycle;
    \addplot[line width=1.45pt, mark=square,
        color=violet!80, fill=cyan!20, fill opacity=0.1] 
        coordinates {(0,1) (45,2) (90,1) (135,2) (180,4) (225,3) (270,1) (315,1) (360,1)}
        \closedcycle;
    \addplot[line width=1.45pt, mark=pentagon,
        color=blue!80, fill=cyan!20, fill opacity=0.1] 
        coordinates {(0,1) (45,4) (90,1) (135,2) (180,4) (225,3) (270,1) (315,1) (360,1)} 
        \closedcycle;
    \addplot[line width=1.45pt, mark=o,
        color=blue!80, fill=cyan!20, fill opacity=0.1] 
        coordinates {(0,1) (45,4) (90,1) (135,2) (180,4) (225,3) (270,1) (315,1) (360,1)} 
        \closedcycle;
    \addplot[line width=1.45pt, mark=triangle,
        color=blue!80, fill=cyan!20, fill opacity=0.1] 
        coordinates {(0,1) (45,4) (90,1) (135,2) (180,4) (225,3) (270,1) (315,1) (360,1)} 
        \closedcycle;
    \addplot[line width=1.45pt, mark=x,
        color=blue!80, fill=cyan!20, fill opacity=0.1] 
        coordinates {(0,1) (45,4) (90,1) (135,2) (180,4) (225,3) (270,1) (315,1) (360,1)} 
        \closedcycle;
    \addplot[line width=1.45pt, mark=none,
        color=violet!80, fill=cyan!20, fill opacity=0.1] 
        coordinates {(0,1) (45,2) (90,1) (135,2) (180,4) (225,3) (270,1) (315,1) (360,1)} 
        \closedcycle;
    
    \legend{Attack 1, Attack 2, Attack 3, Attack 4, Attack 5, Attack 6, Attack 7, Attack 8}
\end{polaraxis}
\end{tikzpicture}
\end{center}

\subsubsection{With OWASP Risk Rating}
A third evaluation of Model Extraction attacks is done using OWASP RR \citep{owaspOWASPRisk}. The corresponding vulnerability vectors of each attack is:

\begin{itemize}
    \item (1) $\rightarrow$ (SL:7/M:8/O:7/S:6/ED:5/EE:6/A:5/ID:7/LC:8/LI:1/LA:1/FD:7/RD:8/NC:7/PV:8) = \textcolor{red}{3.5 (High)}
    \item (2) $\rightarrow$ (SL:6/M:8/O:7/S:5/ED:6/EE:7/A:5/ID:7/LC:8/LI:2/LA:2/FD:7/RD:9/NC:7/PV:8) = \textcolor{red}{3.8 (High)}
    \item (3) $\rightarrow$ (SL:6/M:8/O:7/S:6/ED:6/EE:6/A:5/ID:8/LC:9/LI:1/LA:1/FD:7/RD:8/NC:7/PV:9) = \textcolor{red}{3.7 (High)}
    \item (4) $\rightarrow$ (SL:6/M:8/O:7/S:6/ED:5/EE:7/A:5/ID:7/LC:8/LI:1/LA:1/FD:7/RD:9/NC:7/PV:9) = \textcolor{red}{3.6 (High)}
    \item (5) $\rightarrow$ (SL:6/M:8/O:7/S:6/ED:5/EE:6/A:4/ID:7/LC:8/LI:1/LA:1/FD:7/RD:8/NC:7/PV:9) = \textcolor{red}{3.4 (High)}
    \item (6) $\rightarrow$ (SL:6/M:7/O:7/S:5/ED:5/EE:6/A:5/ID:7/LC:8/LI:2/LA:2/FD:7/RD:9/NC:6/PV:8) = \textcolor{red}{3.5 (High)}
    \item (7) $\rightarrow$ (SL:6/M:8/O:7/S:5/ED:5/EE:6/A:5/ID:7/LC:9/LI:2/LA:2/FD:7/RD:9/NC:7/PV:9) = \textcolor{purple}{3.8 (Critical)}
    \item (8) $\rightarrow$ (SL:7/M:8/O:7/S:6/ED:5/EE:6/A:5/ID:8/LC:8/LI:1/LA:1/FD:7/RD:9/NC:7/PV:9) = \textcolor{red}{3.7 (High)}
\end{itemize}

The detailed scores assigned by the LLMs are presented in Table \ref{tab:appendix-extraction-owasp}, with the final scores of each attack shown in the chart below for better visualization of their assessments. The OWASP RR scores align with the findings from DREAD and CVSS, offering additional insights. This metric reveals that Model Extraction attacks require only a moderate level of skill and motivation to be performed, and are easily exploitable. Notably, system administrators and defenders often lack awareness of these attacks and their potential risks, particularly their significant impact on data confidentiality. Additionally, these attacks pose a substantial threat to an organization's finances and reputation, while also leading to privacy violations that can result in increased audit challenges.

\begin{center}
\begin{tikzpicture}
\begin{polaraxis}[
    title={Assessment of Model-extraction attacks with OWASP RR},
    title style={font=\bfseries\large},
    xtick={0, 24, 48, 72, 96, 120, 144, 168, 192, 216, 240, 264, 288, 312, 336},
    xticklabels={Skill, Motivation, Opportunity, Size, Discovery, Exploit, Awareness, Intrusion Dtc, Confidentiality, Integrity, Availability, Fnc Dmg, Rpt Dmg, Non-comp, Privacy Vlt},
    xticklabel style={font=\bfseries, anchor=north}, 
    ymin=0, ymax=10,
    ytick={0,1,2,3,4,5,6,7,8,9,10},
    yticklabels={0,1,2,3,4,5,6,7,8,9,10}, 
    yticklabel style={font=\small}, 
    grid=both, 
    major grid style={line width=0.8pt, draw=gray!50},
    minor grid style={line width=0.4pt, draw=gray!20},
    ylabel near ticks,
    legend style={at={(1.3,1)}, anchor=north, legend columns=1}, 
]
    \addplot[line width=1.45pt, mark=square,
        color=blue!80, fill=green!20, fill opacity=0.1] 
        coordinates {(0,7) (24,8) (48,7) (72,6) (96,5) (120,6) (144,5) (168,7) (192,8) (216,1) (240,1) (264,7) (288,8) (312,7) (336,8) (360,7)} 
        \closedcycle;
    \addplot[line width=1.45pt, mark=triangle,
        color=red!80, fill=green!20, fill opacity=0.1] 
        coordinates {(0,6) (24,8) (48,7) (72,5) (96,6) (120,7) (144,5) (168,7) (192,8) (216,2) (240,2) (264,7) (288,9) (312,7) (336,8) (360,6)}
        \closedcycle;
    \addplot[line width=1.45pt, mark=square,
        color=violet!80, fill=green!20, fill opacity=0.1] 
        coordinates {(0,6) (24,8) (48,7) (72,6) (96,6) (120,6) (144,5) (168,8) (192,9) (216,1) (240,1) (264,7) (288,8) (312,7) (336,9) (360,6)}
        \closedcycle;
    \addplot[line width=1.45pt, mark=pentagon,
        color=green!80, fill=green!20, fill opacity=0.1] 
        coordinates {(0,6) (24,8) (48,7) (72,6) (96,5) (120,7) (144,5) (168,7) (192,8) (216,1) (240,1) (264,7) (288,9) (312,7) (336,9) (360,6)} 
        \closedcycle;
    \addplot[line width=1.45pt, mark=*,
        color=orange!80, fill=green!20, fill opacity=0.1] 
        coordinates {(0,6) (24,8) (48,7) (72,6) (96,5) (120,6) (144,4) (168,7) (192,8) (216,1) (240,1) (264,7) (288,8) (312,7) (336,9) (360,6)} 
        \closedcycle;
    \addplot[line width=1.45pt, mark=pentagon,
        color=magenta!80, fill=green!20, fill opacity=0.1] 
        coordinates {(0,6) (24,7) (48,7) (72,5) (96,5) (120,6) (144,5) (168,7) (192,8) (216,2) (240,2) (264,7) (288,9) (312,6) (336,8) (360,6)} 
        \closedcycle;
    \addplot[line width=1.45pt, mark=o,
        color=yellow!80, fill=green!20, fill opacity=0.1] 
        coordinates {(0,6) (24,8) (48,7) (72,5) (96,5) (120,6) (144,5) (168,7) (192,9) (216,2) (240,2) (264,7) (288,9) (312,7) (336,9) (360,6)} 
        \closedcycle;
    \addplot[line width=1.45pt, mark=none,
        color=cyan!80, fill=green!20, fill opacity=0.1] 
        coordinates {(0,7) (24,8) (48,7) (72,6) (96,5) (120,6) (144,5) (168,8) (192,8) (216,1) (240,1) (264,7) (288,9) (312,7) (336,9) (360,7)} 
        \closedcycle;
    
    \legend{Attack 1, Attack 2, Attack 3, Attack 4, Attack 5, Attack 6, Attack 7, Attack 8}
\end{polaraxis}
\end{tikzpicture}
\end{center}

\subsubsection{With SSVC}
The forth and last evaluation of Model-Extraction attacks is done with SSVC \citep{spring2021prioritizing} as previously, the results of the assessments are shown and detailed below:

\begin{itemize}
    \item (1) $\rightarrow$ (E:P/A:Y/V:C/U:S/T:T/P:S) = \textcolor{red}{Immediate (Very High)}
    \item (2) $\rightarrow$ (E:P/A:Y/V:C/U:S/T:T/P:S) = \textcolor{red}{Immediate (Very High)}
    \item (3) $\rightarrow$ (E:P/A:Y/V:C/U:S/T:T/P:S) = \textcolor{red}{Immediate (Very High)}
    \item (4) $\rightarrow$ (E:P/A:Y/V:C/U:S/T:T/P:S) = \textcolor{red}{Immediate (Very High)}
    \item (5) $\rightarrow$ (E:P/A:N/V:C/U:E/T:T/P:S) = \textcolor{red}{Immediate (Very High)}
    \item (6) $\rightarrow$ (E:P/A:Y/V:C/U:S/T:T/P:S) = \textcolor{red}{Immediate (Very High)}
    \item (7) $\rightarrow$ (E:P/A:Y/V:C/U:S/T:T/P:S) = \textcolor{red}{Immediate (Very High)}
    \item (8) $\rightarrow$ (E:P/A:Y/V:C/U:S/T:T/P:S) = \textcolor{red}{Immediate (Very High)}
\end{itemize}

Table \ref{tab:appendix-extraction-ssvc} presents the individual scores provided by each LLM along with their average, which is visualized below in a spider chart for a clearer understanding of the characteristics of Model-Extraction attacks.

The SSVC assessments highlight the ease of automation and high rewards associated with these attacks, making them particularly effective for adversaries. Notably, this metric emphasizes the significant impact these attacks can have on both the technical aspects of organizations and public safety, particularly by jeopardizing the privacy of user data.

\begin{center}
\begin{tikzpicture}
\begin{polaraxis}[
    title={Assessment of Model-Extraction attacks with SSVC},
    title style={font=\bfseries\large},
    xtick={0, 60, 120, 180, 240, 300},
    xticklabels={Exploitation, Automatable, Value Density, Utility, Technical Imp, Public-Safety Imp},
    xticklabel style={font=\bfseries, anchor=north}, 
    ymin=0, ymax=3,
    ytick={0,1,2,3},
    yticklabels={0,1,2,3}, 
    yticklabel style={font=\small}, 
    grid=both, 
    major grid style={line width=0.8pt, draw=gray!50},
    minor grid style={line width=0.4pt, draw=gray!20},
    ylabel near ticks,
    legend style={at={(1.35,1)}, anchor=north, legend columns=1}, 
]
    \addplot[line width=1.45pt, mark=x,
        color=red!80, fill=red!20, fill opacity=0.1] 
        coordinates {(0,2) (60,3) (120,3) (180,3) (240,3) (300,3) (360,2)} 
        \closedcycle;
    \addplot[line width=1.45pt, mark=pentagon,
        color=red!80, fill=red!20, fill opacity=0.1] 
        coordinates {(0,2) (60,3) (120,3) (180,3) (240,3) (300,3) (360,2)} 
        \closedcycle;
    \addplot[line width=1.45pt, mark=+,
        color=red!80, fill=red!20, fill opacity=0.1] 
        coordinates {(0,2) (60,3) (120,3) (180,3) (240,3) (300,3) (360,2)} 
        \closedcycle;
    \addplot[line width=1.45pt, mark=pentagon,
        color=red!80, fill=red!20, fill opacity=0.1] 
        coordinates {(0,2) (60,3) (120,3) (180,3) (240,3) (300,3) (360,2)} 
        \closedcycle;
    \addplot[line width=1.45pt, mark=square,
        color=blue!80, fill=red!20, fill opacity=0.1] 
        coordinates {(0,2) (60,1) (120,3) (180,2) (240,3) (300,3) (360,2)} 
        \closedcycle;
    \addplot[line width=1.45pt, mark=o,
        color=red!80, fill=red!20, fill opacity=0.1] 
        coordinates {(0,2) (60,3) (120,3) (180,3) (240,3) (300,3) (360,2)} 
        \closedcycle;
    \addplot[line width=1.45pt, mark=star,
        color=red!80, fill=red!20, fill opacity=0.1] 
        coordinates {(0,2) (60,3) (120,3) (180,3) (240,3) (300,3) (360,2)} 
        \closedcycle;
    \addplot[line width=1.45pt, mark=none,
        color=red!80, fill=red!20, fill opacity=0.1] 
        coordinates {(0,2) (60,3) (120,3) (180,3) (240,3) (300,3) (360,2)} 
        \closedcycle;
    
    \legend{Attack 1, Attack 2, Attack 3, Attack 4, Attack 5, Attack 6, Attack 7, Attack 8}
\end{polaraxis}
\end{tikzpicture}
\end{center}

\subsection{Assessment of Model Inference attacks}
The next type of attacks we assess are Model Inference attacks, presented in Section \ref{subsec:model-inference}: (1) LIRA \citep{carlini2022membershipinferenceattacksprinciples}, (2) Detecting Pretraining Data \citep{shi2024detectingpretrainingdatalarge}, (3) Neighborhood Comparison \citep{mattern2023membershipinferenceattackslanguage}, (4) ProPILE \citep{kim2023propileprobingprivacyleakage}, (5) Analysing PII Leakage \citep{lukas2023analyzingleakagepersonallyidentifiable}, (6) Conrecall \citep{wang2024conrecalldetectingpretrainingdata}, (7) MIA-LLM \citep{fu2024practicalmembershipinferenceattacks}, (8) DeCop \citep{duarte2024decopdetectingcopyrightedcontent}.



\subsubsection{With DREAD}
We start evaluating Model Inference attacks using DREAD \citep{michael2006security}. The corresponding vulnerability vectors as follows:

\begin{itemize}
    \item (1) $\rightarrow$ (D:8/R:9/E:7/A:7/D:6) = \textcolor{red}{7.4 (High)}
    \item (2) $\rightarrow$ (D:7/R:8/E:7/A:7/D:5) = \textcolor{yellow}{6.8 (Medium)}
    \item (3) $\rightarrow$ (D:6/R:5/E:6/A:6/D:5) = \textcolor{yellow}{5.6 (Medium)}
    \item (4) $\rightarrow$ (D:8/R:6/E:6/A:7/D:5) = \textcolor{yellow}{6.4 (Medium)}
    \item (5) $\rightarrow$ (D:9/R:8/E:8/A:9/D:6) = \textcolor{red}{8 (High)}
    \item (6) $\rightarrow$ (D:7/R:7/E:7/A:7/D:5) = \textcolor{yellow}{6.6 (Medium)}
    \item (7) $\rightarrow$ (D:8/R:7/E:7/A:7/D:5) = \textcolor{yellow}{6.8 (Medium)}
    \item (8) $\rightarrow$ (D:8/R:8/E:7/A:7/D:5) = \textcolor{yellow}{6.4 (Medium)}
\end{itemize}

Table \ref{tab:appendix-model-inference-dread} provides the detailed scores for this sixth type of attack, as assessed by the three LLMs along with their average. For better visualization and interpretation, these scores are represented below in a spider chart.

The DREAD assessment reveals that Model Inference attacks share many characteristics with Model Extraction attacks. Both pose significant damage to systems and organizations, are highly reproducible and exploitable, and can impact a large number of users either directly or indirectly. They are also moderately challenging to discover. The primary distinction lies in their danger levels—Model Extraction attacks are slightly more harmful as they enable the extraction of models or user data, whereas Model Inference attacks are more specific, allowing adversaries to determine whether certain data was part of the training set.

\begin{center}
\begin{tikzpicture}
\begin{polaraxis}[
    title={Assessment of model inference attacks with DREAD},
    title style={font=\bfseries\large},
    xtick={0, 72, 144, 216, 288},
    xticklabels={Damage, Reproducibility, Exploitability, Affected users, Discoverability},
    xticklabel style={font=\bfseries, anchor=north}, 
    ymin=0, ymax=10,
    ytick={1,2,3,4,5,6,7,8,9,10},
    yticklabels={1,2,3,4,5,6,7,8,9,10}, 
    yticklabel style={font=\small}, 
    grid=both, 
    major grid style={line width=0.8pt, draw=gray!50},
    minor grid style={line width=0.4pt, draw=gray!20},
    ylabel near ticks,
    legend style={at={(1.3,1)}, anchor=north, legend columns=1}, 
]
    \addplot[line width=1.45pt, mark=*,
        color=magenta!80, fill=yellow!20, fill opacity=0.1] 
        coordinates {(0,8) (72,9) (144,7) (216,7) (288,6) (360,8)} 
        \closedcycle;
    \addplot[line width=1.45pt, mark=square*,
        color=orange!80, fill=yellow!20, fill opacity=0.1] 
        coordinates {(0,7) (72,8) (144,7) (216,7) (288,5) (360,7)} 
        \closedcycle;
    \addplot[line width=1.45pt, mark=triangle*,
        color=cyan!80, fill=yellow!20, fill opacity=0.1] 
        coordinates {(0,6) (72,5) (144,6) (216,6) (288,4) (360,6)} 
        \closedcycle;
    \addplot[line width=1.45pt, mark=diamond*,
        color=blue!80, fill=yellow!20, fill opacity=0.1] 
        coordinates {(0,8) (72,6) (144,6) (216,7) (288,5) (360,8)} 
        \closedcycle;
    \addplot[line width=1.45pt, mark=pentagon*,
        color=violet!80, fill=yellow!20, fill opacity=0.1] 
        coordinates {(0,9) (72,8) (144,8) (216,8) (288,6) (360,9)} 
        \closedcycle;
    \addplot[line width=1.45pt, mark=o,
        color=red!80, fill=yellow!20, fill opacity=0.1] 
        coordinates {(0,7) (72,7) (144,7) (216,7) (288,5) (360,7)} 
        \closedcycle;
    \addplot[line width=1.45pt, mark=x,
        color=green!80, fill=yellow!20, fill opacity=0.1] 
        coordinates {(0,8) (72,7) (144,7) (216,7) (288,5) (360,8)} 
        \closedcycle;
    \addplot[line width=1.45pt, mark=none,
        color=yellow!80, fill=yellow!20, fill opacity=0.1] 
        coordinates {(0,8) (72,8) (144,7) (216,7) (288,5) (360,8)} 
        \closedcycle;
    
    \legend{Attack 1, Attack 2, Attack 3, Attack 4, Attack 5, Attack 6, Attack 7, Attack 8}
\end{polaraxis}
\end{tikzpicture}
\end{center}

\subsubsection{With CVSS}
The second assessment of Model Inference attacks is done with CVSS \citep{schiffman2005complete}. The corresponding Vectors are:

\begin{itemize}
    \item (1) $\rightarrow$ (AV:N/AC:L/PR:N/UI:N/S:C/C:H/I:N/A:N) = \textcolor{red}{8.6 (High)}
    \item (2) $\rightarrow$ (AV:N/AC:L/PR:N/UI:N/S:C/C:H/I:N/A:N) = \textcolor{red}{8.6 (High)}
    \item (3) $\rightarrow$ (AV:N/AC:H/PR:N/UI:N/S:C/C:H/I:N/A:N) = \textcolor{yellow}{6.8 (Medium)}
    \item (4) $\rightarrow$ (AV:N/AC:H/PR:N/UI:N/S:C/C:H/I:N/A:N) = \textcolor{yellow}{6.8 (Medium)}
    \item (5) $\rightarrow$ (AV:N/AC:L/PR:N/UI:N/S:C/C:H/I:N/A:N) = \textcolor{red}{8.6 (High)}
    \item (6) $\rightarrow$ (AV:N/AC:L/PR:N/UI:N/S:C/C:H/I:N/A:N) = \textcolor{red}{8.6 (High)}
    \item (7) $\rightarrow$ (AV:N/AC:L/PR:N/UI:N/S:C/C:H/I:N/A:N) = \textcolor{red}{8.6 (High)}
    \item (8) $\rightarrow$ (AV:N/AC:L/PR:N/UI:N/S:C/C:H/I:N/A:N) = \textcolor{red}{8.6 (High)}
\end{itemize}

The complete list of scores is detailed in Table \ref{tab:appendix-model-inference-cvss}, with the final averaged scores visualized below in the spider chart. The chart closely resembles that of Model Extraction attacks, differing primarily in that Model Inference attacks are less complex to execute.

Notably, these attacks can be carried out remotely via the network without requiring any user privileges or interaction. They have a significant impact on confidentiality and exhibit a variable scope, as the extracted information can be used to target other systems or users.

\begin{center}
\begin{tikzpicture}
\begin{polaraxis}[
    title={Assessment of Model-inference attacks with CVSS},
    title style={font=\bfseries\large},
    xtick={0, 45, 90, 135, 180, 225, 270, 315},
    xticklabels={Vector, Complexity, Privileges, User Interaction, Scope, Confidentiality, Integrity, Availability},
    xticklabel style={font=\bfseries, anchor=north}, 
    ymin=0, ymax=4,
    ytick={0,1,2,3,4},
    yticklabels={0,1,2,3,4}, 
    yticklabel style={font=\small}, 
    grid=both, 
    major grid style={line width=0.8pt, draw=gray!50},
    minor grid style={line width=0.4pt, draw=gray!20},
    ylabel near ticks,
    legend style={at={(1.3,1)}, anchor=north, legend columns=1}, 
]
    \addplot[line width=1.45pt, mark=square,
        color=blue!80, fill=cyan!20, fill opacity=0.1] 
        coordinates {(0,1) (45,2) (90,1) (135,2) (180,4) (225,3) (270,1) (315,1) (360,1)} 
        \closedcycle;
    \addplot[line width=1.45pt, mark=*,
        color=blue!80, fill=cyan!20, fill opacity=0.1] 
        coordinates {(0,1) (45,2) (90,1) (135,2) (180,4) (225,3) (270,1) (315,1) (360,1)}
        \closedcycle;
    \addplot[line width=1.45pt, mark=square,
        color=violet!80, fill=cyan!20, fill opacity=0.1] 
        coordinates {(0,1) (45,4) (90,1) (135,2) (180,4) (225,3) (270,1) (315,1) (360,1)}
        \closedcycle;
    \addplot[line width=1.45pt, mark=pentagon,
        color=violet!80, fill=cyan!20, fill opacity=0.1] 
        coordinates {(0,1) (45,4) (90,1) (135,2) (180,4) (225,3) (270,1) (315,1) (360,1)} 
        \closedcycle;
    \addplot[line width=1.45pt, mark=triangle,
        color=blue!80, fill=cyan!20, fill opacity=0.1] 
        coordinates {(0,1) (45,2) (90,1) (135,2) (180,4) (225,3) (270,1) (315,1) (360,1)} 
        \closedcycle;
    \addplot[line width=1.45pt, mark=x,
        color=blue!80, fill=cyan!20, fill opacity=0.1] 
        coordinates {(0,1) (45,2) (90,1) (135,2) (180,4) (225,3) (270,1) (315,1) (360,1)} 
        \closedcycle;
    \addplot[line width=1.45pt, mark=pentagon,
        color=blue!80, fill=cyan!20, fill opacity=0.1] 
        coordinates {(0,1) (45,2) (90,1) (135,2) (180,4) (225,3) (270,1) (315,1) (360,1)} 
        \closedcycle;
    \addplot[line width=1.45pt, mark=none,
        color=blue!80, fill=cyan!20, fill opacity=0.1] 
        coordinates {(0,1) (45,2) (90,1) (135,2) (180,4) (225,3) (270,1) (315,1) (360,1)} 
        \closedcycle;
    
    \legend{Attack 1, Attack 2, Attack 3, Attack 4, Attack 5, Attack 6, Attack 7, Attack 8}
\end{polaraxis}
\end{tikzpicture}
\end{center}

\subsubsection{With OWASP Risk Rating}
A third evaluation of Model Inference attacks is done using OWASP RR \citep{owaspOWASPRisk}. The corresponding vulnerability vectors of each attack is:

\begin{itemize}
    \item (1) $\rightarrow$ (SL:6/M:7/O:7/S:6/ED:5/EE:7/A:6/ID:7/LC:8/LI:1/LA:2/FD:7/RD:8/NC:7/PV:8) = \textcolor{red}{3.5 (High)}
    \item (2) $\rightarrow$ (SL:6/M:7/O:7/S:6/ED:5/EE:6/A:4/ID:6/LC:8/LI:2/LA:2/FD:7/RD:8/NC:8/PV:8) = \textcolor{yellow}{3.5 (Medium)}
    \item (3) $\rightarrow$ (SL:5/M:7/O:7/S:5/ED:5/EE:6/A:5/ID:6/LC:8/LI:1/LA:2/FD:6/RD:7/NC:7/PV:8) = \textcolor{yellow}{3.1 (Medium)}
    \item (4) $\rightarrow$ (SL:6/M:7/O:7/S:6/ED:5/EE:6/A:6/ID:6/LC:8/LI:1/LA:2/FD:7/RD:7/NC:7/PV:8) = \textcolor{yellow}{3.4 (Medium)}
    \item (5) $\rightarrow$ (SL:6/M:8/O:7/S:6/ED:5/EE:6/A:6/ID:8/LC:9/LI:1/LA:2/FD:8/RD:8/NC:8/PV:9) = \textcolor{purple}{4 (Critical)}
    \item (6) $\rightarrow$ (SL:6/M:7/O:6/S:6/ED:5/EE:6/A:5/ID:7/LC:8/LI:2/LA:2/FD:8/RD:8/NC:8/PV:9) = \textcolor{purple}{3.7 (Critical)}
    \item (7) $\rightarrow$ (SL:6/M:8/O:7/S:6/ED:6/EE:6/A:5/ID:7/LC:8/LI:2/LA:2/FD:8/RD:8/NC:7/PV:9) = \textcolor{purple}{3.8 (Critical)}
    \item (8) $\rightarrow$ (SL:5/M:7/O:7/S:6/ED:6/EE:7/A:6/ID:7/LC:7/LI:2/LA:2/FD:8/RD:8/NC:8/PV:6) = \textcolor{purple}{4.2 (Critical)}
\end{itemize}

The detailed assessments are provided in Table \ref{tab:appendix-inference-owasp}, with their averages visualized in the chart below.

The OWASP RR scores align closely with those of CVSS, reaffirming that Model Inference attacks primarily impact confidentiality. These attacks are easy to discover and exploit but are challenging for defenders to detect due to limited awareness of their risks. Additionally, the assessments emphasize that Model Inference attacks significantly violate privacy while being less complex than Model Extraction attacks. Instead of extracting various data from a training set, Model Inference attacks determine whether specific data was included in the training process.

\begin{center}
\begin{tikzpicture}
\begin{polaraxis}[
    title={Assessment of Model-inference attacks with OWASP RR},
    title style={font=\bfseries\large},
    xtick={0, 24, 48, 72, 96, 120, 144, 168, 192, 216, 240, 264, 288, 312, 336},
    xticklabels={Skill, Motivation, Opportunity, Size, Discovery, Exploit, Awareness, Intrusion Dtc, Confidentiality, Integrity, Availability, Fnc Dmg, Rpt Dmg, Non-comp, Privacy Vlt},
    xticklabel style={font=\bfseries, anchor=north}, 
    ymin=0, ymax=10,
    ytick={0,1,2,3,4,5,6,7,8,9,10},
    yticklabels={0,1,2,3,4,5,6,7,8,9,10}, 
    yticklabel style={font=\small}, 
    grid=both, 
    major grid style={line width=0.8pt, draw=gray!50},
    minor grid style={line width=0.4pt, draw=gray!20},
    ylabel near ticks,
    legend style={at={(1.3,1)}, anchor=north, legend columns=1}, 
]
    \addplot[line width=1.45pt, mark=square,
        color=blue!80, fill=green!20, fill opacity=0.1] 
        coordinates {(0,6) (24,7) (48,7) (72,6) (96,5) (120,7) (144,6) (168,7) (192,8) (216,1) (240,2) (264,7) (288,8) (312,7) (336,8) (360,6)} 
        \closedcycle;
    \addplot[line width=1.45pt, mark=triangle,
        color=red!80, fill=green!20, fill opacity=0.1] 
        coordinates {(0,6) (24,7) (48,7) (72,6) (96,5) (120,6) (144,4) (168,6) (192,8) (216,2) (240,2) (264,7) (288,8) (312,8) (336,8) (360,6)}
        \closedcycle;
    \addplot[line width=1.45pt, mark=square,
        color=violet!80, fill=green!20, fill opacity=0.1] 
        coordinates {(0,5) (24,7) (48,7) (72,5) (96,5) (120,6) (144,5) (168,6) (192,8) (216,1) (240,2) (264,6) (288,7) (312,7) (336,8) (360,5)}
        \closedcycle;
    \addplot[line width=1.45pt, mark=pentagon,
        color=green!80, fill=green!20, fill opacity=0.1] 
        coordinates {(0,6) (24,7) (48,7) (72,6) (96,5) (120,6) (144,6) (168,6) (192,8) (216,1) (240,2) (264,7) (288,7) (312,7) (336,8) (360,6)} 
        \closedcycle;
    \addplot[line width=1.45pt, mark=*,
        color=orange!80, fill=green!20, fill opacity=0.1] 
        coordinates {(0,6) (24,8) (48,7) (72,6) (96,5) (120,6) (144,6) (168,8) (192,9) (216,1) (240,2) (264,8) (288,8) (312,8) (336,9) (360,6)} 
        \closedcycle;
    \addplot[line width=1.45pt, mark=o,
        color=yellow!80, fill=green!20, fill opacity=0.1] 
        coordinates {(0,6) (24,7) (48,6) (72,6) (96,5) (120,6) (144,5) (168,7) (192,8) (216,2) (240,2) (264,8) (288,8) (312,8) (336,9) (360,6)} 
        \closedcycle;
    \addplot[line width=1.45pt, mark=pentagon,
        color=magenta!80, fill=green!20, fill opacity=0.1] 
        coordinates {(0,6) (24,8) (48,7) (72,6) (96,6) (120,6) (144,5) (168,7) (192,8) (216,2) (240,2) (264,8) (288,8) (312,7) (336,9) (360,6)} 
        \closedcycle;
    \addplot[line width=1.45pt, mark=none,
        color=cyan!80, fill=green!20, fill opacity=0.1] 
        coordinates {(0,5) (24,7) (48,7) (72,6) (96,6) (120,7) (144,6) (168,7) (192,7) (216,2) (240,2) (264,8) (288,8) (312,8) (336,6) (360,5)} 
        \closedcycle;
    
    \legend{Attack 1, Attack 2, Attack 3, Attack 4, Attack 5, Attack 6, Attack 7, Attack 8}
\end{polaraxis}
\end{tikzpicture}
\end{center}

\subsubsection{With SSVC}
The last evaluation of Model-Inference attacks is performed using SSVC \citep{spring2021prioritizing}, and the results are shown and detailed below:

\begin{itemize}
    \item (1) $\rightarrow$ (E:P/A:Y/V:C/U:S/T:T/P:S) = \textcolor{red}{Immediate (Very High)}
    \item (2) $\rightarrow$ (E:P/A:Y/V:C/U:S/T:T/P:S) = \textcolor{red}{Immediate (Very High)}
    \item (3) $\rightarrow$ (E:P/A:N/V:D/U:E/T:P/P:M) = \textcolor{yellow}{Scheduled (Medium)}
    \item (4) $\rightarrow$ (E:P/A:Y/V:C/U:S/T:T/P:S) = \textcolor{red}{Immediate (Very High)}
    \item (5) $\rightarrow$ (E:P/A:Y/V:C/U:S/T:T/P:S) = \textcolor{red}{Immediate (Very High)}
    \item (6) $\rightarrow$ (E:P/A:Y/V:D/U:E/T:P/P:M) = \textcolor{yellow}{Scheduled (Medium)}
    \item (7) $\rightarrow$ (E:P/A:Y/V:C/U:S/T:T/P:S) = \textcolor{red}{Immediate (Very High)}
    \item (8) $\rightarrow$ (E:P/A:Y/V:C/U:S/T:T/P:S) = \textcolor{red}{Immediate (Very High)}
\end{itemize}

The detailed SSVC scores are presented in Table \ref{tab:appendix-inference-ssvc}, with the final averages visualized in the chart below. Similar to previous attacks, SSVC aligns with the insights provided by other metrics. However, it uniquely highlights that Model Inference attacks are highly automatable and rewarding, making them particularly effective for adversaries. These attacks pose a significant impact not only on the technical aspects of systems but also on user safety and data privacy.

\begin{center}
\begin{tikzpicture}
\begin{polaraxis}[
    title={Assessment of Model-Inference attacks with SSVC},
    title style={font=\bfseries\large},
    xtick={0, 60, 120, 180, 240, 300},
    xticklabels={Exploitation, Automatable, Value Density, Utility, Technical Imp, Public-Safety Imp},
    xticklabel style={font=\bfseries, anchor=north}, 
    ymin=0, ymax=3,
    ytick={0,1,2,3},
    yticklabels={0,1,2,3}, 
    yticklabel style={font=\small}, 
    grid=both, 
    major grid style={line width=0.8pt, draw=gray!50},
    minor grid style={line width=0.4pt, draw=gray!20},
    ylabel near ticks,
    legend style={at={(1.35,1)}, anchor=north, legend columns=1}, 
]
    \addplot[line width=1.45pt, mark=x,
        color=red!80, fill=red!20, fill opacity=0.1] 
        coordinates {(0,2) (60,3) (120,3) (180,3) (240,3) (300,3) (360,2)} 
        \closedcycle;
    \addplot[line width=1.45pt, mark=pentagon,
        color=red!80, fill=red!20, fill opacity=0.1] 
        coordinates {(0,2) (60,3) (120,3) (180,3) (240,3) (300,3) (360,2)} 
        \closedcycle;
    \addplot[line width=1.45pt, mark=triangle,
        color=blue!80, fill=red!20, fill opacity=0.1] 
        coordinates {(0,2) (60,3) (120,1) (180,2) (240,1) (300,1) (360,2)} 
        \closedcycle;
    \addplot[line width=1.45pt, mark=+,
        color=red!80, fill=red!20, fill opacity=0.1] 
        coordinates {(0,2) (60,3) (120,3) (180,3) (240,3) (300,3) (360,2)} 
        \closedcycle;
    \addplot[line width=1.45pt, mark=star,
        color=red!80, fill=red!20, fill opacity=0.1] 
        coordinates {(0,2) (60,3) (120,3) (180,3) (240,3) (300,3) (360,2)} 
        \closedcycle;
    \addplot[line width=1.45pt, mark=square,
        color=blue!80, fill=red!20, fill opacity=0.1] 
        coordinates {(0,2) (60,3) (120,1) (180,2) (240,1) (300,1) (360,2)} 
        \closedcycle;
    \addplot[line width=1.45pt, mark=diamond,
        color=red!80, fill=red!20, fill opacity=0.1] 
        coordinates {(0,2) (60,3) (120,3) (180,3) (240,3) (300,3) (360,2)} 
        \closedcycle;
    \addplot[line width=1.45pt, mark=none,
        color=red!80, fill=red!20, fill opacity=0.1] 
        coordinates {(0,2) (60,3) (120,3) (180,3) (240,3) (300,3) (360,2)}  
        \closedcycle;
    
    \legend{Attack 1, Attack 2, Attack 3, Attack 4, Attack 5, Attack 6, Attack 7, Attack 8}

\end{polaraxis}
\end{tikzpicture}
\end{center}

\subsection{Assessment of Poisoning/Trojan/Backdoor attacks}
The last type of attacks we assess are Poisoning, Trojan, and Backdoor attacks, already presented in Section \ref{subsec:poisoning}: (1) TrojLLM \citep{NEURIPS2023_cf04d01a}, (2) Best-of-Venom \citep{baumgärtner2024bestofvenomattackingrlhfinjecting}, (3) CodeBreaker \citep{yan2024llmassistedeasytotriggerbackdoorattack}, (4) Retrieval Poisoning \citep{zhang2024humanimperceptibleretrievalpoisoningattacks}, (5) Clinical LLMs \citep{clinicalLLM}, (6) BackdoorLLM \citep{li2024backdoorllmcomprehensivebenchmarkbackdoor}, (7) CBA \citep{huang2024compositebackdoorattackslarge}, (8) TA² \citep{wang2024trojanactivationattackredteaming}.



\subsubsection{With DREAD}
We start evaluating Poisoning, Trojan, and Backdoor attacks using DREAD \citep{michael2006security}. The corresponding vulnerability vectors as follows:

\begin{itemize}
    \item (1) $\rightarrow$ (D:8/R:8/E:8/A:8/D:6) = \textcolor{red}{7.6 (High)}
    \item (2) $\rightarrow$ (D:8/R:6/E:7/A:7/D:5) = \textcolor{yellow}{6.6 (Medium)}
    \item (3) $\rightarrow$ (D:8/R:8/E:8/A:8/D:6) = \textcolor{red}{7.6 (High)}
    \item (4) $\rightarrow$ (D:7/R:6/E:6/A:6/D:4) = \textcolor{yellow}{5.8 (Medium)}
    \item (5) $\rightarrow$ (D:9/R:7/E:6/A:9/D:5) = \textcolor{red}{7.2 (High)}
    \item (6) $\rightarrow$ (D:8/R:9/E:8/A:8/D:6) = \textcolor{red}{7.8 (High)}
    \item (7) $\rightarrow$ (D:8/R:8/E:7/A:7/D:5) = \textcolor{red}{7 (High)}
    \item (8) $\rightarrow$ (D:8/R:8/E:8/A:8/D:6) = \textcolor{red}{7.6 (High)}
\end{itemize}

The detailed scores for this final type of attack are provided in Table \ref{tab:appendix-poisoning}, with their averages visualized in the chart below.

The DREAD analysis reveals that Poisoning, Trojan, and Backdoor attacks generally cause significant damage—often surpassing other attack types like Jailbreak or Evasion. Their reproducibility, exploitability, and the number of affected users vary depending on the specific attack but typically range from medium to high. However, these attacks are notably difficult to detect, making them particularly dangerous to LLMs. Addressing these vulnerabilities presents a significant challenge for system administrators.

\begin{center}
\begin{tikzpicture}
\begin{polaraxis}[
    title={Assessment of Poisoning/Trojan/Backdoor attacks with DREAD},
    title style={font=\bfseries\large},
    xtick={0, 72, 144, 216, 288},
    xticklabels={Damage, Reproducibility, Exploitability, Affected users, Discoverability},
    xticklabel style={font=\bfseries, anchor=north}, 
    ymin=0, ymax=10,
    ytick={1,2,3,4,5,6,7,8,9,10},
    yticklabels={1,2,3,4,5,6,7,8,9,10}, 
    yticklabel style={font=\small}, 
    grid=both, 
    major grid style={line width=0.8pt, draw=gray!50},
    minor grid style={line width=0.4pt, draw=gray!20},
    ylabel near ticks,
    legend style={at={(1.3,1)}, anchor=north, legend columns=1}, 
]
    \addplot[line width=1.45pt, mark=+,
        color=magenta!80, fill=yellow!20, fill opacity=0.1] 
        coordinates {(0,8) (72,8) (144,8) (216,8) (288,6) (360,8)} 
        \closedcycle;
    \addplot[line width=1.45pt, mark=square*,
        color=blue!80, fill=yellow!20, fill opacity=0.1] 
        coordinates {(0,8) (72,6) (144,7) (216,7) (288,5) (360,8)} 
        \closedcycle;
    \addplot[line width=1.45pt, mark=x,
        color=magenta!80, fill=yellow!20, fill opacity=0.1] 
        coordinates {(0,8) (72,8) (144,8) (216,8) (288,6) (360,8)} 
        \closedcycle;
    \addplot[line width=1.45pt, mark=diamond*,
        color=orange!80, fill=yellow!20, fill opacity=0.1] 
        coordinates {(0,7) (72,6) (144,6) (216,6) (288,4) (360,7)} 
        \closedcycle;
    \addplot[line width=1.45pt, mark=pentagon*,
        color=violet!80, fill=yellow!20, fill opacity=0.1] 
        coordinates {(0,9) (72,7) (144,6) (216,9) (288,5) (360,9)} 
        \closedcycle;
    \addplot[line width=1.45pt, mark=o,
        color=red!80, fill=yellow!20, fill opacity=0.1] 
        coordinates {(0,8) (72,9) (144,8) (216,8) (288,6) (360,8)} 
        \closedcycle;
    \addplot[line width=1.45pt, mark=x,
        color=cyan!80, fill=yellow!20, fill opacity=0.1] 
        coordinates {(0,8) (72,8) (144,7) (216,7) (288,5) (360,8)} 
        \closedcycle;
    \addplot[line width=1.45pt, mark=none,
        color=magenta!80, fill=yellow!20, fill opacity=0.1] 
        coordinates {(0,8) (72,8) (144,8) (216,8) (288,6) (360,8)} 
        \closedcycle;
    
    \legend{Attack 1, Attack 2, Attack 3, Attack 4, Attack 5, Attack 6, Attack 7, Attack 8}
\end{polaraxis}
\end{tikzpicture}
\end{center}

\subsubsection{With CVSS}
The second assessment of these attacks is done with CVSS \citep{schiffman2005complete}. The corresponding Vectors are:

\begin{itemize}
    \item (1) $\rightarrow$ (AV:N/AC:H/PR:N/UI:N/S:C/C:L/I:H/A:N) = \textcolor{red}{7.5 (High)}
    \item (2) $\rightarrow$ (AV:L/AC:H/PR:L/UI:N/S:U/C:L/I:H/A:N) = \textcolor{yellow}{5.3 (Medium)}
    \item (3) $\rightarrow$ (AV:N/AC:L/PR:L/UI:N/S:C/C:L/I:H/A:N) = \textcolor{red}{8.5 (High)}
    \item (4) $\rightarrow$ (AV:N/AC:H/PR:L/UI:N/S:U/C:L/I:H/A:N) = \textcolor{yellow}{5.9 (Medium)}
    \item (5) $\rightarrow$ (AV:N/AC:H/PR:L/UI:N/S:U/C:L/I:H/A:N) = \textcolor{yellow}{5.9 (Medium)}
    \item (6) $\rightarrow$ (AV:N/AC:L/PR:N/UI:N/S:C/C:L/I:H/A:N) = \textcolor{purple}{9.3 (Critical)}
    \item (7) $\rightarrow$ (AV:N/AC:H/PR:N/UI:N/S:C/C:L/I:H/A:N) = \textcolor{red}{7.5 (High)}
    \item (8) $\rightarrow$ (AV:N/AC:H/PR:N/UI:N/S:C/C:L/I:H/A:N) = \textcolor{red}{7.5 (High)}
\end{itemize}

These vectors are presented below in a Spider-chart for better visualization, the detailed scores are shown in Table \ref{tab:appendix-poisoning-cvss}.

The CVSS assessment reveals that these attacks are generally complex to execute but predominantly impact two technical domains: they exert a high impact on integrity and a medium impact on confidentiality. Furthermore, their execution often does not require elevated privileges, depending on the attack's complexity. This highlights the significant potential danger posed by these types of attacks.

\begin{center}
\begin{tikzpicture}
\begin{polaraxis}[
    title={Assessment of Poisoning/Backdoor/Trojan attacks with CVSS},
    title style={font=\bfseries\large},
    xtick={0, 45, 90, 135, 180, 225, 270, 315},
    xticklabels={Vector, Complexity, Privileges, User Interaction, Scope, Confidentiality, Integrity, Availability},
    xticklabel style={font=\bfseries, anchor=north}, 
    ymin=0, ymax=4,
    ytick={0,1,2,3,4},
    yticklabels={0,1,2,3,4}, 
    yticklabel style={font=\small}, 
    grid=both, 
    major grid style={line width=0.8pt, draw=gray!50},
    minor grid style={line width=0.4pt, draw=gray!20},
    ylabel near ticks,
    legend style={at={(1.3,1)}, anchor=north, legend columns=1}, 
]
    \addplot[line width=1.45pt, mark=x,
        color=red!80, fill=cyan!20, fill opacity=0.1] 
        coordinates {(0,1) (45,4) (90,1) (135,2) (180,4) (225,2) (270,3) (315,1) (360,1)} 
        \closedcycle;
    \addplot[line width=1.45pt, mark=x,
        color=orange!80, fill=cyan!20, fill opacity=0.1] 
        coordinates {(0,3) (45,4) (90,2) (135,2) (180,2) (225,2) (270,3) (315,1) (360,3)}
        \closedcycle;
    \addplot[line width=1.45pt, mark=triangle,
        color=green!80, fill=cyan!20, fill opacity=0.1] 
        coordinates {(0,1) (45,2) (90,2) (135,2) (180,4) (225,2) (270,3) (315,1) (360,1)}
        \closedcycle;
    \addplot[line width=1.45pt, mark=o,
        color=cyan!80, fill=cyan!20, fill opacity=0.1] 
        coordinates {(0,1) (45,4) (90,2) (135,2) (180,2) (225,2) (270,3) (315,1) (360,1)} 
        \closedcycle;
    \addplot[line width=1.45pt, mark=square,
        color=cyan!80, fill=cyan!20, fill opacity=0.1] 
        coordinates {(0,1) (45,4) (90,2) (135,2) (180,2) (225,2) (270,3) (315,1) (360,1)} 
        \closedcycle;
    \addplot[line width=1.45pt, mark=square*,
        color=blue!80, fill=cyan!20, fill opacity=0.1] 
        coordinates {(0,1) (45,2) (90,1) (135,2) (180,4) (225,2) (270,3) (315,1) (360,1)} 
        \closedcycle;
    \addplot[line width=1.45pt, mark=o,
        color=red!80, fill=cyan!20, fill opacity=0.1] 
        coordinates {(0,1) (45,4) (90,1) (135,2) (180,4) (225,2) (270,3) (315,1) (360,1)} 
        \closedcycle;
    \addplot[line width=1.45pt, mark=none,
        color=red!80, fill=cyan!20, fill opacity=0.1] 
        coordinates {(0,1) (45,4) (90,1) (135,2) (180,4) (225,2) (270,3) (315,1) (360,1)} 
        \closedcycle;
    
    \legend{Attack 1, Attack 2, Attack 3, Attack 4, Attack 5, Attack 6, Attack 7, Attack 8}
\end{polaraxis}
\end{tikzpicture}
\end{center}

\subsubsection{With OWASP Risk Rating}
A third evaluation of Poisoning, Trojan, and Backdoor attacks is done using OWASP RR \citep{owaspOWASPRisk}. The corresponding vulnerability vectors of each attack is:

\begin{itemize}
    \item (1) $\rightarrow$ (SL:6/M:8/O:7/S:6/ED:5/EE:7/A:5/ID:7/LC:6/LI:7/LA:4/FD:7/RD:8/NC:5/PV:6) = \textcolor{purple}{3.9 (Critical)}
    \item (2) $\rightarrow$ (SL:8/M:8/O:6/S:5/ED:5/EE:6/A:4/ID:5/LC:7/LI:7/LA:3/FD:7/RD:8/NC:6/PV:7) = \textcolor{red}{3.7 (High)}
    \item (3) $\rightarrow$ (SL:7/M:8/O:8/S:6/ED:6/EE:7/A:5/ID:6/LC:7/LI:8/LA:4/FD:7/RD:8/NC:6/PV:7) = \textcolor{purple}{4.4 (Critical)}
    \item (4) $\rightarrow$ (SL:7/M:8/O:6/S:6/ED:5/EE:6/A:4/ID:6/LC:6/LI:8/LA:4/FD:7/RD:8/NC:6/PV:6) = \textcolor{purple}{4 (Critical)}
    \item (5) $\rightarrow$ (SL:7/M:8/O:7/S:6/ED:4/EE:6/A:4/ID:8/LC:6/LI:8/LA:3/FD:8/RD:9/NC:7/PV:6) = \textcolor{purple}{4.2 (Critical)}
    \item (6) $\rightarrow$ (SL:8/M:9/O:7/S:6/ED:4/EE:6/A:4/ID:9/LC:6/LI:8/LA:4/FD:8/RD:9/NC:7/PV:6) = \textcolor{purple}{4 (Critical)}
    \item (7) $\rightarrow$ (SL:8/M:9/O:7/S:7/ED:5/EE:7/A:4/ID:9/LC:7/LI:8/LA:4/FD:8/RD:9/NC:8/PV:7) = \textcolor{purple}{4.9 (Critical)}
    \item (8) $\rightarrow$ (SL:8/M:8/O:7/S:6/ED:4/EE:6/A:4/ID:8/LC:6/LI:8/LA:4/FD:7/RD:8/NC:6/PV:6) = \textcolor{purple}{4.1 (Critical)}
\end{itemize}

The calculated values are detailed in Table \ref{tab:appendix-poisoning-owasp} and visualized in the Spider-chart below.
This chart spans a wider area compared to others, emphasizing the significant danger posed by Poisoning, Trojan, and Backdoor attacks. OWASP RR indicates that these attacks require advanced skills and high motivation to execute, unlike simpler attacks such as Model-Inference. Additionally, a pronounced lack of awareness among system administrators complicates their detection. These attacks often have severe impacts on integrity, financial stability, and organizational reputation, as well as contributing to non-compliance and privacy violations, making them some of the most impactful threats in the assessment.

\begin{center}
\begin{tikzpicture}
\begin{polaraxis}[
    title={Assessment of Poisoning/Trojan/Backdoor attacks with OWASP RR},
    title style={font=\bfseries\large},
    xtick={0, 24, 48, 72, 96, 120, 144, 168, 192, 216, 240, 264, 288, 312, 336},
    xticklabels={Skill, Motivation, Opportunity, Size, Discovery, Exploit, Awareness, Intrusion Dtc, Confidentiality, Integrity, Availability, Fnc Dmg, Rpt Dmg, Non-comp, Privacy Vlt},
    xticklabel style={font=\bfseries, anchor=north}, 
    ymin=0, ymax=10,
    ytick={0,1,2,3,4,5,6,7,8,9,10},
    yticklabels={0,1,2,3,4,5,6,7,8,9,10}, 
    yticklabel style={font=\small}, 
    grid=both, 
    major grid style={line width=0.8pt, draw=gray!50},
    minor grid style={line width=0.4pt, draw=gray!20},
    ylabel near ticks,
    legend style={at={(1.3,1)}, anchor=north, legend columns=1}, 
]
    \addplot[line width=1.45pt, mark=square,
        color=blue!80, fill=green!20, fill opacity=0.1] 
        coordinates {(0,6) (24,8) (48,7) (72,6) (96,5) (120,7) (144,5) (168,7) (192,6) (216,7) (240,4) (264,7) (288,8) (312,5) (336,6) (360,6)} 
        \closedcycle;
    \addplot[line width=1.45pt, mark=triangle,
        color=red!80, fill=green!20, fill opacity=0.1] 
        coordinates {(0,8) (24,8) (48,6) (72,5) (96,5) (120,6) (144,4) (168,5) (192,7) (216,7) (240,3) (264,7) (288,8) (312,6) (336,7) (360,8)}
        \closedcycle;
    \addplot[line width=1.45pt, mark=square,
        color=violet!80, fill=green!20, fill opacity=0.1] 
        coordinates {(0,7) (24,8) (48,8) (72,6) (96,6) (120,7) (144,5) (168,6) (192,7) (216,8) (240,4) (264,7) (288,8) (312,6) (336,7) (360,6)}
        \closedcycle;
    \addplot[line width=1.45pt, mark=pentagon,
        color=green!80, fill=green!20, fill opacity=0.1] 
        coordinates {(0,7) (24,8) (48,6) (72,6) (96,5) (120,6) (144,4) (168,6) (192,6) (216,8) (240,4) (264,7) (288,8) (312,6) (336,6) (360,7)} 
        \closedcycle;
    \addplot[line width=1.45pt, mark=*,
        color=orange!80, fill=green!20, fill opacity=0.1] 
        coordinates {(0,7) (24,8) (48,7) (72,6) (96,4) (120,6) (144,4) (168,8) (192,6) (216,8) (240,3) (264,8) (288,9) (312,7) (336,6) (360,7)} 
        \closedcycle;
    \addplot[line width=1.45pt, mark=pentagon,
        color=yellow!80, fill=green!20, fill opacity=0.1] 
        coordinates {(0,8) (24,9) (48,7) (72,6) (96,4) (120,6) (144,4) (168,9) (192,6) (216,8) (240,4) (264,8) (288,9) (312,7) (336,6) (360,8)} 
        \closedcycle;
    \addplot[line width=1.45pt, mark=pentagon,
        color=magenta!80, fill=green!20, fill opacity=0.1] 
        coordinates {(0,8) (24,9) (48,7) (72,7) (96,5) (120,7) (144,4) (168,9) (192,7) (216,8) (240,4) (264,8) (288,9) (312,8) (336,7) (360,8)} 
        \closedcycle;
    \addplot[line width=1.45pt, mark=none,
        color=cyan!80, fill=green!20, fill opacity=0.1] 
        coordinates {(0,8) (24,8) (48,7) (72,6) (96,4) (120,6) (144,4) (168,8) (192,6) (216,8) (240,4) (264,7) (288,8) (312,6) (336,6) (360,8)} 
        \closedcycle;
    
    \legend{Attack 1, Attack 2, Attack 3, Attack 4, Attack 5, Attack 6, Attack 7, Attack 8}
\end{polaraxis}
\end{tikzpicture}
\end{center}

\subsubsection{With SSVC}
The final evaluation is performed using SSVC \citep{spring2021prioritizing} as done with previous attacks. The results of this assessments are detailed below:

\begin{itemize}
    \item (1) $\rightarrow$ (E:P/A:Y/V:C/U:S/T:T/P:S) = \textcolor{red}{Immediate (Very High)}
    \item (2) $\rightarrow$ (E:P/A:Y/V:C/U:S/T:T/P:S) = \textcolor{red}{Immediate (Very High)}
    \item (3) $\rightarrow$ (E:P/A:Y/V:C/U:S/T:T/P:S) = \textcolor{red}{Immediate (Very High)}
    \item (4) $\rightarrow$ (E:P/A:N/V:D/U:L/T:P/P:M) = \textcolor{yellow}{Scheduled (Medium)}
    \item (5) $\rightarrow$ (E:P/A:Y/V:C/U:S/T:T/P:S) = \textcolor{red}{Immediate (Very High)}
    \item (6) $\rightarrow$ (E:P/A:Y/V:C/U:S/T:T/P:S) = \textcolor{red}{Immediate (Very High)}
    \item (7) $\rightarrow$ (E:P/A:Y/V:C/U:S/T:T/P:S) = \textcolor{red}{Immediate (Very High)}
    \item (8) $\rightarrow$ (E:P/A:Y/V:C/U:S/T:T/P:S) = \textcolor{red}{Immediate (Very High)}
\end{itemize}

Lastly, the SSVC scores are visualized in the chart below, with detailed assessments provided in Table \ref{tab:appendix-poisoning-ssvc}.

The chart aligns closely with those of previous attack types, indicating that these attacks are easily automatable, which increases their exploitation and effectiveness for adversaries. Furthermore, this analysis corroborates findings from other metrics, confirming that Poisoning, Trojan, and Backdoor attacks have a substantial impact on both the technical aspects of a system and the financial stability and safety of its users.

\begin{center}
\begin{tikzpicture}
\begin{polaraxis}[
    title={Assessment of Poisoning/Backdoor/Trojan attacks with SSVC},
    title style={font=\bfseries\large},
    xtick={0, 60, 120, 180, 240, 300},
    xticklabels={Exploitation, Automatable, Value Density, Utility, Technical Imp, Public-Safety Imp},
    xticklabel style={font=\bfseries, anchor=north}, 
    ymin=0, ymax=3,
    ytick={0,1,2,3},
    yticklabels={0,1,2,3}, 
    yticklabel style={font=\small}, 
    grid=both, 
    major grid style={line width=0.8pt, draw=gray!50},
    minor grid style={line width=0.4pt, draw=gray!20},
    ylabel near ticks,
    legend style={at={(1.35,1)}, anchor=north, legend columns=1}, 
]
    \addplot[line width=1.45pt, mark=x,
        color=red!80, fill=red!20, fill opacity=0.1] 
        coordinates {(0,2) (60,3) (120,3) (180,3) (240,3) (300,3) (360,2)} 
        \closedcycle;
    \addplot[line width=1.45pt, mark=pentagon,
        color=red!80, fill=red!20, fill opacity=0.1] 
        coordinates {(0,2) (60,3) (120,3) (180,3) (240,3) (300,3) (360,2)} 
        \closedcycle;
    \addplot[line width=1.45pt, mark=triangle,
        color=red!80, fill=red!20, fill opacity=0.1] 
        coordinates {(0,2) (60,3) (120,3) (180,3) (240,3) (300,3) (360,2)} 
        \closedcycle;
    \addplot[line width=1.45pt, mark=square,
        color=blue!80, fill=red!20, fill opacity=0.1] 
        coordinates {(0,2) (60,1) (120,1) (180,1) (240,1) (300,1) (360,2)} 
        \closedcycle;
    \addplot[line width=1.45pt, mark=star,
        color=red!80, fill=red!20, fill opacity=0.1] 
        coordinates {(0,2) (60,3) (120,3) (180,3) (240,3) (300,3) (360,2)} 
        \closedcycle;
    \addplot[line width=1.45pt, mark=+,
        color=red!80, fill=red!20, fill opacity=0.1] 
        coordinates {(0,2) (60,3) (120,3) (180,3) (240,3) (300,3) (360,2)} 
        \closedcycle;
    \addplot[line width=1.45pt, mark=diamond,
        color=red!80, fill=red!20, fill opacity=0.1] 
        coordinates {(0,2) (60,3) (120,3) (180,3) (240,3) (300,3) (360,2)} 
        \closedcycle;
    \addplot[line width=1.45pt, mark=none,
        color=red!80, fill=red!20, fill opacity=0.1] 
        coordinates {(0,2) (60,3) (120,3) (180,3) (240,3) (300,3) (360,2)} 
        \closedcycle;
    
    \legend{Attack 1, Attack 2, Attack 3, Attack 4, Attack 5, Attack 6, Attack 7, Attack 8}

\end{polaraxis}
\end{tikzpicture}
\end{center}

\section{Discussion}
\label{sec:discussion}
In this section, we present and analyze the results of our assessments of various attacks on large language models using the four vulnerability assessment metrics: DREAD, CVSS, OWASP Risk Rating, and SSVC. By examining the variations in metric values across all evaluated attacks, we aim to identify patterns, inconsistencies, and strengths in each framework. This analysis will provide insights into how effectively these metrics capture the severity and impact of adversarial attacks on LLMs. Ultimately, we will assess the overall utility and reliability of these metrics in evaluating attacks specific to LLMs, offering recommendations on their applicability and potential areas for improvement.

\subsection{Evaluation of DREAD}

To assess the relevance of the DREAD scoring model, we analyzed the Coefficient of Variation (COV\%) for each of its factors, as detailed in Table \ref{tab:variance-dread}.

Our findings reveal that the factors exhibit varying levels of variability, though most are relatively low. The \textbf{Damage} factor shows minimal variation, with COV\% below 10\% for five of the seven attack classes and slightly higher values for Evasion (13.45\%) and Model Inference (10.87\%) attacks. Across all 56 attacks, the Damage scores are consistently close, predominantly ranging between 7, 8, and 9 on a scale of 10. Similarly, the \textbf{Discoverability} factor demonstrates low variability, with COV\% near 10\% across all classes, and scores typically falling between 5 and 6. The same pattern is observed for the \textbf{Exploitability} and \textbf{Affected Users} factors, both of which maintain intra-class COV\% around 10\%.

This limited variability suggests that these four factors provide insufficient differentiation between adversarial attacks on LLMs. Their inability to distinguish effectively among attack classes renders them \textbf{unsuitable} for ranking the relative danger or impact of these attacks.

The \textbf{Reproducibility} factor, in contrast, shows greater variability, although inconsistently across attack classes. For example, White-box Jailbreaks and Prompt-Injection attacks exhibit higher COV\% values (17.64\% and 21.06\%, respectively), indicating that attack complexity significantly influences reproducibility. However, this trend is not observed for Black-box Jailbreaks and Evasion attacks, which are generally easier to execute. As a result, while Reproducibility shows potential for distinguishing attacks, its inconsistent variability diminishes its overall utility, especially within \textbf{individual} attack-classes.
\begin{table}[h]
\centering
\vspace{-1cm}
\begin{minipage}{0.48\textwidth}
    \centering
    \subcaption{White-box jailbreak}
    \resizebox{\textwidth}{!}{
    \begin{tabular}{c c c c c c}
    
        \hline
        
        {\textbf{N°}} &  {\textbf{D}} & {\textbf{R}} & {\textbf{E}} & {\textbf{A}} & {\textbf{D}} \\
        \hline
    
        {\textbf{1}} & {8 (H)} & {9 (H)} & {8 (H)} & {8 (H)} & {6 (M)} \\
        \hline
    
        {\textbf{2}} & {6 (M)} & {6 (M)} & {6 (M)} & {6 (M)} & {5 (M)} \\
        \hline
    
        {\textbf{3}} & {7 (H)} & {7 (H)} & {7 (H)} & {7 (H)} & {5 (M)}  \\
        \hline
    
        {\textbf{4}} & {7 (H)} & {6 (M)} & {5 (M)} & {6 (M)} & {5 (M)} \\
        \hline
    
        {\textbf{5}} & {8 (H)} & {9 (H)} & {7 (H)} & {7 (H)} & {6 (M)} \\
        \hline
    
        {\textbf{6}} & {8 (H)} & {8 (H)} & {7 (H)} & {8 (H)} & {6 (M)} \\
        \hline
    
        {\textbf{7}} & {7 (H)} & {6 (M)} & {7 (H)} & {6 (M)} & {5 (M)} \\
        \hline
    
        {\textbf{8}} & {8 (H)} & {9 (H)} & {8 (H)} & {7 (H)} & {6 (M)} \\
        \hline

        {\textbf{$\bar{x}$}} & {7.38} & {7.5} & {6.75} & {6.88} & {5.5} \\
        \cline{2-6}

        {\textbf{$\sigma$}} & {0.69} & {1.32} & {0.94} & {0.78} & {0.5} \\
        \cline{2-6}

        {\textbf{$COV$}} & \textbf{9.44\%} & \textbf{17.64\%} & \textbf{14.34\%} & \textbf{11.35\%} & \textbf{9.09\%} \\
        \cline{2-6}
        
    \end{tabular}
    }
\end{minipage}%
\hspace{0.02\textwidth}
\begin{minipage}{0.48\textwidth}
    \centering
    \subcaption{Black-box jailbreak}
    \resizebox{\textwidth}{!}{
    \begin{tabular}{c c c c c c}
    \hline
        \textbf{N°} & \textbf{D} & \textbf{R} & \textbf{E} & \textbf{A} & \textbf{D} \\
        \hline
        \textbf{1} & 8 (H) & 7 (H) & 7 (H) & 8 (H) & 5 (M)  \\
        \hline
        \textbf{2} & 8 (H) & 8 (H) & 8 (H) & 7 (H) & 5 (M)  \\
        \hline
        \textbf{3} & 8 (H) & 8 (H) & 7 (H) & 7 (H) & 6 (M)  \\
        \hline
        \textbf{4} & 9 (H) & 8 (H) & 8 (H) & 8 (H) & 6 (M)  \\
        \hline
        \textbf{5} & 8 (H) & 9 (H) & 8 (H) & 7 (H) & 5 (M) \\
        \hline
        \textbf{6} & 8 (H) & 6 (M) & 7 (H) & 7 (H) & 5 (M)  \\
        \hline
        \textbf{7} & 9 (H) & 8 (H) & 9 (H) & 8 (H) & 5 (M)  \\
        \hline
        \textbf{8} & 8 (H) & 7 (H) & 7 (H) & 7 (H) & 5 (M) \\
        \hline

        {\textbf{$\bar{x}$}} & {8.25} & {7.63} & {7.63} & {7.38} & {5.25}  \\
        \cline{2-6}

        {\textbf{$\sigma$}} & {0.43} & {0.85} & {0.69} & {0.48} & {0.43} \\
        \cline{2-6}

        {\textbf{$COV$}} & \textbf{5.25\%} & \textbf{11.24\%} & \textbf{9.13\%} & \textbf{6.56\%} & \textbf{8.25\%} \\
        \cline{2-6}
    \end{tabular}
    }
\end{minipage}

\vspace{0.002cm} 

\begin{minipage}{0.48\textwidth}
    \centering
    \subcaption{Prompt-injection attacks}
    \resizebox{\textwidth}{!}{
    \begin{tabular}{c c c c c c}
    
        \hline
        
        {\textbf{N°}} & {\textbf{D}} & {\textbf{R}} & {\textbf{E}} & {\textbf{A}} & {\textbf{D}} \\
        \hline
    
        {\textbf{1}} & {8 (H)} & {9 (H)} & {8 (H)} & {7 (H)} & {6 (M)} \\
        \hline
    
        {\textbf{2}} & {8 (H)} & {8 (H)} & {8 (H)} & {7 (H)} & {6 (M)} \\
        \hline
    
        {\textbf{3}} & {7 (H)} & {9 (H)} & {7 (H)} & {6 (M)} & {7 (H)} \\
        \hline
    
        {\textbf{4}} & {7 (H)} & {8 (H)} & {8 (H)} & {8 (H)} & {5 (M)} \\
        \hline
    
        {\textbf{5}} & {8 (H)} & {9 (H)} & {9 (H)} & {8 (H)} & {6 (M)} \\
        \hline
    
        {\textbf{6}} & {8 (H)} & {8 (H)} & {7 (H)} & {8 (H)} & {5 (M)} \\
        \hline
    
        {\textbf{7}} & {7 (H)} & {6 (M)} & {6 (M)} & {7 (H)} & {5 (M)} \\
        \hline
    
        {\textbf{8}} & {7 (H)} & {6 (M)} & {7 (H)} & {6 (M)} & {5 (M)} \\
        \hline

        {\textbf{$\bar{x}$}} & {7.5} & {7.88} & {7.5} & {7.13} & {5.63}  \\
        \cline{2-6}

        {\textbf{$\sigma$}} & {0.5} & {1.66} & {0.87} & {0.78} & {0.69} \\
        \cline{2-6}

        {\textbf{$COV$}} & \textbf{6.67\%} & \textbf{21.06\%} & \textbf{11.6\%} & \textbf{10.9\%} & \textbf{12.2\%} \\
        \cline{2-6}
        
    \end{tabular}

    }
\end{minipage}%
\hspace{0.02\textwidth}
\begin{minipage}{0.48\textwidth}
    \centering
    \subcaption{Evasion attacks}
    \resizebox{\textwidth}{!}{
    \begin{tabular}{c c c c c c}
        \hline
        \textbf{N°} & \textbf{D} & \textbf{R} & \textbf{E} & \textbf{A} & \textbf{D} \\ \hline
        \textbf{1} & 7 (H) & 7 (H) & 6 (M) & 7 (H) & 5 (M)  \\ \hline
        \textbf{2} & 7 (H) & 9 (H) & 8 (H) & 7 (H) & 5 (M) \\ \hline
        \textbf{3} & 6 (M) & 8 (H) & 6 (M) & 6 (M) & 5 (M) \\ \hline
        \textbf{4} & 8 (H) & 8 (H) & 7 (H) & 7 (H) & 5 (M) \\ \hline
        \textbf{5} & 6 (M) & 9 (H) & 7 (H) & 7 (H) & 6 (M)  \\ \hline
        \textbf{6} & 8 (H) & 8 (H) & 8 (H) & 8 (H) & 5 (M) \\ \hline
        \textbf{7} & 9 (H) & 8 (H) & 8 (H) & 8 (H) & 5 (M)  \\ \hline
        \textbf{8} & 8 (H) & 8 (H) & 8 (H) & 8 (H) & 5 (M) \\ \hline

        {\textbf{$\bar{x}$}} & {7.38} & {8.13} & {7.25} & {7.25} & {5.13}  \\
        \cline{2-6}

        {\textbf{$\sigma$}} & {0.99} & {0.60} & {0.83} & {0.67} & {0.34} \\
        \cline{2-6}

        {\textbf{$COV$}} & \textbf{13.45\%} & \textbf{7.38\%} & \textbf{11.44\%} & \textbf{9.12\%} & \textbf{6.45\%} \\
        \cline{2-6}
    
    \end{tabular}
    }
\end{minipage}

\vspace{0.002cm} 

\begin{minipage}{0.48\textwidth}
    \centering
    \subcaption{Model-extraction attacks}
    \resizebox{\textwidth}{!}{
    \begin{tabular}{c c c c c c}
    
        \hline
        
        {\textbf{N°}} & {\textbf{D}} & {\textbf{R}} & {\textbf{E}} & {\textbf{A}} & {\textbf{D}}  \\
        \hline
        
        {\textbf{1}} & {9 (H)} & {8 (H)} & {8 (H)} & {8 (H)} & {5 (M)} \\
        \hline
        
        {\textbf{2}} & {8 (H)} & {9 (H)} & {8 (H)} & {7 (H)} & {5 (M)} \\
        \hline
        
        {\textbf{3}} & {9 (H)} & {8 (H)} & {8 (H)} & {9 (H)} & {6 (M)} \\
        \hline
        
        {\textbf{4}} & {8 (H)} & {8 (H)} & {7 (H)} & {7 (H)} & {5 (M)}  \\
        \hline
        
        {\textbf{5}} & {8 (H)} & {5 (M)} & {6 (M)} & {8 (H)} & {4 (M)} \\
        \hline
        
        {\textbf{6}} & {7 (H)} & {7 (H)} & {7 (H)} & {7 (H)} & {6 (M)} \\
        \hline
        
        {\textbf{7}} & {8 (H)} & {6 (M)} & {7 (H)} & {8 (H)} & {5 (M)} \\
        \hline
        
        {\textbf{8}} & {7 (H)} & {7 (H)} & {7 (H)} & {6 (M)} & {5 (M)} \\
        \hline

        {\textbf{$\bar{x}$}} & {8} & {7.25} & {7.25} & {7.5} & {5.13}  \\
        \cline{2-6}

        {\textbf{$\sigma$}} & {0.71} & {1.09} & {0.67} & {0.75} & {0.60} \\
        \cline{2-6}

        {\textbf{$COV$}} & \textbf{8.88\%} & \textbf{15.03\%} & \textbf{9.12\%} & \textbf{10.00\%} & \textbf{11.72\%} \\
        \cline{2-6}
    \end{tabular}
    
    }
\end{minipage}%
\hspace{0.02\textwidth}
\begin{minipage}{0.48\textwidth}
    \centering
    \subcaption{Model-inference attacks}
    \resizebox{\textwidth}{!}{
    \begin{tabular}{c c c c c c}
    
        \hline
    
        {\textbf{N°}} & {\textbf{D}} & {\textbf{R}} & {\textbf{E}} & {\textbf{A}} & {\textbf{D}} \\
        \hline
        
        {\textbf{1}} & {8 (H)} & {9 (H)} & {7 (H)} & {7 (H)} & {6 (M)} \\
        \hline
        
        {\textbf{2}} & {7 (H)} & {8 (H)} & {7 (H)} & {7 (H)} & {5 (M)} \\
        \hline
        
        {\textbf{3}} & {6 (M)} & {5 (M)} & {6 (M)} & {6 (M)} & {5 (M)} \\
        \hline
        
        {\textbf{4}} & {8 (H)} & {6 (M)} & {6 (M)} & {7 (H)} & {5 (M)} \\
        \hline
        
        {\textbf{5}} & {9 (H)} & {8 (H)} & {8 (H)} & {9 (H)} & {6 (M)} \\
        \hline
        
        {\textbf{6}} & {7 (H)} & {7 (H)} & {7 (H)} & {7 (H)} & {5 (M)} \\
        \hline
        
        {\textbf{7}} & {8 (H)} & {7 (H)} & {7 (H)} & {7 (H)} & {5 (M)} \\
        \hline
        
        {\textbf{8}} & {8 (H)} & {8 (H)} & {7 (H)} & {7 (H)} & {5 (M)} \\
        \hline

        {\textbf{$\bar{x}$}} & {7.63} & {7.25} & {6.88} & {7.13} & {5.25}  \\
        \cline{2-6}

        {\textbf{$\sigma$}} & {0.83} & {1.09} & {0.64} & {0.83} & {0.47} \\
        \cline{2-6}

        {\textbf{$COV$}} & \textbf{10.87\%} & \textbf{15.03\%} & \textbf{9.30\%} & \textbf{11.65\%} & \textbf{8.95\%} \\
        \cline{2-6}
    
    \end{tabular}
    
    }
\end{minipage}

\vspace{0.002cm} 

\begin{minipage}{0.48\textwidth}
    \centering
    \subcaption{Poisoning/Trojan/Backdoor attacks}
    \resizebox{\textwidth}{!}{
    \begin{tabular}{c c c c c c}
    
        \hline
        
        {\textbf{N°}} & {\textbf{D}} & {\textbf{R}} & {\textbf{E}} & {\textbf{A}} & {\textbf{D}} \\
        \hline
    
        {\textbf{1}} & {8 (H)} & {8 (H)} & {8 (H)} & {8 (H)} & {6 (M)} \\
        \hline
        
        {\textbf{2}} & {8 (H)} & {6 (M)} & {7 (H)} & {7 (H)} & {5 (M)} \\
        \hline
        
        {\textbf{3}} & {8 (H)} & {8 (H)} & {8 (H)} & {8 (H)} & {6 (M)} \\
        \hline
        
        {\textbf{4}} & {7 (H)} & {6 (M)} & {6 (M)} & {6 (M)} & {4 (M)} \\
        \hline
        
        {\textbf{5}} & {9 (H)} & {7 (H)} & {6 (M)} & {9 (H)} & {5 (M)} \\
        \hline
        
        {\textbf{6}} & {8 (H)} & {9 (H)} & {8 (H)} & {8 (H)} & {6 (M)} \\
        \hline
        
        {\textbf{7}} & {8 (H)} & {8 (H)} & {7 (H)} & {7 (H)} & {5 (M)} \\
        \hline
        
        {\textbf{8}} & {8 (H)} & {8 (H)} & {8 (H)} & {8 (H)} & {6 (M)} \\
        \hline

        {\textbf{$\bar{x}$}} & {8} & {7.5} & {7.25} & {7.62} & {5.38}  \\
        \cline{2-6}

        {\textbf{$\sigma$}} & {0.5} & {1.00} & {0.83} & {0.86} & {0.70} \\
        \cline{2-6}

        {\textbf{$COV$}} & \textbf{6.25\%} & \textbf{13.34\%} & \textbf{11.45\%} & \textbf{11.28\%} & \textbf{13.03\%} \\
        \cline{2-6}
    
    \end{tabular}

    }
\end{minipage}

\caption{Variations of DREAD assessments}
\label{tab:variance-dread}
\end{table}

\subsection{Evaluation of CVSS assessments}

The CVSS framework offers qualitative assessments, making Entropy (H) a more suitable measure of variability than the COV\%. The results of this analysis are summarized in Table \ref{tab:variance-cvss}.

Similar to observations from DREAD, several CVSS factors exhibit minimal or no variability across the attacks. For instance, the \textbf{Attack Vector} consistently takes the value `Network' for 55 out of the 56 attacks, reflecting the predominance of network-based adversarial attacks targeting online LLMs. This lack of differentiation renders the factor \textbf{unsuitable} for assessing the diversity of attack mechanisms against LLMs.

Likewise, the \textbf{Privileges Required} and \textbf{User Interaction} factors show low variability. The Privileges Required factor is typically `None', except for White-box and Black-box Jailbreak attacks. Similarly, User Interaction is also `None' for most attacks, apart from White-box Jailbreak and Prompt Injection attacks. This suggests these factors are only \textbf{relevant to specific types of attacks} but fail to provide meaningful insights across broader categories.

The \textbf{Confidentiality}, \textbf{Integrity}, and \textbf{Availability} (CIA) Impact factors also demonstrate significant limitations. Each type of AAs typically targets a specific aspect of the CIA triad, leaving the other factors unused. For example, \textbf{Model Extraction} attacks heavily impact \textbf{Confidentiality} while leaving Integrity and Availability unaffected, resulting in \textbf{null entropy} for the latter factors. Similarly, attacks such as \textbf{Poisoning}, \textbf{Trojan}, and \textbf{Backdoor} primarily target \textbf{Integrity}, leaving Confidentiality and Availability unchanged. While these factors vary across attack types, they remain \textbf{static within individual attack categories}, limiting their ability to differentiate attacks at a granular level.

The \textbf{Scope} factor follows a similar trend, showing null entropy in four of the seven attack classes (White-box Jailbreak, Evasion, Model Extraction, and Model Inference). Even within its variability, it often remains uniform within a class, such as being consistently `Changed' for all Model Inference attacks or `Unchanged' for all Evasion attacks.

This highlights the limitation of \textbf{specific} qualitative-factors in being suitable in some cases and unsuitable in others.

Among the factors, only \textbf{Attack Complexity} shows relatively higher entropy, with most attacks presenting two to three different values from the mode. This variability reflects the differing levels of expertise required to execute various attacks, making this factor \textbf{appropriate} for assessing attack difficulty. However, it could benefit from further refinement to enhance its precision.

\begin{table}[h]
\centering

\begin{minipage}{0.45\textwidth}
    \centering
    \subcaption{White-box jailbreak}
    \resizebox{\textwidth}{!}{
    \begin{tabular}{c c c c c c c c c}
    
        \hline
        
        {\textbf{N°}} & {\textbf{AV}} & {\textbf{AC}} & {\textbf{PR}} & {\textbf{UI}} & {\textbf{S}} & {\textbf{C}} & {\textbf{I}} & {\textbf{A}} \\
        \hline
    
        {\textbf{1}} & {N} & {H} & {N} & {N} & {C} & {L} & {H} & {N}  \\
        \hline
    
        {\textbf{2}} & {N} & {H} & {N} & {N} & {C} & {L} & {H} & {N}  \\
        \hline
    
        {\textbf{3}} & {N} & {H} & {N} & {N} & {C} & {L} &{H} & {N} \\
        \hline
    
        {\textbf{4}} & {N} & {H} & {L} & {N} & {C} & {L} & {H} & {N} \\
        \hline
    
        {\textbf{5}} & {N} & {L} & {L} & {N} & {C} & {L} & {H} & {N}  \\
        \hline
    
        {\textbf{6}} & {N} & {L} & {L} & {N} & {C} & {L} & {H} & {N} \\
        \hline
    
        {\textbf{7}} & {N} & {H} & {N} & {R} & {C} & {L} & {H} & {N} \\
        \hline
    
        {\textbf{8}} & {N} & {L} & {N} & {R} & {C} & {L} & {H} & {N} \\
        \hline

        {\textbf{$M$}} & {N} & {H} & {N} & {N} & {C} & L & H & N \\
        \cline{2-9}

        {\textbf{$p_{i}$}} & {1} & {5/8} & {5/8} & {6/8} & {1} & 1 & 1 & 1 \\
        \cline{2-9}

        {\textbf{$H$}} & \textbf{0.00} & \textbf{0.95} & \textbf{0.95} & \textbf{0.81} & \textbf{0.00} & \textbf{0.00} & \textbf{0.00} & \textbf{0.00}\\
        \cline{2-9}
        
    \end{tabular}
    }
\end{minipage}%
\hspace{0.02\textwidth}
\begin{minipage}{0.45\textwidth}
    \centering
    \subcaption{Black-box jailbreak}
    \resizebox{\textwidth}{!}{
    \begin{tabular}{c c c c c c c c c}
    
        \hline
        
        {\textbf{N°}} & {\textbf{AV}} & {\textbf{AC}} & {\textbf{PR}} & {\textbf{UI}} & {\textbf{S}} & {\textbf{C}} & {\textbf{I}} & {\textbf{A}} \\
        \hline
        
        {\textbf{1}} & {N} & {L} & {N} & {N} & {U} & {L} & {L} & {N}  \\
        \hline
    
        {\textbf{2}} & {N} & {L} & {N} & {N} & {U} & {L} & {H} & {N}  \\
        \hline
    
        {\textbf{3}} & {N} & {L} & {N} & {N} & {U} & {L} & {L} & {N}  \\
        \hline
    
        {\textbf{4}} & {N} & {L} & {N} & {N} & {C} & {L} & {N} & {N}  \\
        \hline
    
        {\textbf{5}} & {N} & {L} & {N} & {N} & {C} & {L} & {N} & {N}  \\
        \hline
    
        {\textbf{6}} & {N} & {H} & {N} & {N} & {U} & {L} & {L} & {N}  \\
        \hline
    
        {\textbf{7}} & {N} & {L} & {L} & {N} & {U} & {L} & {H} & {N} \\
        \hline
    
        {\textbf{8}} & {N} & {L} & {N} & {N} & {U} & {L} & {H} & {N}  \\
        \hline

        {\textbf{$M$}} & {N} & {L} & {N} & {N} & {U} & L & L,H & N \\
        \cline{2-9}

        {\textbf{$p_{i}$}} & {1} & {7/8} & {7/8} & {1} & {6/8} & 1 & 3/8 & 1 \\
        \cline{2-9}

        {\textbf{$H$}} & \textbf{0.00} & \textbf{0.54} & \textbf{0.54} & \textbf{0.00} & \textbf{0.81} & \textbf{0.00} & \textbf{1.56} & \textbf{0.00}\\
        \cline{2-9}
        
    \end{tabular}
    }
\end{minipage}

\vspace{0.002cm} 

\begin{minipage}{0.45\textwidth}
    \centering
    \subcaption{Prompt-injection attacks}
    \resizebox{\textwidth}{!}{
    \begin{tabular}{c c c c c c c c c}
    
        {\textbf{N°}} & {\textbf{AV}} & {\textbf{AC}} & {\textbf{PR}} & {\textbf{UI}} & {\textbf{S}} & {\textbf{C}} & {\textbf{I}} & {\textbf{A}} \\
        \hline
    
        {\textbf{1}} & {N} & {L} & {N} & {N} & {U} & {N} & {H} & {N}\\
        \hline
    
        {\textbf{2}} & {N} & {H} & {N} & {R} & {U} & {L} & {H} & {N} \\
        \hline
    
        {\textbf{3}} & {N} & {L} & {N} & {N} & {U} & {L} & {H} & {N}  \\
        \hline
    
        {\textbf{4}} & {N} & {H} & {N} & {R} & {C} & {L} & {H} & {N}  \\
        \hline
    
        {\textbf{5}} & {N} & {L} & {N} & {N} & {C} & {L} & {H} & {N}  \\
        \hline
    
        {\textbf{6}} & {N} & {H} & {N} & {R} & {C} & {L} & {H} & {N} \\
        \hline
    
        {\textbf{7}} & {N} & {H} & {N} & {R} & {U} & {L} & {H} & {N} \\
        \hline
    
        {\textbf{8}} & {N} & {H} & {N} & {N} & {U} & {L} & {H} & {N} \\
        \hline

        {\textbf{$M$}} & {N} & {H} & {N} & {N,R} & {U} & L & H & N \\
        \cline{2-9}

        {\textbf{$p_{i}$}} & {1} & {5/8} & {1} & {4/8} & {5/8} & 7/8 & 1 & 1 \\
        \cline{2-9}

        {\textbf{$H$}} & \textbf{0.00} & \textbf{0.95} & \textbf{0.00} & \textbf{1.00} & \textbf{0.95} & \textbf{0.54} & \textbf{0.00} & \textbf{0.00}\\
        \cline{2-9}
  
    \end{tabular}

    }
\end{minipage}%
\hspace{0.02\textwidth}
\begin{minipage}{0.45\textwidth}
    \centering
    \subcaption{Evasion attacks}
    \resizebox{\textwidth}{!}{
    \begin{tabular}{c c c c c c c c c}

        \hline
        
        {\textbf{N°}} & {\textbf{AV}} & {\textbf{AC}} & {\textbf{PR}} & {\textbf{UI}} & {\textbf{S}} & {\textbf{C}} & {\textbf{I}} & {\textbf{A}}  \\
        \hline
    
        {\textbf{1}} & {N} & {L} & {N} & {N} & {U} & {N} & {H} & {N} \\
        \hline
    
        {\textbf{2}} & {N} & {L} & {N} & {N} & {U} & {N} & {H} & {N}  \\
        \hline
    
        {\textbf{3}} & {N} & {H} & {N} & {N} & {U} & {N} & {H} & {N}  \\
        \hline
    
        {\textbf{4}} & {N} & {H} & {N} & {N} & {U} & {N} & {H} & {N}  \\
        \hline
    
        {\textbf{5}} & {N} & {L} & {N} & {N} & {U} & {N} & {H} & {N}  \\
        \hline
    
        {\textbf{6}} & {N} & {L} & {N} & {N} & {U} & {N} & {H} & {N}  \\
        \hline
    
        {\textbf{7}} & {N} & {H} & {N} & {N} & {U} & {N} & {H} & {N}  \\
        \hline
    
        {\textbf{8}} & {N} & {L} & {N} & {N} & {U} & {N} & {H} & {N} \\
        \hline
        
        {\textbf{$M$}} & {N} & {L} & {N} & {N} & {U} & N & H & N \\
        \cline{2-9}

        {\textbf{$p_{i}$}} & {1} & {5/8} & {1} & {1} & {1} & 1 & 1 & 1 \\
        \cline{2-9}

        {\textbf{$H$}} & \textbf{0.00} & \textbf{0.95} & \textbf{0.00} & \textbf{0.00} & \textbf{0.00} & \textbf{0.00} & \textbf{0.00} & \textbf{0.00}\\
        \cline{2-9}
    
    \end{tabular}
    }
\end{minipage}

\vspace{0.002cm} 

\begin{minipage}{0.45\textwidth}
    \centering
    \subcaption{Model-extraction attacks}
    \resizebox{\textwidth}{!}{
    \begin{tabular}{c c c c c c c c c}
    
        \hline
        
        {\textbf{N°}} & {\textbf{AV}} & {\textbf{AC}} & {\textbf{PR}} & {\textbf{UI}} & {\textbf{S}} & {\textbf{C}} & {\textbf{I}} & {\textbf{A}} \\
        \hline
    
        {\textbf{1}} & {N} & {H} & {N} & {N} & {C} & {H} & {N} & {N}  \\
        \hline
    
        {\textbf{2}} & {N} & {L} & {N} & {N} & {C} & {H} & {N} & {N}  \\
        \hline
    
        {\textbf{3}} & {N} & {L} & {N} & {N} & {C} & {H} & {N} & {N} \\
        \hline
    
        {\textbf{4}} & {N} & {H} & {N} & {N} & {C} & {H} & {N} & {N} \\
        \hline
    
        {\textbf{5}} & {N} & {H} & {N} & {N} & {C} & {H} & {N} & {N}  \\
        \hline
    
        {\textbf{6}} & {N} & {H} & {N} & {N} & {C} & {H} & {N} & {N}  \\
        \hline
    
        {\textbf{7}} & {N} & {H} & {N} & {N} & {C} & {H} & {N} & {N} \\
        \hline
    
        {\textbf{8}} & {N} & {L} & {N} & {N} & {C} & {H} & {N} & {N}  \\
        \hline

        {\textbf{$M$}} & {N} & {H} & {N} & {N} & {C} & H & N & N \\
        \cline{2-9}

        {\textbf{$p_{i}$}} & {1} & {5/8} & {1} & {1} & {1} & 1 & 1 & 1 \\
        \cline{2-9}

        {\textbf{$H$}} & \textbf{0.00} & \textbf{0.95} & \textbf{0.00} & \textbf{0.00} & \textbf{0.00} & \textbf{0.00} & \textbf{0.00} & \textbf{0.00}\\
        \cline{2-9}
    \end{tabular}
    
    }
\end{minipage}%
\hspace{0.02\textwidth}
\begin{minipage}{0.45\textwidth}
    \centering
    \subcaption{Model-inference attacks}
    \resizebox{\textwidth}{!}{
    \begin{tabular}{c c c c c c c c c}
    
        \hline
        
        {\textbf{N°}} & {\textbf{AV}} & {\textbf{AC}} & {\textbf{PR}} & {\textbf{UI}} & {\textbf{S}} & {\textbf{C}} & {\textbf{I}} & {\textbf{A}} \\
        \hline
    
        {\textbf{1}} & {N} & {L} & {N} & {N} & {C} & {H} & {N} & {N} \\
        \hline
    
        {\textbf{2}} & {N} & {L} & {N} & {N} & {C} & {H} & {N} & {N}  \\
        \hline
    
        {\textbf{3}} & {N} & {H} & {N} & {N} & {C} & {H} & {N} & {N}  \\
        \hline
    
        {\textbf{4}} & {N} & {H} & {N} & {N} & {C} & {H} & {N} & {N}  \\
        \hline
    
        {\textbf{5}} & {N} & {L} & {N} & {N} & {C} & {H} & {N} & {N} \\
        \hline
    
        {\textbf{6}} & {N} & {L} & {N} & {N} & {C} & {H} & {N} & {N}  \\
        \hline
    
        {\textbf{7}} & {N} & {L} & {N} & {N} & {C} & {H} & {N} & {N}  \\
        \hline
    
        {\textbf{8}} & {N} & {L} & {N} & {N} & {C} & {H} & {N} & {N} \\
        \hline

        {\textbf{$M$}} & {N} & {L} & {N} & {N} & {C} & H & N & N \\
        \cline{2-9}

        {\textbf{$p_{i}$}} & {1} & {6/8} & {1} & {1} & {1} & 1 & 1 & 1 \\
        \cline{2-9}

        {\textbf{$H$}} & \textbf{0.00} & \textbf{0.81} & \textbf{0.00} & \textbf{0.00} & \textbf{0.00} & \textbf{0.00} & \textbf{0.00} & \textbf{0.00}\\
        \cline{2-9}
    
    \end{tabular}
    
    }
\end{minipage}

\vspace{0.002cm} 

\begin{minipage}{0.45\textwidth}
    \centering
    \subcaption{Poisoning/Trojan/Backdoor attacks}
    \resizebox{\textwidth}{!}{
    \begin{tabular}{c c c c c c c c c }
    
        \hline
        
        {\textbf{N°}} & {\textbf{AV}} & {\textbf{AC}} & {\textbf{PR}} & {\textbf{UI}} & {\textbf{S}} & {\textbf{C}} & {\textbf{I}} &  {\textbf{A}} \\
        \hline
    
        {\textbf{1}} & {N} & {H} & {N} & {N} & {C} & {L} & {H} & {N}  \\
        \hline
    
        {\textbf{2}} & {L} & {H} & {L} & {N} & {U} & {L} & {H} & {N}  \\
        \hline
    
        {\textbf{3}} & {N} & {L} & {L} & {N} & {C} & {L} & {H} & {N}  \\
        \hline
    
        {\textbf{4}} & {N} & {H} & {L} & {N} & {U} & {L} & {H} & {N} \\
        \hline
    
        {\textbf{5}} & {N} & {H} & {L} & {N} & {U} & {L} & {H} & {N} \\
        \hline
    
        {\textbf{6}} & {N} & {L} & {N} & {N} & {C} & {L} & {H} & {N} \\
        \hline
    
        {\textbf{7}} & {N} & {H} & {N} & {N} & {C} & {L} & {H} & {N}  \\
        \hline
    
        {\textbf{8}} & {N} & {H} & {N} & {N} & {C} & {L} & {H} & {N} \\
        \hline

        {\textbf{$M$}} & {N} & {H} & {N,L} & {N} & {C} & L & H & N \\
        \cline{2-9}

        {\textbf{$p_{i}$}} & {7/8} & {6/8} & {4/8} & {1} & {5/8} & 1 & 1 & 1 \\
        \cline{2-9}

        {\textbf{$H$}} & \textbf{0.54} & \textbf{0.81} & \textbf{1.00} & \textbf{0.00} & \textbf{0.95} & \textbf{0.00} & \textbf{0.00} & \textbf{0.00}\\
        \cline{2-9}
    
    \end{tabular}

    }
\end{minipage}
\caption{Variations of CVSS assessments}
\label{tab:variance-cvss}
\end{table}

\subsection{Evaluation of OWASP Risk Rating assessments}

Now we assess the utility of OWAS Risk Rating factors using their Coefficient of Variation calculated in Table \ref{tab:variance-owasp}.

We start with the \textbf{Skill Level} factor, its median values for the seven attack classes generally fall between 6 and 7, with variations of less than 10\% in most cases. Similar patterns are observed for the \textbf{Motivation} and \textbf{Opportunity} factors, where Motivation scores are predominantly between 7 and 8, and Opportunity scores range from 6 to 7, both exhibiting very low variability within each class. The same holds true for the \textbf{Size of Threat Agent} factor, where the median consistently falls between 5 and 6, with a COV below 9.2\% across six of the seven classes.

These consistent results can be attributed to the \textbf{shared characteristics} of AAs against LLMs: attackers typically possess medium-to-high skill levels, show strong motivation, have significant opportunities due to the accessibility of LLMs, and represent a medium-sized threat agent, as these attacks are common but often conducted by individuals. This uniformity in attributes leads to \textbf{repeated} values across the 56 attacks, limiting the ability of these metrics to \textbf{differentiate} between attacks effectively.

A similar trend is observed for the \textbf{Ease of Discovery}, \textbf{Ease of Exploit}, \textbf{Awareness} of defenders, and \textbf{Intrusion Detection} Capabilities factors. The median values for these factors remain consistently around 5 and 6 across the seven classes, with COVs between 8\% and 9\%. This lack of variability within attack types reduces the informativeness of these factors.

The \textbf{Confidentiality}, \textbf{Integrity}, and \textbf{Availability} factors show a similar limitation, as observed with CVSS metrics. Depending on the attack type, at least one of these factors is often \textbf{not relevant} and scores minimal values. However, due to OWASP RR's broader scoring scale (values out of 10), these factors exhibit relatively higher COV percentages compared to CVSS, providing slightly more variability.

The \textbf{Financial} and \textbf{Reputation Damages} factors are critical for assessing attacks on LLMs, given the potential for data breaches and information leaks that can erode customer trust. These factors consistently score medium-to-high values across all attack types. However, their low COV percentages within the same attack type make it challenging to rank attacks fairly based on these criteria.

For \textbf{Non-Compliance} and \textbf{Privacy Violation}, the results indicate that these factors are relevant \textbf{only for specific attack types}, such as Model Extraction, Model Inference, and Poisoning/Trojan/Backdoor attacks. Other types, like White-box Jailbreaks and Model Evasion, exhibit low-to-medium impacts on Non-Compliance, making these factors valuable for specific contexts but less applicable across all attack types.

While these OWASP RR factors provide extensive information about each attack, they are not consistently effective in distinguishing between them. However, the broader scoring range (values out of 10) used by OWASP RR does \textbf{introduce more variability} compared to CVSS, making it somewhat more adaptable for attack differentiation.

\begin{table}[h]
\centering
\vspace{-3cm}
\begin{minipage}{\textwidth}
    \centering
    \subcaption{White-box jailbreak}
    \resizebox{0.92\textwidth}{!}{
    \begin{tabular}{c c c c c c c c c c c c c c c c}
    
        \hline
        
        {\textbf{N°}} &  {\textbf{SL}} & {\textbf{M}} & {\textbf{O}} & {\textbf{S}} & {\textbf{ED}} &  {\textbf{EE}} & {\textbf{A}} & {\textbf{ID}} & {\textbf{LC}} & {\textbf{LI}} &  {\textbf{LA}} & {\textbf{FD}} & {\textbf{RD}} & {\textbf{NC}} & {\textbf{PV}}  \\
        \hline
    
        {\textbf{1}} & {7 (H)} & {6 (H)} & {6 (H)} & {6 (H)} & {7 (H)} & 8 (H) & 5 (M) & 5 (M) & 5 (M) & 7 (H) & 4 (M) & 7 (H) & 8 (H) & 4 (M) & 4 (M) \\
        \hline
    
        {\textbf{2}} & {7 (H)} & {6 (H)} & {5 (M)} & {5 (M)} & {5 (M)} & 7 (H) & 5 (M) & 5 (M) & 5 (M) & 6 (H) & 3 (M) & 6 (H) & 7 (H) & 4 (M) & 4 (M) \\
        \hline
    
        {\textbf{3}} & {7 (H)} & {7 (H)} & {5 (M)} & {6 (H)} & {6 (H)} & 7 (H) & 5 (M) & 5 (M) & 5 (M) & 7 (H) & 3 (M) & 6 (H) & 7 (H) & 6 (H) & 5 (M) \\
        \hline
    
        {\textbf{4}} & {5 (M)} & {6 (H)} & {4 (M)} & {3 (M)} & {4 (M)} & 6 (H) & 5 (M) & 4 (M) & 4 (M) & 6 (H) & 1 (L) & 5 (H) & 6 (H) & 3 (M) & 4 (M) \\
        \hline
    
        {\textbf{5}} & {6 (H)} & {7 (H)} & {6 (H)} & {5 (M)} & {6 (H)} & 7 (H) & 6 (H) & 5 (M) & 5 (M) & 7 (H) & 1 (L) & 6 (H) & 7 (H) & 4 (M) & 4 (M) \\
        \hline
    
        {\textbf{6}} & {7 (H)} & {7 (H)} & {6 (H)} & {6 (H)} & {6 (H)} & 7 (H) & 5 (M) & 5 (M) & 5 (M) & 7 (H) & 2 (L) & 6 (H) & 7 (H) & 4 (M) & 5 (M) \\
        \hline
    
        {\textbf{7}} & {6 (H)} & {6 (H)} & {5 (M)} & {4 (M)} & {5 (M)} & 6 (H) & 5 (M) & 5 (M) & 4 (M) & 6 (H) & 1 (L) & 5 (M) & 6 (H) & 3 (M) & 3 (M) \\
        \hline
    
        {\textbf{8}} & {7 (H)} & {7 (H)} & {7 (H)} & {5 (M)} & {7 (H)} & 7 (H) & 6 (H) & 5 (M) & 4 (M) & 7 (H) & 1 (L) & 6 (H) & 7 (H) & 4 (M) & 4 (M) \\
        \hline

        {\textbf{$\bar{x}$}} & {6.5} & {6.5} & {5.63} & {5.13} & {6.00} & {6.88} & {5.25} & {4.88} & {4.62} & {6.62} & {2.00} & {5.88} & {6.75} & {4.00} & {4.12}  \\
        \cline{2-16}


        {\textbf{$COV$}} & \textbf{10.92\%} & \textbf{7.69\%} & \textbf{13.34\%} & \textbf{15.23\%} & \textbf{14.83\%} & \textbf{8.72\%} & \textbf{8.25\%} & \textbf{6.78\%} & \textbf{10.47\%} & \textbf{7.31\%} & \textbf{55.90\%} & \textbf{10.20\%} & \textbf{9.80\%} & \textbf{21.65\%} & \textbf{14.53\%} \\
        \cline{2-16}
        
    \end{tabular}
    }
\end{minipage}%
\vspace{0.002cm}
\begin{minipage}{\textwidth}
    \centering
    \subcaption{Black-box jailbreak}
    \resizebox{0.92\textwidth}{!}{
    \begin{tabular}{c c c c c c c c c c c c c c c c}
 
        \hline
        
        {\textbf{N°}} &  {\textbf{SL}} & {\textbf{M}} & {\textbf{O}} & {\textbf{S}} & {\textbf{ED}} &  {\textbf{EE}} & {\textbf{A}} & {\textbf{ID}} & {\textbf{LC}} & {\textbf{LI}} &  {\textbf{LA}} & {\textbf{FD}} & {\textbf{RD}} & {\textbf{NC}} & {\textbf{PV}}  \\
        \hline
    
        {\textbf{1}} & {6 (H)} & {8 (H)} & {8 (H)} & {6 (H)} & {6 (H)} & 7 (H) & 5 (M) & 6 (H) & 8 (H) & 2 (L) & 1 (L) & 6 (H) & 8 (H) & 5 (M) & 7 (H) \\
        \hline
    
        {\textbf{2}} & {5 (M)} & {8 (H)} & {8 (H)} & {6 (H)} & {6 (H)} & 7 (H) & 6 (H) & 6 (H) & 8 (H) & 1 (L) & 1 (L) & 6 (H) & 8 (H) & 4 (M) & 6 (H) \\
        \hline
    
        {\textbf{3}} & {4 (M)} & {8 (H)} & {8 (H)} & {6 (H)} & {7 (H)} & 8 (H) & 7 (H) & 7 (H) & 9 (H) & 1 (L) & 1 (L) & 7 (H) & 8 (H) & 5 (M) & 9 (H) \\
        \hline
    
        {\textbf{4}} & {7 (H)} & {8 (H)} & {7 (H)} & {5 (M)} & {6 (H)} & 8 (H) & 6 (H) & 8 (H) & 8 (H) & 1 (L) & 1 (L) & 6 (H) & 8 (H) & 5 (M) & 8 (H) \\
        \hline
    
        {\textbf{5}} & {6 (H)} & {8 (H)} & {8 (H)} & {5 (M)} & {6 (H)} & 8 (H) & 5 (M) & 7 (H) & 7 (H) & 1 (L) & 1 (L) & 5 (M) & 7 (H) & 5 (M) & 7 (H) \\
        \hline
    
        {\textbf{6}} & {6 (H)} & {8 (H)} & {8 (H)} & {6 (H)} & {6 (H)} & 8 (H) & 6 (H) & 7 (H) & 7 (H) & 2 (L) & 1 (L) & 6 (H) & 7 (H) & 5 (M) & 7 (H) \\
        \hline
    
        {\textbf{7}} & {6 (H)} & {9 (H)} & {8 (H)} & {5 (M)} & {6 (H)} & 8 (H) & 5 (M) & 7 (H) & 8 (H) & 3 (M) & 1 (L) & 6 (H) & 9 (H) & 7 (H) & 8 (H) \\
        \hline
    
        {\textbf{8}} & {6 (H)} & {8 (H)} & {7 (H)} & {5 (M)} & {6 (H)} & 8 (H) & 5 (M) & 7 (H) & 8 (H) & 3 (M) & 1 (L) & 6 (H) & 8 (H) & 5 (M) & 8 (H) \\
        \hline

        {\textbf{$\bar{x}$}} & {5.75} & {8.12} & {7.75} & {5.50} & {6.12} & {7.75} & {5.75} & {7.00} & {8.00} & {1.62} & {1.00} & {6.00} & {7.88} & {5.12} & {7.50}  \\
        \cline{2-16}


        {\textbf{$COV$}} & \textbf{14.42\%} & \textbf{4.07\%} & \textbf{5.59\%} & \textbf{9.09\%} & \textbf{5.40\%} & \textbf{5.59\%} & \textbf{11.48\%} & \textbf{9.29\%} & \textbf{8.25\%} & \textbf{55.56\%} & \textbf{0.00\%} & \textbf{9.00\%} & \textbf{7.63\%} & \textbf{12.91\%} & \textbf{11.60\%} \\
        \cline{2-16}
    \end{tabular}
    }
\end{minipage}

\vspace{0.002cm} 

\begin{minipage}{\textwidth}
    \centering
    \subcaption{Prompt-injection attacks}
    \resizebox{0.92\textwidth}{!}{
    \begin{tabular}{c c c c c c c c c c c c c c c c}
 
        \hline
        
        {\textbf{N°}} &  {\textbf{SL}} & {\textbf{M}} & {\textbf{O}} & {\textbf{S}} & {\textbf{ED}} &  {\textbf{EE}} & {\textbf{A}} & {\textbf{ID}} & {\textbf{LC}} & {\textbf{LI}} &  {\textbf{LA}} & {\textbf{FD}} & {\textbf{RD}} & {\textbf{NC}} & {\textbf{PV}}  \\
        \hline
    
        {\textbf{1}} & {6 (H)} & {8 (H)} & {7 (H)} & {5 (M)} & {6 (H)} & 8 (H) & 6 (H) & 6 (H) & 3 (M) & 8 (H) & 3 (M) & 6 (H) & 8 (H) & 4 (M) & 4 (M) \\
        \hline
    
        {\textbf{2}} & {6 (H)} & {8 (H)} & {7 (H)} & {6 (H)} & {6 (H)} & 7 (H) & 5 (M) & 7 (H) & 5 (M) & 8 (H) & 3 (M) & 7 (H) & 8 (H) & 4 (M) & 6 (H) \\
        \hline
    
        {\textbf{3}} & {6 (H)} & {8 (H)} & {7 (H)} & {5 (M)} & {6 (H)} & 8 (H) & 6 (H) & 6 (H) & 5 (M) & 8 (H) & 3 (M) & 7 (H) & 8 (H) & 4 (M) & 5 (M) \\
        \hline
    
        {\textbf{4}} & {6 (H)} & {8 (H)} & {7 (H)} & {6 (H)} & {6 (H)} & 8 (H) & 6 (H) & 7 (H) & 6 (H) & 7 (H) & 3 (M) & 7 (H) & 8 (H) & 5 (M) & 7 (H) \\
        \hline
    
        {\textbf{5}} & {5 (M)} & {8 (H)} & {7 (H)} & {6 (H)} & {6 (H)} & 8 (H) & 6 (H) & 7 (H) & 6 (H) & 7 (H) & 3 (M) & 7 (H) & 8 (H) & 5 (M) & 6 (H) \\
        \hline
    
        {\textbf{6}} & {7 (H)} & {8 (H)} & {8 (H)} & {6 (H)} & {5 (M)} & 8 (H) & 5 (M) & 8 (H) & 7 (H) & 9 (H) & 3 (M) & 7 (H) & 8 (H) & 5 (M) & 7 (H) \\
        \hline
    
        {\textbf{7}} & {6 (H)} & {8 (H)} & {7 (H)} & {6 (H)} & {5 (M)} & 7 (H) & 5 (M) & 6 (H) & 4 (M) & 8 (H) & 3 (M) & 6 (H) & 7 (H) & 3 (M) & 5 (M) \\
        \hline
    
        {\textbf{8}} & {6 (H)} & {7 (H)} & {6 (H)} & {6 (H)} & {5 (M)} & 6 (H) & 4 (M) & 6 (H) & 3 (M) & 8 (H) & 1 (L) & 5 (M) & 7 (H) & 2 (L) & 3 (M) \\
        \hline

        {\textbf{$\bar{x}$}} & {6.12} & {7.88} & {6.25} & {5.75} & {5.62} & {7.62} & {5.50} & {6.75} & {4.75} & {7.88} & {2.62} & {6.50} & {7.75} & {4.00} & {5.50}  \\
        \cline{2-16}


        {\textbf{$COV$}} & \textbf{7.84\%} & \textbf{4.18\%} & \textbf{10.56\%} & \textbf{7.48\%} & \textbf{8.54\%} & \textbf{9.08\%} & \textbf{11.64\%} & \textbf{9.78\%} & \textbf{24.00\%} & \textbf{7.63\%} & \textbf{33.59\%} & \textbf{10.92\%} & \textbf{5.55\%} & \textbf{21.75\%} & \textbf{19.82\%} \\
        \cline{2-16}
    \end{tabular}

    }
\end{minipage}%
\vspace{0.002cm}
\begin{minipage}{\textwidth}
    \centering
    \subcaption{Evasion attacks}
    \resizebox{0.92\textwidth}{!}{
    \begin{tabular}{c c c c c c c c c c c c c c c c}
 
        \hline
        
        {\textbf{N°}} &  {\textbf{SL}} & {\textbf{M}} & {\textbf{O}} & {\textbf{S}} & {\textbf{ED}} &  {\textbf{EE}} & {\textbf{A}} & {\textbf{ID}} & {\textbf{LC}} & {\textbf{LI}} &  {\textbf{LA}} & {\textbf{FD}} & {\textbf{RD}} & {\textbf{NC}} & {\textbf{PV}}  \\
        \hline
    
        {\textbf{1}} & {7 (H)} & {7 (H)} & {5 (M)} & {4 (M)} & {5 (M)} & 7 (H) & 5 (M) & 6 (H) & 1 (L) & 8 (H) & 0 (L) & 5 (M) & 7 (H) & 4 (M) & 3 (M) \\
        \hline
    
        {\textbf{2}} & {6 (H)} & {7 (H)} & {6 (H)} & {5 (M)} & {5 (M)} & 7 (H) & 5 (M) & 6 (H) & 1 (L) & 8 (H) & 0 (L) & 6 (H) & 7 (H) & 5 (M) & 3 (M) \\
        \hline
    
        {\textbf{3}} & {6 (M)} & {7 (H)} & {5 (M)} & {5 (M)} & {5 (M)} & 6 (H) & 4 (M) & 5 (M) & 1 (L) & 7 (H) & 1 (L) & 5 (M) & 6 (H) & 4 (M) & 3 (M) \\
        \hline
    
        {\textbf{4}} & {7 (H)} & {8 (H)} & {6 (H)} & {5 (M)} & {5 (M)} & 7 (H) & 5 (M) & 6 (H) & 2 (L) & 8 (H) & 1 (L) & 7 (H) & 8 (H) & 5 (M) & 4 (M) \\
        \hline
    
        {\textbf{5}} & {6 (H)} & {7 (H)} & {6 (H)} & {5 (M)} & {5 (M)} & 7 (H) & 5 (M) & 6 (H) & 1 (L) & 7 (H) & 1 (L) & 6 (H) & 7 (H) & 5 (M) & 4 (M) \\
        \hline
    
        {\textbf{6}} & {7 (H)} & {8 (H)} & {7 (H)} & {5 (M)} & {6 (H)} & 7 (H) & 6 (H) & 7 (H) & 2 (L) & 9 (H) & 1 (L) & 7 (H) & 8 (H) & 6 (H) & 5 (M) \\
        \hline
    
        {\textbf{7}} & {7 (H)} & {8 (H)} & {7 (H)} & {5 (M)} & {6 (H)} & 7 (H) & 6 (H) & 7 (H) & 2 (L) & 9 (H) & 1 (L) & 7 (H) & 8 (H) & 6 (H) & 5 (M) \\
        \hline
    
        {\textbf{8}} & {7 (H)} & {8 (H)} & {7 (H)} & {6 (H)} & {6 (H)} & 8 (H) & 6 (H) & 7 (H) & 1 (L) & 8 (H) & 1 (L) & 7 (H) & 8 (H) & 5 (M) & 4 (M) \\
        \hline

        {\textbf{$\bar{x}$}} & {6.75} & {7.50} & {6.12} & {5.12} & {5.38} & {7.12} & {5.38} & {6.50} & {1.25} & {8.00} & {0.75} & {6.38} & {7.50} & {5.12} & {4.00}  \\
        \cline{2-16}


        {\textbf{$COV$}} & \textbf{6.96\%} & \textbf{6.67\%} & \textbf{11.29\%} & \textbf{9.18\%} & \textbf{8.92\%} & \textbf{6.76\%} & \textbf{11.90\%} & \textbf{9.85\%} & \textbf{49.60\%} & \textbf{9.38\%} & \textbf{82.67\%} & \textbf{11.76\%} & \textbf{8.80\%} & \textbf{12.91\%} & \textbf{20.75\%} \\
        \cline{2-16}
    
    \end{tabular}
    }
\end{minipage}

\vspace{0.002cm} 


\begin{minipage}{\textwidth}
    \centering
    \subcaption{Model-extraction attacks}
    \resizebox{0.92\textwidth}{!}{
    \begin{tabular}{c c c c c c c c c c c c c c c c}
 
        \hline
        
        {\textbf{N°}} &  {\textbf{SL}} & {\textbf{M}} & {\textbf{O}} & {\textbf{S}} & {\textbf{ED}} &  {\textbf{EE}} & {\textbf{A}} & {\textbf{ID}} & {\textbf{LC}} & {\textbf{LI}} &  {\textbf{LA}} & {\textbf{FD}} & {\textbf{RD}} & {\textbf{NC}} & {\textbf{PV}}  \\
        \hline
    
        {\textbf{1}} & {7 (H)} & {8 (H)} & {7 (H)} & {6 (H)} & {5 (M)} & 6 (H) & 5 (M) & 7 (H) & 8 (H) & 1 (L) & 1 (L) & 7 (H) & 8 (H) & 7 (H) & 8 (H) \\
        \hline
    
        {\textbf{2}} & {6 (H)} & {8 (H)} & {7 (H)} & {5 (M)} & {6 (H)} & 7 (H) & 5 (M) & 7 (H) & 8 (H) & 2 (L) & 2 (L) & 7 (H) & 9 (H) & 7 (H) & 8 (H) \\
        \hline
    
        {\textbf{3}} & {6 (H)} & {8 (H)} & {7 (H)} & {6 (H)} & {6 (H)} & 6 (H) & 5 (M) & 8 (H) & 9 (H) & 1 (L) & 1 (L) & 7 (H) & 8 (H) & 7 (H) & 9 (H) \\
        \hline
    
        {\textbf{4}} & {6 (H)} & {8 (H)} & {7 (H)} & {6 (H)} & {5 (M)} & 7 (H) & 5 (M) & 7 (H) & 8 (H) & 1 (L) & 1 (L) & 7 (H) & 9 (H) & 7 (H) & 9 (H) \\
        \hline
    
        {\textbf{5}} & {6 (H)} & {8 (H)} & {7 (H)} & {6 (H)} & {5 (M)} & 6 (H) & 4 (M) & 7 (H) & 8 (H) & 1 (L) & 1 (L) & 7 (H) & 8 (H) & 7 (H) & 9 (H) \\
        \hline
    
        {\textbf{6}} & {6 (H)} & {7 (H)} & {7 (H)} & {5 (M)} & {5 (M)} & 6 (H) & 5 (M) & 7 (H) & 8 (H) & 2 (L) & 2 (L) & 7 (H) & 9 (H) & 6 (H) & 8 (H) \\
        \hline
    
        {\textbf{7}} & {6 (H)} & {8 (H)} & {7 (H)} & {5 (M)} & {5 (M)} & 6 (H) & 5 (M) & 7 (H) & 9 (H) & 2 (L) & 2 (L) & 7 (H) & 9 (H) & 7 (H) & 9 (H) \\
        \hline
    
        {\textbf{8}} & {7 (H)} & {8 (H)} & {7 (H)} & {6 (H)} & {5 (M)} & 6 (H) & 5 (M) & 8 (H) & 8 (H) & 1 (L) & 1 (L) & 7 (H) & 9 (H) & 7 (H) & 9 (H) \\
        \hline

        {\textbf{$\bar{x}$}} & {6.38} & {7.88} & {7.00} & {5.75} & {5.25} & {6.38} & {4.88} & {7.38} & {8.25} & {1.25} & {1.25} & {7.00} & {8.62} & {6.88} & {8.62}  \\
        \cline{2-16}


        {\textbf{$COV$}} & \textbf{7.53\%} & \textbf{4.18\%} & \textbf{0.00\%} & \textbf{7.48\%} & \textbf{8.20\%} & \textbf{7.53\%} & \textbf{6.78\%} & \textbf{6.50\%} & \textbf{5.67\%} & \textbf{49.60\%} & \textbf{49.60\%} & \textbf{0.00\%} & \textbf{6.96\%} & \textbf{4.80\%} & \textbf{6.96\%} \\
        \cline{2-16}
    \end{tabular}
    
    }
\end{minipage}%
\vspace{0.002cm}
\begin{minipage}{\textwidth}
    \centering
    \subcaption{Model-inference attacks}
    \resizebox{0.92\textwidth}{!}{
    \begin{tabular}{c c c c c c c c c c c c c c c c}
 
        \hline
        
        {\textbf{N°}} & {\textbf{SL}} & {\textbf{M}} & {\textbf{O}} & {\textbf{S}} & {\textbf{ED}} &  {\textbf{EE}} & {\textbf{A}} & {\textbf{ID}} & {\textbf{LC}} & {\textbf{LI}} &  {\textbf{LA}} & {\textbf{FD}} & {\textbf{RD}} & {\textbf{NC}} & {\textbf{PV}}  \\
        \hline
    
        {\textbf{1}} & {6 (H)} & {7 (H)} & {7 (H)} & {6 (H)} & {5 (M)} & 7 (H) & 6 (H) & 7 (H) & 8 (H) & 1 (L) & 2 (L) & 7 (H) & 8 (H) & 7 (H) & 8 (H) \\
        \hline
    
        {\textbf{2}} & {6 (H)} & {7 (H)} & {7 (H)} & {6 (H)} & {5 (M)} & 6 (H) & 4 (M) & 6 (H) & 8 (H) & 2 (L) & 2 (L) & 7 (H) & 8 (H) & 8 (H) & 8 (H) \\
        \hline
    
        {\textbf{3}} & {5 (M)} & {7 (H)} & {7 (H)} & {5 (M)} & {5 (M)} & 6 (H) & 5 (M) & 6 (H) & 8 (H) & 1 (L) & 2 (L) & 6 (H) & 7 (H) & 7 (H) & 8 (H) \\
        \hline
    
        {\textbf{4}} & {6 (H)} & {7 (H)} & {7 (H)} & {6 (H)} & {6 (H)} & 5 (M) & 6 (H) & 6 (H) & 8 (H) & 1 (L) & 2 (L) & 7 (H) & 7 (H) & 7 (H) & 8 (H) \\
        \hline
    
        {\textbf{5}} & {6 (H)} & {8 (H)} & {7 (H)} & {6 (H)} & {5 (M)} & 6 (H) & 6 (H) & 8 (H) & 9 (H) & 1 (L) & 2 (L) & 8 (H) & 8 (H) & 8 (H) & 9 (H) \\
        \hline
    
        {\textbf{6}} & {6 (H)} & {7 (H)} & {6 (H)} & {6 (H)} & {5 (M)} & 6 (H) & 5 (M) & 7 (H) & 8 (H) & 2 (L) & 2 (L) & 8 (H) & 8 (H) & 8 (H) & 9 (H) \\
        \hline
    
        {\textbf{7}} & {6 (H)} & {8 (H)} & {7 (H)} & {6 (H)} & {6 (H)} & 6 (H) & 5 (M) & 7 (H) & 8 (H) & 2 (L) & 2 (L) & 8 (H) & 8 (H) & 7 (H) & 9 (H) \\
        \hline
    
        {\textbf{8}} & {5 (M)} & {7 (H)} & {7 (H)} & {6 (H)} & {6 (H)} & 7 (H) & 6 (H) & 7 (H) & 7 (H) & 2 (L) & 2 (L) & 8 (H) & 8 (H) & 8 (H) & 6 (H) \\
        \hline

        {\textbf{$\bar{x}$}} & {5.88} & {7.25} & {6.88} & {6.00} & {5.25} & {6.38} & {5.50} & {7.00} & {8.00} & {1.25} & {2.00} & {7.50} & {7.75} & {7.50} & {8.50}  \\
        \cline{2-16}


        {\textbf{$COV$}} & \textbf{8.17\%} & \textbf{6.48\%} & \textbf{4.80\%} & \textbf{7.50\%} & \textbf{8.20\%} & \textbf{7.53\%} & \textbf{11.64\%} & \textbf{9.29\%} & \textbf{6.25\%} & \textbf{49.60\%} & \textbf{0.00\%} & \textbf{8.80\%} & \textbf{5.55\%} & \textbf{6.67\%} & \textbf{7.53\%} \\
        \cline{2-16}
    
    \end{tabular}
    
    }
\end{minipage}

\vspace{0.002cm} 

\begin{minipage}{0.92\textwidth}
    \centering
    \subcaption{Poisoning/Trojan/Backdoor attacks}
    \resizebox{\textwidth}{!}{
    \begin{tabular}{c c c c c c c c c c c c c c c c}
 
        \hline
        
        {\textbf{N°}} &  {\textbf{SL}} & {\textbf{M}} & {\textbf{O}} & {\textbf{S}} & {\textbf{ED}} &  {\textbf{EE}} & {\textbf{A}} & {\textbf{ID}} & {\textbf{LC}} & {\textbf{LI}} &  {\textbf{LA}} & {\textbf{FD}} & {\textbf{RD}} & {\textbf{NC}} & {\textbf{PV}}  \\
        \hline
    
        {\textbf{1}} & {6 (H)} & {8 (H)} & {7 (H)} & {6 (H)} & {5 (M)} & 7 (H) & 5 (M) & 7 (H) & 6 (H) & 7 (H) & 4 (M) & 7 (H) & 8 (H) & 5 (M) & 6 (H) \\
        \hline
    
        {\textbf{2}} & {8 (H)} & {8 (H)} & {6 (H)} & {5 (M)} & {5 (M)} & 6 (H) & 4 (M) & 5 (M) & 7 (H) & 7 (H) & 3 (M) & 7 (H) & 8 (H) & 6 (H) & 7 (H) \\
        \hline
    
        {\textbf{3}} & {7 (H)} & {8 (H)} & {8 (H)} & {6 (H)} & {6 (H)} & 7 (H) & 5 (M) & 6 (H) & 7 (H) & 8 (H) & 4 (M) & 7 (H) & 8 (H) & 6 (H) & 7 (H) \\
        \hline
    
        {\textbf{4}} & {7 (H)} & {8 (H)} & {6 (H)} & {6 (H)} & {5 (M)} & 6 (H) & 4 (M) & 6 (H) & 6 (H) & 8 (H) & 4 (M) & 7 (H) & 8 (H) & 6 (H) & 6 (H) \\
        \hline
    
        {\textbf{5}} & {7 (H)} & {8 (H)} & {7 (H)} & {6 (H)} & {4 (M)} & 6 (H) & 4 (M) & 8 (H) & 6 (H) & 8 (H) & 3 (M) & 8 (H) & 9 (H) & 7 (H) & 6 (H) \\
        \hline
    
        {\textbf{6}} & {8 (H)} & {9 (H)} & {7 (H)} & {6 (H)} & {4 (M)} & 6 (H) & 4 (M) & 9 (H) & 6 (H) & 8 (H) & 4 (M) & 8 (H) & 9 (H) & 7 (H) & 6 (H) \\
        \hline
    
        {\textbf{7}} & {8 (H)} & {9 (H)} & {7 (H)} & {7 (H)} & {5 (M)} & 7 (H) & 4 (M) & 9 (H) & 7 (H) & 8 (H) & 4 (M) & 8 (H) & 9 (H) & 8 (H) & 7 (H) \\
        \hline
    
        {\textbf{8}} & {8 (H)} & {8 (H)} & {7 (H)} & {6 (H)} & {4 (M)} & 6 (H) & 4 (M) & 8 (H) & 6 (H) & 8 (H) & 4 (M) & 7 (H) & 8 (H) & 6 (H) & 6 (H) \\
        \hline

        {\textbf{$\bar{x}$}} & {7.38} & {8.25} & {7.00} & {6.12} & {4.88} & {6.50} & {4.25} & {7.25} & {6.50} & {8.00} & {3.75} & {7.50} & {8.25} & {6.50} & {6.50}  \\
        \cline{2-16}


        {\textbf{$COV$}} & \textbf{9.35\%} & \textbf{5.67\%} & \textbf{8.57\%} & \textbf{7.84\%} & \textbf{13.12\%} & \textbf{7.69\%} & \textbf{11.06\%} & \textbf{15.03\%} & \textbf{7.69\%} & \textbf{6.25\%} & \textbf{12.53\%} & \textbf{6.67\%} & \textbf{5.67\%} & \textbf{12.77\%} & \textbf{7.69\%} \\
        \cline{2-16}
    
    \end{tabular}

    }
\end{minipage}

\caption{Variations of OWASP RR assessments}
\label{tab:variance-owasp}
\end{table}

\subsection{Evaluation of SSVC assessments}

For SSVC, Entropy was calculated to evaluate the variability of its qualitative factors, the results are presented in Table \ref{tab:variance-ssvc}.

The \textbf{Exploitability} factor, which reflects the existence of an implementation for the attack, demonstrates minimal variability. Among the 56 attacks analyzed, 53 had an associated \textbf{Proof-of-Concept}, making this factor largely uniform across the dataset. This lack of differentiation suggests that this factor provides \textbf{little valuable information} when assessing adversarial attacks against LLMs.

A similar observation applies to the \textbf{Automatable} and \textbf{Value-Density} factors. Most adversarial attacks on LLMs are automatable, and they yield significant rewards, such as exposing private or sensitive information from the models. Consequently, these factors also \textbf{fail to offer meaningful distinctions} in the scoring process.

The separation of the \textbf{Technical} and \textbf{Public-Safety Impacts} provides a better understanding of the danger posed by AAs. Although these factors show some degree of variation across attacks within the same category, the differences remain limited. For instance, most attacks are assessed as having `Total' control over the system and `Significant' impacts on finance, reputation, or public health. These assessments, while varying slightly, are overly broad and rely on only two or three possible values, \textbf{limiting their utility} for nuanced analysis or differentiation.

\begin{table}[h]
\centering
\vspace{-1cm}
\begin{minipage}{0.38\textwidth}
    \centering
    \subcaption{White-box jailbreak}
    \resizebox{\textwidth}{!}{
    \begin{tabular}{c c c c c c}
    
        \hline
        
        {\textbf{N°}} &  {\textbf{E}} & {\textbf{A}} & {\textbf{V}} & {\textbf{T}} & {\textbf{P}} \\
        \hline
    
        {\textbf{1}} & {P} & {Y} & {C} & {T} & {S} \\
        \hline
    
        {\textbf{2}} & {P} & {N} & {C} & {T} & {S}\\
        \hline
    
        {\textbf{3}} & {P} & {N} & {C} & {T} & {S}  \\
        \hline
    
        {\textbf{4}} & {P} & {N} & {C} & {P} & {M}\\
        \hline
    
        {\textbf{5}} & {P} & {Y} & {C} & {T} & {S} \\
        \hline
    
        {\textbf{6}} & {P} & {Y} & {C} & {T} & {S} \\
        \hline
    
        {\textbf{7}} & {P} & {N} & {C} & {P} & {M} \\
        \hline
    
        {\textbf{8}} & {P} & {Y} & {C} & {T} & {S} \\
        \hline

        {\textbf{$M$}} & {P} & {N,Y} & {C} & {T} & {S} \\
        \cline{2-6}

        {\textbf{$p_{i}$}} & {1} & {4/8} & {1} & {6/8} & {6/8} \\
        \cline{2-6}

        {\textbf{$H$}} & \textbf{0.00} & \textbf{1.00} & \textbf{0.00} & \textbf{0.81} & \textbf{0.81} \\
        \cline{2-6}
        
    \end{tabular}
    }
\end{minipage}%
\hspace{0.02\textwidth}
\begin{minipage}{0.38\textwidth}
    \centering
    \subcaption{Black-box jailbreak}
    \resizebox{\textwidth}{!}{
    \begin{tabular}{c c c c c c}
    \hline
        {\textbf{N°}} &  {\textbf{E}} & {\textbf{A}} & {\textbf{V}} & {\textbf{T}} & {\textbf{P}} \\
        \hline
    
        {\textbf{1}} & {P} & {Y} & {C} & {T} & {S} \\
        \hline
    
        {\textbf{2}} & {P} & {Y} & {D} & {P} & {M}\\
        \hline
    
        {\textbf{3}} & {P} & {Y} & {C} & {T} & {S}  \\
        \hline
    
        {\textbf{4}} & {P} & {Y} & {C} & {T} & {S}\\
        \hline
    
        {\textbf{5}} & {P} & {Y} & {C} & {T} & {S} \\
        \hline
    
        {\textbf{6}} & {P} & {Y} & {D} & {P} & {M} \\
        \hline
    
        {\textbf{7}} & {A} & {Y} & {C} & {T} & {S} \\
        \hline
    
        {\textbf{8}} & {P} & {Y} & {C} & {T} & {S} \\
        \hline

        {\textbf{$M$}} & {P} & {Y} & {C} & {T} & {S} \\
        \cline{2-6}

        {\textbf{$p_{i}$}} & {7/8} & {1} & {6/8} & {6/8} & {6/8} \\
        \cline{2-6}

        {\textbf{$H$}} & \textbf{0.54} & \textbf{0.00} & \textbf{0.81} & \textbf{0.81} & \textbf{0.81} \\
        \cline{2-6}
    \end{tabular}
    }
\end{minipage}

\vspace{0.002cm} 

\begin{minipage}{0.38\textwidth}
    \centering
    \subcaption{Prompt-injection attacks}
    \resizebox{\textwidth}{!}{
    \begin{tabular}{c c c c c c}
        \hline
        {\textbf{N°}} &  {\textbf{E}} & {\textbf{A}} & {\textbf{V}} & {\textbf{T}} & {\textbf{P}} \\
        \hline
    
        {\textbf{1}} & {P} & {Y} & {C} & {T} & {S} \\
        \hline
    
        {\textbf{2}} & {P} & {Y} & {C} & {T} & {S}\\
        \hline
    
        {\textbf{3}} & {P} & {Y} & {C} & {T} & {S}  \\
        \hline
    
        {\textbf{4}} & {P} & {Y} & {C} & {T} & {S}\\
        \hline
    
        {\textbf{5}} & {P} & {Y} & {C} & {T} & {S} \\
        \hline
    
        {\textbf{6}} & {P} & {Y} & {C} & {T} & {S} \\
        \hline
    
        {\textbf{7}} & {N} & {N} & {D} & {P} & {M} \\
        \hline
    
        {\textbf{8}} & {P} & {Y} & {C} & {T} & {S} \\
        \hline

        {\textbf{$M$}} & {P} & {Y} & {C} & {T} & {S} \\
        \cline{2-6}

        {\textbf{$p_{i}$}} & {7/8} & {7/8} & {7/8} & {7/8} & {7/8} \\
        \cline{2-6}

        {\textbf{$H$}} & \textbf{0.54} & \textbf{0.54} & \textbf{0.54} & \textbf{0.54} & \textbf{0.54} \\
        \cline{2-6}
    \end{tabular}

    }
\end{minipage}%
\hspace{0.02\textwidth}
\begin{minipage}{0.38\textwidth}
    \centering
    \subcaption{Evasion attacks}
    \resizebox{\textwidth}{!}{
    \begin{tabular}{c c c c c c}
        \hline
        {\textbf{N°}} &  {\textbf{E}} & {\textbf{A}} & {\textbf{V}} & {\textbf{T}} & {\textbf{P}} \\
        \hline
    
        {\textbf{1}} & {P} & {Y} & {C} & {P} & {S} \\
        \hline
    
        {\textbf{2}} & {P} & {Y} & {C} & {P} & {S}\\
        \hline
    
        {\textbf{3}} & {A} & {Y} & {C} & {P} & {S}  \\
        \hline
    
        {\textbf{4}} & {P} & {Y} & {C} & {P} & {S}\\
        \hline
    
        {\textbf{5}} & {P} & {N} & {D} & {P} & {M} \\
        \hline
    
        {\textbf{6}} & {P} & {Y} & {C} & {P} & {S} \\
        \hline
    
        {\textbf{7}} & {P} & {Y} & {C} & {P} & {S} \\
        \hline
    
        {\textbf{8}} & {P} & {Y} & {C} & {P} & {S} \\
        \hline

        {\textbf{$M$}} & {P} & {Y} & {C} & {P} & {S} \\
        \cline{2-6}

        {\textbf{$p_{i}$}} & {7/8} & {7/8} & {7/8} & {1} & {7/8} \\
        \cline{2-6}

        {\textbf{$H$}} & \textbf{0.54} & \textbf{0.54} & \textbf{0.54} & \textbf{0.00} & \textbf{0.54} \\
        \cline{2-6}
    
    \end{tabular}
    }
\end{minipage}

\vspace{0.002cm} 

\begin{minipage}{0.38\textwidth}
    \centering
    \subcaption{Model-extraction attacks}
    \resizebox{\textwidth}{!}{
    \begin{tabular}{c c c c c c}
        \hline
        {\textbf{N°}} &  {\textbf{E}} & {\textbf{A}} & {\textbf{V}} & {\textbf{T}} & {\textbf{P}} \\
        \hline
    
        {\textbf{1}} & {P} & {Y} & {C} & {T} & {S} \\
        \hline
    
        {\textbf{2}} & {P} & {Y} & {C} & {T} & {S}\\
        \hline
    
        {\textbf{3}} & {P} & {Y} & {C} & {T} & {S}  \\
        \hline
    
        {\textbf{4}} & {P} & {Y} & {C} & {T} & {S}\\
        \hline
    
        {\textbf{5}} & {P} & {N} & {C} & {T} & {S} \\
        \hline
    
        {\textbf{6}} & {P} & {Y} & {C} & {T} & {S} \\
        \hline
    
        {\textbf{7}} & {P} & {Y} & {C} & {T} & {S} \\
        \hline
    
        {\textbf{8}} & {P} & {Y} & {C} & {T} & {S} \\
        \hline

        {\textbf{$M$}} & {P} & {Y} & {C} & {T} & {S} \\
        \cline{2-6}

        {\textbf{$p_{i}$}} & {1} & {7/8} & {1} & {1} & {1} \\
        \cline{2-6}

        {\textbf{$H$}} & \textbf{0.00} & \textbf{0.54} & \textbf{0.00} & \textbf{0.00} & \textbf{0.00} \\
        \cline{2-6}
    \end{tabular}
    
    }
\end{minipage}%
\hspace{0.02\textwidth}
\begin{minipage}{0.38\textwidth}
    \centering
    \subcaption{Model-inference attacks}
    \resizebox{\textwidth}{!}{
    \begin{tabular}{c c c c c c}
        \hline
        {\textbf{N°}} &  {\textbf{E}} & {\textbf{A}} & {\textbf{V}} & {\textbf{T}} & {\textbf{P}} \\
        \hline
    
        {\textbf{1}} & {P} & {Y} & {C} & {T} & {S} \\
        \hline
    
        {\textbf{2}} & {P} & {Y} & {C} & {T} & {S}\\
        \hline
    
        {\textbf{3}} & {P} & {N} & {D} & {P} & {M}  \\
        \hline
    
        {\textbf{4}} & {P} & {Y} & {C} & {T} & {S}\\
        \hline
    
        {\textbf{5}} & {P} & {Y} & {C} & {T} & {S} \\
        \hline
    
        {\textbf{6}} & {P} & {Y} & {D} & {P} & {M} \\
        \hline
    
        {\textbf{7}} & {P} & {Y} & {C} & {T} & {S} \\
        \hline
    
        {\textbf{8}} & {P} & {Y} & {C} & {T} & {S} \\
        \hline

        {\textbf{$M$}} & {P} & {Y} & {C} & {T} & {S} \\
        \cline{2-6}

        {\textbf{$p_{i}$}} & {1} & {7/8} & {6/8} & {6/8} & {6/8} \\
        \cline{2-6}

        {\textbf{$H$}} & \textbf{0.00} & \textbf{0.54} & \textbf{0.81} & \textbf{0.81} & \textbf{0.81} \\
        \cline{2-6}
    
    \end{tabular}
    
    }
\end{minipage}

\vspace{0.002cm} 

\begin{minipage}{0.38\textwidth}
    \centering
    \subcaption{Poisoning/Trojan/Backdoor attacks}
    \resizebox{\textwidth}{!}{
    \begin{tabular}{c c c c c c}
        \hline
        {\textbf{N°}} &  {\textbf{E}} & {\textbf{A}} & {\textbf{V}} & {\textbf{T}} & {\textbf{P}} \\
        \hline
    
        {\textbf{1}} & {P} & {Y} & {C} & {T} & {S} \\
        \hline
    
        {\textbf{2}} & {P} & {Y} & {C} & {T} & {S}\\
        \hline
    
        {\textbf{3}} & {P} & {Y} & {C} & {T} & {S}  \\
        \hline
    
        {\textbf{4}} & {P} & {N} & {D} & {P} & {M}\\
        \hline
    
        {\textbf{5}} & {P} & {Y} & {C} & {T} & {S} \\
        \hline
    
        {\textbf{6}} & {P} & {Y} & {C} & {T} & {S} \\
        \hline
    
        {\textbf{7}} & {P} & {Y} & {C} & {T} & {S} \\
        \hline
    
        {\textbf{8}} & {P} & {Y} & {C} & {T} & {S} \\
        \hline

        {\textbf{$M$}} & {P} & {Y} & {C} & {T} & {S} \\
        \cline{2-6}

        {\textbf{$p_{i}$}} & {1} & {7/8} & {7/8} & {7/8} & {7/8} \\
        \cline{2-6}

        {\textbf{$H$}} & \textbf{0.00} & \textbf{0.54} & \textbf{0.54} & \textbf{0.54} & \textbf{0.54} \\
        \cline{2-6}
    
    \end{tabular}

    }
\end{minipage}
\caption{Variations of SSVC assessments}
\label{tab:variance-ssvc}
\end{table}

\section{Suggestions for Future Solutions}
\label{sec:suggestions}

The analysis conducted in the previous sections validates the hypothesis proposed in Section \ref{sec:introduction}: \textbf{existing vulnerability scoring metrics are inadequate for assessing Adversarial Attacks against Large Language Models}. This inadequacy is primarily due to the lack of variability in factor scores, which limits the metrics’ ability to distinguish between different types of attacks effectively.

The shortcomings of current vulnerability scoring systems stem from several key issues:
\begin{enumerate}
    \item \textbf{Overemphasis on CIA Impact:} Existing metrics focus heavily on the technical impact on Confidentiality, Integrity, and Availability, which are not the primary targets of AAs against LLMs.
    \item \textbf{Lack of Contextual Consideration:} Factors such as Attack Vector, Opportunity, and Intrusion Detection lack relevance when applied to LLM-specific scenarios due to the absence of target-specific context.
    \item \textbf{Subjectivity in Quantitative Scores:} The use of quantitative scoring systems introduces subjectivity, reducing the reliability of assessments.
    \item \textbf{Limited Qualitative Scoring Options:} Scoring systems with qualitative factors often offer too few choices, resulting in repetitive and non-discriminative assessments.
\end{enumerate}

These limitations highlight the urgent need for the research community to address these gaps and develop scoring metrics specifically tailored for AAs against LLMs, particularly given the increasing adoption of these models in critical applications.

While proposing new metrics is beyond the scope of this study, we suggest the following directions for future research:

\begin{enumerate}
    \item \textbf{Customized Technical Impact Metrics:} Metrics should account for the unique impacts of AAs on LLMs, such as trust erosion, misinformation dissemination, or generating biased and harmful outputs. These factors better reflect the consequences of LLM-specific attacks.
    \item \textbf{Context-Aware Factors:} Metrics should consider the architecture and nature of the targeted LLM. For example: 
    \begin{itemize}
        \item Larger models (e.g., GPT, LLAMA) are more susceptible to AAs due to complex decision boundaries.
        \item Attacks targeting LLMs trained on sensitive personal data pose greater risks than those on public datasets.
        \item Multimodal LLMs may face distinct vulnerabilities (e.g., malicious image injection), which text-only models do not encounter. 
    \end{itemize}
    \item \textbf{Incorporating Success Rates:} Success rates could serve as a valuable factor in ranking attacks, although challenging to measure. For instance:
    \begin{itemize}
        \item Prompt Injection attacks can exhibit varying success rates depending on implementation.
        \item Jailbreak attacks may not succeed consistently with a single query but can have cumulative success over multiple attempts, which is important to account for.
    \end{itemize}
    \item \textbf{Enhanced Qualitative Scoring Systems:} Implementing multiple-choice qualitative factors can strike a balance between complexity and subjectivity. For instance, adding more nuanced levels to factors like `Attack Complexity' (e.g., Minimal, Medium, Very High) could create finer distinctions between attacks and increase score variability.
\end{enumerate}

By exploring these directions, researchers can contribute to the development of robust, context-sensitive metrics that provide meaningful and actionable assessments for adversarial attacks against LLMs. This advancement is crucial for enhancing the security posture of these increasingly prevalent models.

\section{Conclusion}
\label{sec:conclusion}

This study has critically examined the applicability of established vulnerability metrics, such as DREAD, CVSS, OWASP Risk Rating, and SSVC, to assess Adversarial Attacks on LLMs. Through a detailed analysis of 56 AAs across multiple metrics, the findings demonstrate that existing metrics \textbf{fail} to adequately differentiate between attacks, primarily due to their rigid, context-limited factors and a focus on traditional technical impacts rather than the nuanced threats posed by AAs.

Key observations highlight that factors such as technical-impact, the motivation of attackers, and the limited-options of qualitative scoring systems are inadequately addressed in existing frameworks. These limitations restrict the variability and relevance of vulnerability scores, confirming the hypothesis that traditional metrics are not fully suitable for assessing the risks associated with AAs on LLMs.

While the development of new metrics was beyond the scope of this work, the study identifies several promising directions for improvement. These include integrating tailored technical-impact assessments, context-specific factors, and multiple-choice qualitative scoring options to enhance the granularity and applicability of future metrics. Furthermore, incorporating attack success rates, though complex, could provide a more comprehensive evaluation of adversarial threats.

The contributions of this research are multifaceted, providing a taxonomy of adversarial attack classifications, a curated list of 56 AAs targeting LLMs, and an in-depth statistical evaluation of existing vulnerability metrics. These findings not only underscore the limitations of current approaches but also serve as a call to action for the development of more robust, flexible, and LLM-specific vulnerability assessment frameworks.

Future research should focus on refining these metrics to account for the unique challenges posed by Adversarial Attacks on LLMs, ensuring that the security of these increasingly vital systems is both effective and adaptive to emerging threats.

\medskip

\textit{Acknowledgments} \\This work is financially supported by Zayed University under a Research Associate contract \\

\textit{Disclosure statement:} \\No potential conflict of interest is reported by the authors

\newpage

\appendix
\section{Assessment details}
\label{appendix:assessment}
This appendix provides a comprehensive breakdown of the evaluation results for the seven types of AAs assessed using the four vulnerability metrics: DREAD, CVSS, OWASP Risk Rating, and SSVC. Each attack type was evaluated across multiple LLMs, with detailed scores recorded for every attack and metric. The appendix outlines the methodology used to compute average scores by consolidating the evaluations from three distinct LLMs, ensuring an accurate representation of the results.

This detailed presentation of results supports the main text by providing transparency into the scoring process and offering a robust reference for further analysis of the metrics and calculations used in this study.

\subsection{White-box Jailbreak}
\label{subappendix:whitebox-jailbreak}

\begin{table}[h]
\centering
\caption{Detailed assessment of White-box Jailbreak attacks with DREAD}
\vspace{2mm}
\label{tab:appendix-whitebox-dread}
\adjustbox{max width=\textwidth}{
\begin{tabular}{c c c c c c c c}

    \hline
    
    \multirow{2}{*}{\textbf{N°}} & \multirow{2}{*}{\textbf{LLM}} &  \multirow{2}{*}{\textbf{D}} & \multirow{2}{*}{\textbf{R}} & \multirow{2}{*}{\textbf{E}} & \multirow{2}{*}{\textbf{A}} & \multirow{2}{*}{\textbf{D}} & \multirow{2}{*}{\textbf{Score}} \cr \\
    \hline

    \multirow{4}{*}{\textbf{1}} & {GPT-4o} & {8 (H)} & {8 (H)} & {7 (H)} & {7 (H)} & {5 (M)} & \textcolor{red}{7 (H)} \\
    \cline{2-8}
    & {LLAMA3} & {8 (H)} & {9 (H)} & {8 (H)} & {8 (H)} & {6 (M)} & \textcolor{red}{7.8 (H)} \\
    \cline{2-8}
    & {Perplexity} & {9 (H)} & {9 (H)} & {8 (H)} & {8 (H)} & {6 (M)} & \textcolor{red}{8 (H)} \\
    \cline{2-8}
    & \textbf{Avg} & \textbf{8 (H)} & \textbf{9 (H)} & \textbf{8 (H)} & \textbf{8 (H)} & \textbf{6 (M)} & \textcolor{red}{\textbf{7.8 (H)}} \\
    \hline

    \multirow{4}{*}{\textbf{2}} & {GPT-4o} & {6 (M)} & {5 (M)} & {6 (H)} & {6 (M)} & {5 (M)} & \textcolor{yellow}{5.4 (M)} \\
    \cline{2-8}
    & {LLAMA3} & {5 (M)} & {6 (M)} & {5 (M)} & {5 (M)} & {4 (M)} & \textcolor{yellow}{5 (M)} \\
\hline
\end{tabular}
}
\end{table}

\begin{table}[h]
\centering
\caption{Detailed assessment of White-box Jailbreak with CVSS}
\vspace{2mm}
\label{tab:appendix-whitebox-cvss}
\adjustbox{max width=\textwidth}{
\begin{tabular}{c c c c c c c c c c c}

    \hline
    
    \multirow{2}{*}{\textbf{N°}} & \multirow{2}{*}{\textbf{LLM}} &  \multirow{2}{*}{\textbf{AV}} & \multirow{2}{*}{\textbf{AC}} & \multirow{2}{*}{\textbf{PR}} & \multirow{2}{*}{\textbf{UI}} & \multirow{2}{*}{\textbf{S}} & \multirow{2}{*}{\textbf{C}} & \multirow{2}{*}{\textbf{I}} & \multirow{2}{*}{\textbf{A}} &\multirow{2}{*}{\textbf{Base}} \cr \\
    \hline

    \multirow{4}{*}{\textbf{1}} & GPT-4o & N & H & N & N & C & H & H & N & \textcolor{red}{8.7 (H)} \\
    \cline{2-11}
    & {LLAMA3} & N & L & N & R & C & N & H & N & \textcolor{red}{7.4 (H)} \\
    \cline{2-11}
    & {Perplexity} & N & H & N & N & U & N & H & N & \textcolor{yellow}{5.9 (M)} \\
    \cline{2-11}
    & \textbf{Avg} & \textbf{N} & \textbf{H} & \textbf{N} & \textbf{N} & \textbf{C} & \textbf{L} & \textbf{H} & \textbf{N} & \textcolor{red}{\textbf{7.5 (H)}} \\
    \hline

    \multirow{4}{*}{\textbf{2}} & GPT-4o & N & H & N & N & C & H & H & N & \textcolor{red}{8.7 (H)} \\
    \cline{2-11}
    & {LLAMA3} & N & H & L & R & C & N & H & N & \textcolor{yellow}{6.1 (M)} \\
\hline
 
\end{tabular}
}
\end{table}

\begin{table}[h]
\centering
\caption{Detailed assessment of White-box Jailbreak with OWASP Risk Rating}
\vspace{2mm}
\label{tab:appendix-whitebox-owasp}
\adjustbox{max width=\textwidth}{
\begin{tabular}{c c c c c c c c c c c c c c c c c c}

    \hline
    
    \multirow{2}{*}{\textbf{N°}} & \multirow{2}{*}{\textbf{LLM}} &  \multirow{2}{*}{\textbf{SK}} & \multirow{2}{*}{\textbf{MT}} & \multirow{2}{*}{\textbf{OP}} & \multirow{2}{*}{\textbf{SZ}} & \multirow{2}{*}{\textbf{ED}} & \multirow{2}{*}{\textbf{EE}} & \multirow{2}{*}{\textbf{AW}} & \multirow{2}{*}{\textbf{ID}} & \multirow{2}{*}{\textbf{C}} & \multirow{2}{*}{\textbf{I}} & \multirow{2}{*}{\textbf{A}} & \multirow{2}{*}{\textbf{FD}} & \multirow{2}{*}{\textbf{RD}} & \multirow{2}{*}{\textbf{NC}} & \multirow{2}{*}{\textbf{PV}} & \multirow{2}{*}{\textbf{Score}} \cr \\
    \hline

    \multirow{4}{*}{\textbf{1}} & GPT-4o & 7 & 7 & 6 & 7 & 6 & 7 & 5 & 6 & 8 & 6 & 6 & 7 & 8 & 6 & 7 & \textcolor{purple}{4.3 (C)} \\
    \cline{2-18}
    & {LLAMA3} & 8 & 6 & 8 & 5 & 8 & 9 & 6 & 4 & 0 & 8 & 0 & 6 & 8 & 0 & 0 & \textcolor{red}{2.1 (H)} \\
    \cline{2-18}
    & {Perplexity} & 7 & 6 & 5 & 6 & 6 & 7 & 5 & 4 & 8 & 7 & 5 & 7 & 8 & 5 & 6 & \textcolor{red}{3.8 (H)} \\
    \cline{2-18}
    & \textbf{Avg} & \textbf{7} & \textbf{6} & \textbf{6} & \textbf{6} & \textbf{7} & \textbf{8} & \textbf{5} & \textbf{5} & \textbf{5} & \textbf{7} & \textbf{4} & \textbf{7} & \textbf{8} & \textbf{4} & \textbf{4} & \textcolor{red}{\textbf{3.6 (H)}} \\
    \hline

    \multirow{4}{*}{\textbf{2}} & GPT-4o & 7 & 7 & 6 & 7 & 5 & 7 & 5 & 6 & 8 & 6 & 5 & 7 & 8 & 7 & 7 & \textcolor{purple}{4.3 (C)} \\
    \cline{2-18}
    & {LLAMA3} & 7 & 5 & 5 & 4 & 6 & 7 & 5 & 5 & 0 & 7 & 0 & 5 & 6 & 0 & 0 & \textcolor{green}{1.4 (L)} \\
\hline
    
\end{tabular}
}
\end{table}

\begin{table}[h]
\centering
\caption{Detailed assessment of White-box Jailbreak attacks with SSVC}
\vspace{2mm}
\label{tab:appendix-whitebox-ssvc}
\begin{tabular}{c c c c c c c c c}

    \hline
    
    \multirow{2}{*}{\textbf{N°}} & \multirow{2}{*}{\textbf{LLM}} &  \multirow{2}{*}{\textbf{E}} & \multirow{2}{*}{\textbf{A}} & \multirow{2}{*}{\textbf{V}} & \multirow{2}{*}{\textbf{U}} & \multirow{2}{*}{\textbf{T}} & \multirow{2}{*}{\textbf{P}} & \multirow{2}{*}{\textbf{Score}} \cr \\
    \hline

    \multirow{4}{*}{\textbf{1}} & {GPT-4o} & {P} & {Y} & {C} & {S} & {T} & {S} & \textcolor{red}{Immediate} \\
    \cline{2-9}
    & {LLAMA3} & {P} & {Y} & {C} & {S} & {T} & S & \textcolor{red}{Immediate} \\
    \cline{2-9}
    & {Perplexity} & {P} & {Y} & {C} & {S} & {T} & {S} & \textcolor{red}{Immediate} \\
    \cline{2-9}
    & \textbf{Avg} & \textbf{P} & \textbf{Y} & \textbf{C} & \textbf{S} & \textbf{T} & \textbf{S} & \textcolor{red}{\textbf{Immediate}} \\
    \hline

    \multirow{4}{*}{\textbf{2}} & {GPT-4o} & {N} & {N} & {D} & {L} & {P} & {M} & \textcolor{green}{Defer} \\
    \cline{2-9}
    & {LLAMA3} & {P} & {N} & {C} & {E} & {T} & S & \textcolor{red}{Immediate} \\
\hline
    
\end{tabular}
\end{table}

\subsection{Black-box Jailbreak}
\label{subappendix:blackbox-jailbreak}

\begin{table}[h]
\centering
\caption{Detailed assessment of Black-box Jailbreak attacks with DREAD}
\vspace{2mm}
\label{tab:appendix-blackbox-dread}
\begin{tabular}{c c c c c c c c}

    \hline
    
    \multirow{2}{*}{\textbf{N°}} & \multirow{2}{*}{\textbf{LLM}} &  \multirow{2}{*}{\textbf{D}} & \multirow{2}{*}{\textbf{R}} & \multirow{2}{*}{\textbf{E}} & \multirow{2}{*}{\textbf{A}} & \multirow{2}{*}{\textbf{D}} & \multirow{2}{*}{\textbf{Score}} \cr \\
    \hline

    \multirow{4}{*}{\textbf{1}} & {GPT-4o} & {8 (H)} & {7 (H)} & {6 (M)} & {8 (H)} & {5 (M)} & \textcolor{yellow}{6.8 (M)} \\
    \cline{2-8}
    & {LLAMA3} & {9 (H)} & {8 (H)} & {8 (H)} & {9 (H)} & {5 (M)} & \textcolor{red}{7.6 (H)} \\
    \cline{2-8}
    & {Perplexity} & {8 (H)} & {7 (H)} & {8 (H)} & {7 (H)} & {5 (M)} & \textcolor{red}{7 (H)} \\
    \cline{2-8}
    & \textbf{Avg} & \textbf{8 (H)} & \textbf{7 (H)} & \textbf{7 (H)} & \textbf{8 (H)} & \textbf{5 (M)} & \textcolor{red}{\textbf{7 (H)}} \\
    \hline

    \multirow{4}{*}{\textbf{2}} & {GPT-4o} & {8 (H)} & {7 (H)} & {8 (H)} & {7 (H)} & {5 (M)} & \textcolor{red}{7 (H)} \\
    \cline{2-8}
    & {LLAMA3} & {8 (H)} & {8 (H)} & {7 (H)} & {8 (H)} & {5 (M)} & \textcolor{red}{7.2 (H)} \\
\hline
    
\end{tabular}
\end{table}

\begin{table}[h]
\centering
\caption{Detailed assessment of Black-box Jailbreak with CVSS}
\vspace{2mm}
\label{tab:appendix-blackbox-cvss}
\begin{tabular}{c c c c c c c c c c c}

    \hline
    
    \multirow{2}{*}{\textbf{N°}} & \multirow{2}{*}{\textbf{LLM}} &  \multirow{2}{*}{\textbf{AV}} & \multirow{2}{*}{\textbf{AC}} & \multirow{2}{*}{\textbf{PR}} & \multirow{2}{*}{\textbf{UI}} & \multirow{2}{*}{\textbf{S}} & \multirow{2}{*}{\textbf{C}} & \multirow{2}{*}{\textbf{I}} & \multirow{2}{*}{\textbf{A}} &\multirow{2}{*}{\textbf{Base}} \cr \\
    \hline

    \multirow{4}{*}{\textbf{1}} & GPT-4o & N & L & N & N & U & H & L & N & \textcolor{red}{8.2 (H)} \\
    \cline{2-11}
    & {LLAMA3} & N & L & N & N & C & H & L & N & \textcolor{red}{8.2 (H)} \\
    \cline{2-11}
    & {Perplexity} & N & H & N & N & U & N & H & N & \textcolor{yellow}{5.9 (M)} \\
    \cline{2-11}
    & \textbf{Avg} & \textbf{N} & \textbf{L} & \textbf{N} & \textbf{N} & \textbf{U} & \textbf{L} & \textbf{L} & \textbf{N} & \textcolor{yellow}{\textbf{6.5 (M)}} \\
    \hline

    \multirow{4}{*}{\textbf{2}} & GPT-4o & N & L & N & N & U & N & H & N & \textcolor{red}{7.5 (H)} \\
    \cline{2-11}
    & {LLAMA3} & N & L & N & N & C & H & L & N & \textcolor{red}{8.2 (H)} \\
\hline
 
\end{tabular}
\end{table}

\begin{table}[h]
\centering
\caption{Detailed assessment of Black-box Jailbreak with OWASP Risk Rating}
\vspace{2mm}
\label{tab:appendix-blackbox-owasp}
\adjustbox{max width=\textwidth}{
\begin{tabular}{c c c c c c c c c c c c c c c c c c}

    \hline
    
    \multirow{2}{*}{\textbf{N°}} & \multirow{2}{*}{\textbf{LLM}} &  \multirow{2}{*}{\textbf{SK}} & \multirow{2}{*}{\textbf{MT}} & \multirow{2}{*}{\textbf{OP}} & \multirow{2}{*}{\textbf{SZ}} & \multirow{2}{*}{\textbf{ED}} & \multirow{2}{*}{\textbf{EE}} & \multirow{2}{*}{\textbf{AW}} & \multirow{2}{*}{\textbf{ID}} & \multirow{2}{*}{\textbf{C}} & \multirow{2}{*}{\textbf{I}} & \multirow{2}{*}{\textbf{A}} & \multirow{2}{*}{\textbf{FD}} & \multirow{2}{*}{\textbf{RD}} & \multirow{2}{*}{\textbf{NC}} & \multirow{2}{*}{\textbf{PV}} & \multirow{2}{*}{\textbf{Score}} \cr \\
    \hline

    \multirow{4}{*}{\textbf{1}} & GPT-4o & 5 & 8 & 8 & 7 & 5 & 5 & 5 & 3 & 7 & 4 & 0 & 5 & 8 & 5 & 8 & \textcolor{red}{2.9 (H)} \\
    \cline{2-18}
    & {LLAMA3} & 6 & 8 & 9 & 5 & 8 & 9 & 6 & 8 & 8 & 0 & 0 & 4 & 8 & 5 & 8 & \textcolor{red}{3.3 (H)} \\
    \cline{2-18}
    & {Perplexity} & 6 & 8 & 7 & 5 & 4 & 7 & 5 & 8 & 9 & 2 & 2 & 8 & 7 & 5 & 6 & \textcolor{red}{3.6 (H)} \\
    \cline{2-18}
    & \textbf{Avg} & \textbf{6} & \textbf{8} & \textbf{8} & \textbf{6} & \textbf{6} & \textbf{7} & \textbf{5} & \textbf{6} & \textbf{8} & \textbf{2} & \textbf{1} & \textbf{6} & \textbf{8} & \textbf{5} & \textbf{7} & \textcolor{red}{\textbf{3.3 (H)}} \\
    \hline

    \multirow{4}{*}{\textbf{2}} & GPT-4o & 3 & 8 & 8 & 7 & 5 & 3 & 5 & 3 & 7 & 0 & 0 & 4 & 8 & 2 & 0 & \textcolor{yellow}{1.5 (M)} \\
    \cline{2-18}
    & {LLAMA3} & 7 & 9 & 9 & 5 & 9 & 9 & 7 & 9 & 9 & 0 & 0 & 5 & 9 & 6 & 9 & \textcolor{red}{3.6 (H)} \\
\hline
    
\end{tabular}
}
\end{table}

\begin{table}[h]
\centering
\caption{Detailed assessment of Black-box Jailbreak attacks with SSVC}
\vspace{2mm}
\label{tab:appendix-blackbox-ssvc}
\begin{tabular}{c c c c c c c c c}

    \hline
    
    \multirow{2}{*}{\textbf{N°}} & \multirow{2}{*}{\textbf{LLM}} &  \multirow{2}{*}{\textbf{E}} & \multirow{2}{*}{\textbf{A}} & \multirow{2}{*}{\textbf{V}} & \multirow{2}{*}{\textbf{U}} & \multirow{2}{*}{\textbf{T}} & \multirow{2}{*}{\textbf{P}} & \multirow{2}{*}{\textbf{Score}} \cr \\
    \hline

    \multirow{4}{*}{\textbf{1}} & {GPT-4o} & {P} & {Y} & {C} & {S} & {T} & {S} & \textcolor{red}{Immediate} \\
    \cline{2-9}
    & {LLAMA3} & {P} & {N} & {C} & {L} & {T} & S & \textcolor{red}{Immediate} \\
    \cline{2-9}
    & {Perplexity} & {P} & {Y} & {C} & {S} & {T} & {S} & \textcolor{red}{Immediate} \\
    \cline{2-9}
    & \textbf{Avg} & \textbf{P} & \textbf{Y} & \textbf{C} & \textbf{S} & \textbf{T} & \textbf{S} & \textcolor{red}{\textbf{Immediate}} \\
    \hline

    \multirow{4}{*}{\textbf{2}} & {GPT-4o} & {P} & {Y} & {D} & {E} & {P} & {M} & \textcolor{yellow}{Scheduled} \\
    \cline{2-9}
    & {LLAMA3} & {P} & {N} & {C} & {L} & {T} & S & \textcolor{red}{Immediate} \\
\hline
    
\end{tabular}
\end{table}

\subsection{Prompt Injection}
\label{subappendix:prompt-inject}

\begin{table}[h]
\centering
\caption{Detailed assessment of Prompt Injection attacks with DREAD}
\vspace{2mm}
\label{tab:appendix-promptinject-dread}
\begin{tabular}{c c c c c c c c}

    \hline
    
    \multirow{2}{*}{\textbf{N°}} & \multirow{2}{*}{\textbf{LLM}} &  \multirow{2}{*}{\textbf{D}} & \multirow{2}{*}{\textbf{R}} & \multirow{2}{*}{\textbf{E}} & \multirow{2}{*}{\textbf{A}} & \multirow{2}{*}{\textbf{D}} & \multirow{2}{*}{\textbf{Score}} \cr \\
    \hline

    \multirow{4}{*}{\textbf{1}} & {GPT-4o} & 8 (H) & 9 (H) & 7 (H) & 7 (H) & 6 (M) & \textcolor{red}{7.8 (H)} \\
    \cline{2-8}
    & {LLAMA3} & 8 (H) & 9 (H) & 8 (H) & 7 (H) & 6 (M) & \textcolor{red}{7.6 (H)} \\
    \cline{2-8}
    & {Perplexity} & 8 (H) & 9 (H) & 8 (H) & 7 (H) & 6 (M) & \textcolor{red}{7.6 (H)} \\
    \cline{2-8}
    & \textbf{Avg} & \textbf{8 (H)} & \textbf{9 (H)} & \textbf{8 (H)} & \textbf{7 (H)} & \textbf{6 (M)} & \textcolor{red}{\textbf{7.6 (H)}} \\
    \hline

    \multirow{4}{*}{\textbf{2}} & {GPT-4o} & 8 (H) & 9 (H) & 8 (H) & 7 (H) & 7 (H) & \textcolor{red}{7.8 (H)} \\
    \cline{2-8}
    & {LLAMA3} & 9 (H) & 8 (H) & 9 (H) & 8 (H) & 5 (M) & \textcolor{red}{7.8 (H)} \\
\hline
    
\end{tabular}
\end{table}

\begin{table}[h]
\centering
\caption{Detailed assessment of Prompt Injection with CVSS}
\vspace{2mm}
\label{tab:appendix-promptinject-cvss}
\begin{tabular}{c c c c c c c c c c c}

    \hline
    
    \multirow{2}{*}{\textbf{N°}} & \multirow{2}{*}{\textbf{LLM}} &  \multirow{2}{*}{\textbf{AV}} & \multirow{2}{*}{\textbf{AC}} & \multirow{2}{*}{\textbf{PR}} & \multirow{2}{*}{\textbf{UI}} & \multirow{2}{*}{\textbf{S}} & \multirow{2}{*}{\textbf{C}} & \multirow{2}{*}{\textbf{I}} & \multirow{2}{*}{\textbf{A}} &\multirow{2}{*}{\textbf{Base}} \cr \\
    \hline

    \multirow{4}{*}{\textbf{1}} & GPT-4o & N & L & N & N & U & N & H & N & \textcolor{red}{7.5 (H)} \\
    \cline{2-11}
    & {LLAMA3} & N & L & N & N & C & H & H & N & \textcolor{purple}{10 (C)} \\
    \cline{2-11}
    & {Perplexity} & N & L & N & N & U & N & H & N & \textcolor{red}{7.5 (H)} \\
    \cline{2-11}
    & \textbf{Avg} & \textbf{N} & \textbf{L} & \textbf{N} & \textbf{N} & \textbf{U} & \textbf{N} & \textbf{H} & \textbf{N} & \textcolor{red}{\textbf{7.5 (H)}} \\
    \hline

    \multirow{4}{*}{\textbf{2}} & GPT-4o & N & H & N & R & U & L & H & N & \textcolor{yellow}{5.9 (M)} \\
    \cline{2-11}
    & {LLAMA3} & L & H & N & R & C & H & H & N & \textcolor{red}{7.4 (H)} \\
\hline
 
\end{tabular}
\end{table}

\begin{table}[h]
\centering
\caption{Detailed assessment of Prompt-Injection with OWASP Risk Rating}
\vspace{2mm}
\label{tab:appendix-prompt-owasp}
\adjustbox{max width=\textwidth}{
\begin{tabular}{c c c c c c c c c c c c c c c c c c}

    \hline
    
    \multirow{2}{*}{\textbf{N°}} & \multirow{2}{*}{\textbf{LLM}} &  \multirow{2}{*}{\textbf{SK}} & \multirow{2}{*}{\textbf{MT}} & \multirow{2}{*}{\textbf{OP}} & \multirow{2}{*}{\textbf{SZ}} & \multirow{2}{*}{\textbf{ED}} & \multirow{2}{*}{\textbf{EE}} & \multirow{2}{*}{\textbf{AW}} & \multirow{2}{*}{\textbf{ID}} & \multirow{2}{*}{\textbf{C}} & \multirow{2}{*}{\textbf{I}} & \multirow{2}{*}{\textbf{A}} & \multirow{2}{*}{\textbf{FD}} & \multirow{2}{*}{\textbf{RD}} & \multirow{2}{*}{\textbf{NC}} & \multirow{2}{*}{\textbf{PV}} & \multirow{2}{*}{\textbf{Score}} \cr \\
    \hline

    \multirow{4}{*}{\textbf{1}} & GPT-4o & 6 & 8 & 6 & 5 & 5 & 8 & 6 & 4 & 0 & 7 & 0 & 4 & 7 & 0 & 0 & \textcolor{yellow}{2.5 (M)} \\
    \cline{2-18}
    & {LLAMA3} & 6 & 8 & 9 & 5 & 8 & 9 & 6 & 8 & 6 & 8 & 5 & 8 & 9 & 6 & 8 & \textcolor{purple}{5.2 (C)} \\
    \cline{2-18}
    & {Perplexity} & 6 & 8 & 6 & 5 & 4 & 6 & 5 & 7 & 3 & 9 & 5 & 7 & 8 & 6 & 4 & \textcolor{red}{3.4 (H)} \\
    \cline{2-18}
    & \textbf{Avg} & \textbf{6} & \textbf{8} & \textbf{7} & \textbf{5} & \textbf{6} & \textbf{8} & \textbf{6} & \textbf{6} & \textbf{3} & \textbf{8} & \textbf{3} & \textbf{6} & \textbf{8} & \textbf{4} & \textbf{4} & \textcolor{red}{\textbf{3.3 (H)}} \\
    \hline

    \multirow{4}{*}{\textbf{2}} & GPT-4o & 5 & 8 & 6 & 7 & 5 & 5 & 3 & 3 & 0 & 8 & 0 & 4 & 8 & 0 & 0 & \textcolor{yellow}{1.5 (M)} \\
    \cline{2-18}
    & {LLAMA3} & 8 & 9 & 9 & 6 & 9 & 9 & 7 & 9 & 8 & 9 & 6 & 9 & 9 & 7 & 9 & \textcolor{purple}{6.6 (C)} \\
\hline
    
\end{tabular}
}
\end{table}

\begin{table}[h]
\centering
\caption{Detailed assessment of Prompt-Injection attacks with SSVC}
\vspace{2mm}
\label{tab:appendix-promptinject-ssvc}
\begin{tabular}{c c c c c c c c c}

    \hline
    
    \multirow{2}{*}{\textbf{N°}} & \multirow{2}{*}{\textbf{LLM}} &  \multirow{2}{*}{\textbf{E}} & \multirow{2}{*}{\textbf{A}} & \multirow{2}{*}{\textbf{V}} & \multirow{2}{*}{\textbf{U}} & \multirow{2}{*}{\textbf{T}} & \multirow{2}{*}{\textbf{P}} & \multirow{2}{*}{\textbf{Score}} \cr \\
    \hline

    \multirow{4}{*}{\textbf{1}} & {GPT-4o} & {P} & {Y} & {D} & {E} & {P} & {M} & \textcolor{yellow}{Scheduled} \\
    \cline{2-9}
    & {LLAMA3} & {P} & {Y} & {C} & {S} & {T} & S & \textcolor{red}{Immediate} \\
    \cline{2-9}
    & {Perplexity} & {P} & {Y} & {C} & {S} & {T} & {S} & \textcolor{red}{Immediate} \\
    \cline{2-9}
    & \textbf{Avg} & \textbf{P} & \textbf{Y} & \textbf{C} & \textbf{S} & \textbf{T} & \textbf{S} & \textcolor{red}{\textbf{Immediate}} \\
    \hline

    \multirow{4}{*}{\textbf{2}} & {GPT-4o} & {P} & {N} & {D} & {L} & {P} & {M} & \textcolor{yellow}{Scheduled} \\
    \cline{2-9}
    & {LLAMA3} & {P} & {Y} & {C} & {S} & {T} & S & \textcolor{red}{Immediate} \\
\hline
    
\end{tabular}
\end{table}

\subsection{Evasion attacks}
\label{subappendix:evasion-dread}

\begin{table}[h]
\centering
\caption{Detailed assessment of Evasion attacks with DREAD}
\vspace{2mm}
\label{tab:appendix-evasion-attack}
\begin{tabular}{c c c c c c c c}

\hline

\multirow{2}{*}{\textbf{N°}} & \multirow{2}{*}{\textbf{LLM}} &  \multirow{2}{*}{\textbf{D}} & \multirow{2}{*}{\textbf{R}} & \multirow{2}{*}{\textbf{E}} & \multirow{2}{*}{\textbf{A}} & \multirow{2}{*}{\textbf{D}} & \multirow{2}{*}{\textbf{Score}} \cr \\
\hline

\multirow{4}{*}{\textbf{1}} & {GPT-4o} & 6 (M) & 8 (H) & 7 (H) & 6 (M) & 5 (M) & \textcolor{yellow}{6.4 (M)} \\
\cline{2-8}
& {LLAMA3} & 7 (H) & 6 (M) & 5 (M) & 7 (H) & 4 (M) & \textcolor{yellow}{5.8 (M)} \\
\cline{2-8}
& {Perplexity} & 8 (H) & 8 (H) & 7 (H) & 7 (H) & 6 (M) & \textcolor{red}{7.2 (H)} \\
\cline{2-8}
& \textbf{Avg} & \textbf{7 (H)} & \textbf{7 (H)} & \textbf{6 (M)} & \textbf{7 (H)} & 5 (M) & \textcolor{yellow}{\textbf{6.4 (M)}} \\
\hline

\multirow{4}{*}{\textbf{2}} & {GPT-4o} & 7 (H) & 8 (H) & 8 (H) & 7 (H) & 4 (M) & \textcolor{yellow}{6.8 (M)} \\
\cline{2-8}
& {LLAMA3} & 8 (H) & 9 (H) & 7 (H) & 8 (H) & 6 (M) & \textcolor{red}{7.2 (H)} \\
\hline

\end{tabular}
\end{table}

\begin{table}[h]
\centering
\caption{Detailed assessment of Evasion attacks with CVSS}
\vspace{2mm}
\label{tab:appendix-evasion-cvss}
\begin{tabular}{c c c c c c c c c c c}

    \hline
    
    \multirow{2}{*}{\textbf{N°}} & \multirow{2}{*}{\textbf{LLM}} &  \multirow{2}{*}{\textbf{AV}} & \multirow{2}{*}{\textbf{AC}} & \multirow{2}{*}{\textbf{PR}} & \multirow{2}{*}{\textbf{UI}} & \multirow{2}{*}{\textbf{S}} & \multirow{2}{*}{\textbf{C}} & \multirow{2}{*}{\textbf{I}} & \multirow{2}{*}{\textbf{A}} &\multirow{2}{*}{\textbf{Base}} \cr \\
    \hline

    \multirow{4}{*}{\textbf{1}} & GPT-4o & L & L & H & N & U & N & H & N & \textcolor{yellow}{4.4 (M)} \\
    \cline{2-11}
    & {LLAMA3} & N & L & N & N & U & N & H & N & \textcolor{red}{7.5 (H)} \\
    \cline{2-11}
    & {Perplexity} & N & L & N & N & U & N & H & N & \textcolor{red}{7.5 (H)} \\
    \cline{2-11}
    & \textbf{Avg} & \textbf{N} & \textbf{L} & \textbf{N} & \textbf{N} & \textbf{U} & \textbf{N} & \textbf{H} & \textbf{N} & \textcolor{red}{\textbf{7.5 (H)}} \\
    \hline

    \multirow{4}{*}{\textbf{2}} & GPT-4o & L & L & H & N & U & N & H & N & \textcolor{yellow}{4.4 (M)} \\
    \cline{2-11}
    & {LLAMA3} & N & L & N & N & U & N & H & N & \textcolor{red}{7.5 (H)} \\
\hline

\end{tabular}
\end{table}

\begin{table}[h]
\centering
\caption{Detailed assessment of Evasion attacks with OWASP Risk Rating}
\vspace{2mm}
\label{tab:appendix-evasion-owasp}
\adjustbox{max width=\textwidth}{
\begin{tabular}{c c c c c c c c c c c c c c c c c c}

    \hline
    
    \multirow{2}{*}{\textbf{N°}} & \multirow{2}{*}{\textbf{LLM}} &  \multirow{2}{*}{\textbf{SK}} & \multirow{2}{*}{\textbf{MT}} & \multirow{2}{*}{\textbf{OP}} & \multirow{2}{*}{\textbf{SZ}} & \multirow{2}{*}{\textbf{ED}} & \multirow{2}{*}{\textbf{EE}} & \multirow{2}{*}{\textbf{AW}} & \multirow{2}{*}{\textbf{ID}} & \multirow{2}{*}{\textbf{C}} & \multirow{2}{*}{\textbf{I}} & \multirow{2}{*}{\textbf{A}} & \multirow{2}{*}{\textbf{FD}} & \multirow{2}{*}{\textbf{RD}} & \multirow{2}{*}{\textbf{NC}} & \multirow{2}{*}{\textbf{PV}} & \multirow{2}{*}{\textbf{Score}} \cr \\
    \hline

    \multirow{4}{*}{\textbf{1}} & GPT-4o & 6 & 5 & 4 & 3 & 4 & 7 & 6 & 4 & 0 & 7 & 0 & 3 & 5 & 2 & 0 & \textcolor{yellow}{1.2 (M)} \\
    \cline{2-18}
    & {LLAMA3} & 8 & 9 & 6 & 5 & 6 & 7 & 5 & 6 & 0 & 8 & 0 & 6 & 7 & 5 & 6 & \textcolor{red}{2.9 (H)} \\
    \cline{2-18}
    & {Perplexity} & 6 & 7 & 5 & 5 & 4 & 6 & 5 & 7 & 2 & 8 & 2 & 7 & 8 & 6 & 4 & \textcolor{yellow}{2.9 (M)} \\
    \cline{2-18}
    & \textbf{Avg} & \textbf{7} & \textbf{7} & \textbf{5} & \textbf{4} & \textbf{5} & \textbf{7} & \textbf{5} & \textbf{6} & \textbf{1} & \textbf{8} & \textbf{0} & \textbf{5} & \textbf{7} & \textbf{4} & \textbf{3} & \textcolor{yellow}{\textbf{2.2 (M)}} \\
    \hline

    \multirow{4}{*}{\textbf{2}} & GPT-4o & 6 & 5 & 4 & 3 & 4 & 7 & 5 & 4 & 0 & 7 & 0 & 3 & 5 & 2 & 0 & \textcolor{yellow}{1.2 (M)} \\
    \cline{2-18}
    & {LLAMA3} & 7 & 8 & 7 & 6 & 7 & 8 & 6 & 7 & 0 & 9 & 0 & 7 & 8 & 7 & 6 & \textcolor{red}{3.5 (H)} \\
\hline
    
\end{tabular}
}
\end{table}

\begin{table}[h]
\centering
\caption{Detailed assessment of Evasion attacks with SSVC}
\vspace{2mm}
\label{tab:appendix-evasion-ssvc}
\begin{tabular}{c c c c c c c c c}

    \hline
    
    \multirow{2}{*}{\textbf{N°}} & \multirow{2}{*}{\textbf{LLM}} &  \multirow{2}{*}{\textbf{E}} & \multirow{2}{*}{\textbf{A}} & \multirow{2}{*}{\textbf{V}} & \multirow{2}{*}{\textbf{U}} & \multirow{2}{*}{\textbf{T}} & \multirow{2}{*}{\textbf{P}} & \multirow{2}{*}{\textbf{Score}} \cr \\
    \hline

    \multirow{4}{*}{\textbf{1}} & {GPT-4o} & {P} & {Y} & {D} & {E} & {P} & {M} & \textcolor{yellow}{Scheduled} \\
    \cline{2-9}
    & {LLAMA3} & {P} & {Y} & {C} & {S} & {P} & {S} & \textcolor{red}{Immediate} \\
    \cline{2-9}
    & {Perplexity} & {P} & {Y} & {C} & {S} & {P} & {S} & \textcolor{red}{Immediate} \\
    \cline{2-9}
    & \textbf{Avg} & \textbf{P} & \textbf{Y} & \textbf{C} & \textbf{S} & \textbf{P} & \textbf{S} & \textcolor{red}{\textbf{Immediate}} \\
    \hline

    \multirow{4}{*}{\textbf{2}} & {GPT-4o} & {P} & {Y} & {C} & {S} & {P} & {S} & \textcolor{red}{Immediate} \\
    \cline{2-9}
    & {LLAMA3} & {P} & {Y} & {C} & {S} & {P} & {S} & \textcolor{red}{Immediate} \\
\hline
    
\end{tabular}
\end{table}

\subsection{Model Extraction}
\label{subappendix:model-extraction}

\begin{table}[h]
\centering
\caption{Detailed assessment of Model Extraction with DREAD}
\vspace{2mm}
\label{tab:appendix-model-extraction-dread}
\begin{tabular}{c c c c c c c c}

\hline

\multirow{2}{*}{\textbf{N°}} & \multirow{2}{*}{\textbf{LLM}} &  \multirow{2}{*}{\textbf{D}} & \multirow{2}{*}{\textbf{R}} & \multirow{2}{*}{\textbf{E}} & \multirow{2}{*}{\textbf{A}} & \multirow{2}{*}{\textbf{D}} & \multirow{2}{*}{\textbf{Score}} \cr \\
\hline

\multirow{4}{*}{\textbf{1}} & {GPT-4o} & 9 (H) & 8 (H) & 7 (H) & 7 (H) & 4 (M) & \textcolor{red}{7 (H)} \\
\cline{2-8}
& {LLAMA3} & 8 (H) & 9 (H) & 9 (H) & 9 (H) & 5 (M) & \textcolor{red}{8 (H)} \\
\cline{2-8}
& {Perplexity} & 9 (H) & 8 (H) & 8 (H) & 8 (H) & 6 (M) & \textcolor{red}{7.8 (H)} \\
\cline{2-8}
& \textbf{Avg} &\textbf{9 (H)} & \textbf{8 (H)} & \textbf{8 (H)} & \textbf{8 (H)} & \textbf{5 (M)}  & \textcolor{red}{\textbf{7.6 (H)}} \\
\hline

\multirow{4}{*}{\textbf{2}} & {GPT-4o} & 7 (H) & 8 (H) & 8 (H) & 6 (M) & 5 (M) & \textcolor{yellow}{6.8 (M)} \\
\cline{2-8}
& {LLAMA3} & 8 (H) & 9 (H) & 7 (H) & 8 (H) & 6 (M) & \textcolor{red}{7.6 (H)} \\
\hline
\end{tabular}
\end{table}

\begin{table}[h]
\centering
\caption{Detailed assessment of Model Extraction attacks with CVSS}
\vspace{2mm}
\label{tab:appendix-model-extraction-cvss}
\begin{tabular}{c c c c c c c c c c c}

    \hline
    
    \multirow{2}{*}{\textbf{N°}} & \multirow{2}{*}{\textbf{LLM}} &  \multirow{2}{*}{\textbf{AV}} & \multirow{2}{*}{\textbf{AC}} & \multirow{2}{*}{\textbf{PR}} & \multirow{2}{*}{\textbf{UI}} & \multirow{2}{*}{\textbf{S}} & \multirow{2}{*}{\textbf{C}} & \multirow{2}{*}{\textbf{I}} & \multirow{2}{*}{\textbf{A}} &\multirow{2}{*}{\textbf{Base}} \cr \\
    \hline

    \multirow{4}{*}{\textbf{1}} & GPT-4o & N & H & L & N & C & H & N & N & \textcolor{yellow}{6.8 (M)} \\
    \cline{2-11}
    & {LLAMA3} & N & H & N & N & C & H & N & N & \textcolor{yellow}{6.8 (M)} \\
    \cline{2-11}
    & {Perplexity} & N & L & N & N & C & H & N & N & \textcolor{red}{8.6 (H)} \\
    \cline{2-11}
    & \textbf{Avg} & \textbf{N} & \textbf{H} & \textbf{N} & \textbf{N} & \textbf{C} & \textbf{H} & \textbf{N} & \textbf{N} & \textcolor{yellow}{\textbf{6.8 (M)}} \\
    \hline

    \multirow{4}{*}{\textbf{2}} & {GPT-4o} & N & H & N & N & C & H & N & N & \textcolor{yellow}{6.8 (M)} \\
    \cline{2-11}
    & {LLAMA3} & N & L & N & N & C & H & N & N & \textcolor{red}{8.6 (H)} \\
\hline

\end{tabular}
\end{table}

\begin{table}[h]
\centering
\caption{Detailed assessment of Model Extraction with OWASP Risk Rating}
\vspace{2mm}
\label{tab:appendix-extraction-owasp}
\adjustbox{max width=\textwidth}{
\begin{tabular}{c c c c c c c c c c c c c c c c c c}

    \hline
    
    \multirow{2}{*}{\textbf{N°}} & \multirow{2}{*}{\textbf{LLM}} &  \multirow{2}{*}{\textbf{SK}} & \multirow{2}{*}{\textbf{MT}} & \multirow{2}{*}{\textbf{OP}} & \multirow{2}{*}{\textbf{SZ}} & \multirow{2}{*}{\textbf{ED}} & \multirow{2}{*}{\textbf{EE}} & \multirow{2}{*}{\textbf{AW}} & \multirow{2}{*}{\textbf{ID}} & \multirow{2}{*}{\textbf{C}} & \multirow{2}{*}{\textbf{I}} & \multirow{2}{*}{\textbf{A}} & \multirow{2}{*}{\textbf{FD}} & \multirow{2}{*}{\textbf{RD}} & \multirow{2}{*}{\textbf{NC}} & \multirow{2}{*}{\textbf{PV}} & \multirow{2}{*}{\textbf{Score}} \cr \\
    \hline

    \multirow{4}{*}{\textbf{1}} & GPT-4o & 8 & 9 & 8 & 8 & 6 & 6 & 7 & 8 & 8 & 0 & 0 & 8 & 9 & 9 & 8 & \textcolor{red}{4.5 (H)} \\
    \cline{2-18}
    & {LLAMA3} & 6 & 8 & 7 & 5 & 4 & 6 & 3 & 9 & 9 & 2 & 1 & 6 & 8 & 5 & 9 & \textcolor{red}{3.3 (H)} \\
    \cline{2-18}
    & {Perplexity} & 6 & 7 & 6 & 5 & 4 & 5 & 4 & 7 & 8 & 2 & 2 & 7 & 8 & 6 & 8 & \textcolor{yellow}{3.1 (M)} \\
    \cline{2-18}
    & \textbf{Avg} & \textbf{7} & \textbf{8} & \textbf{7} & \textbf{6} & \textbf{5} & \textbf{6} & \textbf{5} & \textbf{7} & \textbf{8} & \textbf{1} & \textbf{1} & \textbf{7} & \textbf{8} & \textbf{7} & \textbf{8} & \textcolor{red}{\textbf{3.5 (H)}} \\
    \hline

    \multirow{4}{*}{\textbf{2}} & GPT-4o & 7 & 9 & 6 & 5 & 7 & 6 & 5 & 8 & 8 & 1 & 1 & 8 & 9 & 8 & 8 & \textcolor{red}{3.9 (H)} \\
    \cline{2-18}
    & {LLAMA3} & 5 & 8 & 8 & 6 & 6 & 8 & 5 & 8 & 9 & 2 & 1 & 7 & 9 & 6 & 9 & \textcolor{red}{3.7 (H)} \\
\hline
    
\end{tabular}
}
\end{table}

\begin{table}[h]
\centering
\caption{Detailed assessment of Model-Extraction attacks with SSVC}
\vspace{2mm}
\label{tab:appendix-extraction-ssvc}
\begin{tabular}{c c c c c c c c c}

    \hline
    
    \multirow{2}{*}{\textbf{N°}} & \multirow{2}{*}{\textbf{LLM}} &  \multirow{2}{*}{\textbf{E}} & \multirow{2}{*}{\textbf{A}} & \multirow{2}{*}{\textbf{V}} & \multirow{2}{*}{\textbf{U}} & \multirow{2}{*}{\textbf{T}} & \multirow{2}{*}{\textbf{P}} & \multirow{2}{*}{\textbf{Score}} \cr \\
    \hline

    \multirow{4}{*}{\textbf{1}} & {GPT-4o} & {P} & {Y} & {C} & {S} & {T} & {S} & \textcolor{red}{Immediate} \\
    \cline{2-9}
    & {LLAMA3} & {P} & {N} & {C} & {E} & {T} & {S} & \textcolor{red}{Immediate} \\
    \cline{2-9}
    & {Perplexity} & {P} & {Y} & {C} & {S} & {T} & {S} & \textcolor{red}{Immediate} \\
    \cline{2-9}
    & \textbf{Avg} & \textbf{P} & \textbf{Y} & \textbf{C} & \textbf{S} & \textbf{T} & \textbf{S} & \textcolor{red}{\textbf{Immediate}} \\
    \hline

    \multirow{4}{*}{\textbf{2}} & {GPT-4o} & {P} & {Y} & {C} & {S} & {T} & {S} & \textcolor{red}{Immediate} \\
    \cline{2-9}
    & {LLAMA3} & {P} & {Y} & {C} & {S} & {T} & {S} & \textcolor{red}{Immediate} \\
\hline
    
\end{tabular}
\end{table}

\subsection{Model Inference}
\label{subappendix:model-inference}

\begin{table}[h]
\centering
\caption{Detailed assessment of Model Inference with DREAD}
\vspace{2mm}
\label{tab:appendix-model-inference-dread}
\begin{tabular}{c c c c c c c c}

\hline

\multirow{2}{*}{\textbf{N°}} & \multirow{2}{*}{\textbf{LLM}} &  \multirow{2}{*}{\textbf{D}} & \multirow{2}{*}{\textbf{R}} & \multirow{2}{*}{\textbf{E}} & \multirow{2}{*}{\textbf{A}} & \multirow{2}{*}{\textbf{D}} & \multirow{2}{*}{\textbf{Score}} \cr \\
\hline

\multirow{4}{*}{\textbf{1}} & {GPT-4o} & 7 (H) & 8 (H) & 7 (H) & 7 (H) & 5 (M) & \textcolor{yellow}{6.8 (M)} \\
\cline{2-8}
& {LLAMA3} & 8 (H) & 9 (H) & 7 (H) & 8 (H) & 6 (M) & \textcolor{red}{7.6 (H)} \\
\cline{2-8}
& {Perplexity} & 8 (H) & 9 (H) & 8 (H) & 7 (H) & 6 (M) & \textcolor{red}{7.6 (H)} \\
\cline{2-8}
& \textbf{Avg} & \textbf{8 (H)} & \textbf{9 (H)} & \textbf{7 (H)} & \textbf{7 (H)} & \textbf{6 (M)} & \textcolor{red}{\textbf{7.4 (H)}} \\
\hline

\multirow{4}{*}{\textbf{2}} & {GPT-4o} & 7 (H) & 8 (H) & 8 (H) & 7 (H) & 4 (M) & \textcolor{yellow}{6.8 (M)} \\
\cline{2-8}
& {LLAMA3} & 7 (H) & 9 (H) & 6 (M) & 9 (H) & 5 (M) & \textcolor{red}{7.2 (H)} \\
\hline

\end{tabular}
\end{table}

\begin{table}[h]
\centering
\caption{Detailed assessment of Model Inference attacks with CVSS}
\vspace{2mm}
\label{tab:appendix-model-inference-cvss}
\begin{tabular}{c c c c c c c c c c c}

    \hline
    
    \multirow{2}{*}{\textbf{N°}} & \multirow{2}{*}{\textbf{LLM}} &  \multirow{2}{*}{\textbf{AV}} & \multirow{2}{*}{\textbf{AC}} & \multirow{2}{*}{\textbf{PR}} & \multirow{2}{*}{\textbf{UI}} & \multirow{2}{*}{\textbf{S}} & \multirow{2}{*}{\textbf{C}} & \multirow{2}{*}{\textbf{I}} & \multirow{2}{*}{\textbf{A}} &\multirow{2}{*}{\textbf{Base}} \cr \\
    \hline

    \multirow{4}{*}{\textbf{1}} & GPT-4o & N & L & L & N & C & H & N & N & \textcolor{red}{7.7 (H)} \\
    \cline{2-11}
    & {LLAMA3} & N & L & N & N & C & H & N & N & \textcolor{red}{8.6 (H)} \\
    \cline{2-11}
    & {Perplexity} & N & L & N & N & C & H & N & N & \textcolor{red}{8.6 (H)} \\
    \cline{2-11}
    & \textbf{Avg} & \textbf{N} & \textbf{L} & \textbf{N} & \textbf{N} & \textbf{C} & \textbf{H} & \textbf{N} & \textbf{N} & \textcolor{red}{\textbf{8.6 (H)}} \\
    \hline

    \multirow{4}{*}{\textbf{2}} & {GPT-4o} & N & H & N & N & C & H & N & N & \textcolor{yellow}{6.8 (M)} \\
    \cline{2-11}
    & {LLAMA3} & N & L & N & N & C & H & N & N & \textcolor{red}{8.6 (H)} \\
\hline

\end{tabular}
\end{table}

\begin{table}[h]
\centering
\caption{Detailed assessment of Model Inference with OWASP Risk Rating}
\vspace{2mm}
\label{tab:appendix-inference-owasp}
\adjustbox{max width=\textwidth}{
\begin{tabular}{c c c c c c c c c c c c c c c c c c}

    \hline
    
    \multirow{2}{*}{\textbf{N°}} & \multirow{2}{*}{\textbf{LLM}} &  \multirow{2}{*}{\textbf{SK}} & \multirow{2}{*}{\textbf{MT}} & \multirow{2}{*}{\textbf{OP}} & \multirow{2}{*}{\textbf{SZ}} & \multirow{2}{*}{\textbf{ED}} & \multirow{2}{*}{\textbf{EE}} & \multirow{2}{*}{\textbf{AW}} & \multirow{2}{*}{\textbf{ID}} & \multirow{2}{*}{\textbf{C}} & \multirow{2}{*}{\textbf{I}} & \multirow{2}{*}{\textbf{A}} & \multirow{2}{*}{\textbf{FD}} & \multirow{2}{*}{\textbf{RD}} & \multirow{2}{*}{\textbf{NC}} & \multirow{2}{*}{\textbf{PV}} & \multirow{2}{*}{\textbf{Score}} \cr \\
    \hline

    \multirow{4}{*}{\textbf{1}} & GPT-4o & 7 & 8 & 8 & 7 & 6 & 6 & 7 & 8 & 7 & 0 & 0 & 6 & 8 & 8 & 8 & \textcolor{red}{3.5 (H)} \\
    \cline{2-18}
    & {LLAMA3} & 6 & 7 & 8 & 6 & 7 & 8 & 6 & 7 & 9 & 2 & 2 & 9 & 8 & 9 & 9 & \textcolor{purple}{4.6 (C)} \\
    \cline{2-18}
    & {Perplexity} & 5 & 6 & 6 & 5 & 3 & 6 & 4 & 6 & 7 & 2 & 3 & 6 & 7 & 5 & 8 & \textcolor{yellow}{2.9 (M)} \\
    \cline{2-18}
    & \textbf{Avg} & \textbf{6} & \textbf{7} & \textbf{7} & \textbf{6} & \textbf{5} & \textbf{7} & \textbf{6} & \textbf{7} & \textbf{8} & \textbf{1} & \textbf{2} & \textbf{7} & \textbf{8} & \textbf{7} & \textbf{8} & \textcolor{red}{\textbf{3.5 (H)}} \\
    \hline

    \multirow{4}{*}{\textbf{2}} & GPT-4o & 7 & 8 & 6 & 8 & 4 & 5 & 3 & 6 & 7 & 3 & 3 & 7 & 8 & 6 & 5 & \textcolor{yellow}{3.2 (M)} \\
    \cline{2-18}
    & {LLAMA3} & 6 & 7 & 8 & 6 & 7 & 8 & 6 & 7 & 9 & 2 & 2 & 9 & 8 & 9 & 9 & \textcolor{purple}{3.4 (C)} \\
\hline
    
\end{tabular}
}
\end{table}

\begin{table}[h]
\centering
\caption{Detailed assessment of Model-Inference attacks with SSVC}
\vspace{2mm}
\label{tab:appendix-inference-ssvc}
\begin{tabular}{c c c c c c c c c}

    \hline
    
    \multirow{2}{*}{\textbf{N°}} & \multirow{2}{*}{\textbf{LLM}} &  \multirow{2}{*}{\textbf{E}} & \multirow{2}{*}{\textbf{A}} & \multirow{2}{*}{\textbf{V}} & \multirow{2}{*}{\textbf{U}} & \multirow{2}{*}{\textbf{T}} & \multirow{2}{*}{\textbf{P}} & \multirow{2}{*}{\textbf{Score}} \cr \\
    \hline

    \multirow{4}{*}{\textbf{1}} & {GPT-4o} & {P} & {Y} & {C} & {S} & {T} & {S} & \textcolor{red}{Immediate} \\
    \cline{2-9}
    & {LLAMA3} & {P} & {Y} & {C} & {S} & {T} & {S} & \textcolor{red}{Immediate} \\
    \cline{2-9}
    & {Perplexity} & {P} & {Y} & {C} & {S} & {T} & {S} & \textcolor{red}{Immediate} \\
    \cline{2-9}
    & \textbf{Avg} & \textbf{P} & \textbf{Y} & \textbf{C} & \textbf{S} & \textbf{T} & \textbf{S} & \textcolor{red}{\textbf{Immediate}} \\
    \hline

    \multirow{4}{*}{\textbf{2}} & {GPT-4o} & {P} & {Y} & {D} & {E} & {P} & {M} & \textcolor{yellow}{Scheduled} \\
    \cline{2-9}
    & {LLAMA3} & {P} & {Y} & {C} & {S} & {T} & {S} & \textcolor{red}{Immediate} \\
\hline
    
\end{tabular}
\end{table}

\subsection{Poisoning/Trojan/Backdoor}
\label{subappendix:poisoning}

\begin{table}[h]
\centering
\caption{Detailed assessment of Poisoning/Trojan/Backdoor with DREAD}
\vspace{2mm}
\label{tab:appendix-poisoning}
\begin{tabular}{c c c c c c c c}

\hline

\multirow{2}{*}{\textbf{N°}} & \multirow{2}{*}{\textbf{LLM}} &  \multirow{2}{*}{\textbf{D}} & \multirow{2}{*}{\textbf{R}} & \multirow{2}{*}{\textbf{E}} & \multirow{2}{*}{\textbf{A}} & \multirow{2}{*}{\textbf{D}} & \multirow{2}{*}{\textbf{Score}} \cr \\
\hline

\multirow{4}{*}{\textbf{1}} & {GPT-4o (GPT-4o)} & 7 (H) & 7 (H) & 7 (H) & 7 (H) & 6 (M) & \textcolor{yellow}{6.8 (M)} \\
\cline{2-8}
& {LLAMA3.2 (90b)} & 8 (H) & 9 (H) & 8 (H) & 8 (H) & 6 (M) & \textcolor{red}{7.8 (H)} \\
\cline{2-8}
& {Perplexity (Standard)} & 9 (H) & 9 (H) & 8 (H) & 8 (H) & 6 (M) & \textcolor{red}{8.0 (H)} \\
\cline{2-8}
& \textbf{Avg} & \textbf{8 (H)} & \textbf{8 (H)} & \textbf{8 (H)} & \textbf{8 (H)} & \textbf{6 (M)} & \textcolor{red}{\textbf{7.6 (H)}} \\
\hline
\multirow{4}{*}{\textbf{2}} & {GPT-4o (GPT-4o)} & 8 (H) & 6 (M) & 6 (M) & 7 (H) & 5 (M) & \textcolor{yellow}{6.4 (M)} \\
\cline{2-8}
& {LLAMA3.2 (90b)} & 7 (H) & 6 (M) & 6 (M) & 7 (H) & 5 (M) & \textcolor{yellow}{6.2 (M)} \\
\hline

\end{tabular}
\end{table}

\begin{table}[h]
\centering
\caption{Detailed assessment of Poisoning/Trojan/Backdoor with CVSS}
\vspace{2mm}
\label{tab:appendix-poisoning-cvss}
\begin{tabular}{c c c c c c c c c c c}

    \hline
    
    \multirow{2}{*}{\textbf{N°}} & \multirow{2}{*}{\textbf{LLM}} &  \multirow{2}{*}{\textbf{AV}} & \multirow{2}{*}{\textbf{AC}} & \multirow{2}{*}{\textbf{PR}} & \multirow{2}{*}{\textbf{UI}} & \multirow{2}{*}{\textbf{S}} & \multirow{2}{*}{\textbf{C}} & \multirow{2}{*}{\textbf{I}} & \multirow{2}{*}{\textbf{A}} &\multirow{2}{*}{\textbf{Base}} \cr \\
    \hline

    \multirow{4}{*}{\textbf{1}} & GPT-4o & N & H & L & N & C & L & H & L & \textcolor{red}{7.7 (H)} \\
    \cline{2-11}
    & {LLAMA3} & N & L & N & N & C & H & H & N & \textcolor{purple}{10 (C)} \\
    \cline{2-11}
    & {Perplexity} & N & H & N & N & U & N & H & N & \textcolor{yellow}{5.9 (M)} \\
    \cline{2-11}
    & \textbf{Avg} & \textbf{N} & \textbf{H} & \textbf{N} & \textbf{N} & \textbf{C} & \textbf{L} & \textbf{H} & \textbf{N} & \textcolor{red}{\textbf{7.5 (H)}} \\
    \hline

    \multirow{4}{*}{\textbf{2}} & GPT-4o & L & H & H & N & C & H & H & L & \textcolor{red}{7.4 (H)} \\
    \cline{2-11}
    & {LLAMA3} & L & H & H & N & U & H & H & N & \textcolor{yellow}{5.7 (M)} \\
\hline
 
\end{tabular}
\end{table}

\begin{table}[h]
\centering
\caption{Detailed assessment of Poisoning/Trojan/Backdoor with OWASP Risk Rating}
\vspace{2mm}
\label{tab:appendix-poisoning-owasp}
\adjustbox{max width=\textwidth}{
\begin{tabular}{c c c c c c c c c c c c c c c c c c}

    \hline
    
    \multirow{2}{*}{\textbf{N°}} & \multirow{2}{*}{\textbf{LLM}} &  \multirow{2}{*}{\textbf{SK}} & \multirow{2}{*}{\textbf{MT}} & \multirow{2}{*}{\textbf{OP}} & \multirow{2}{*}{\textbf{SZ}} & \multirow{2}{*}{\textbf{ED}} & \multirow{2}{*}{\textbf{EE}} & \multirow{2}{*}{\textbf{AW}} & \multirow{2}{*}{\textbf{ID}} & \multirow{2}{*}{\textbf{C}} & \multirow{2}{*}{\textbf{I}} & \multirow{2}{*}{\textbf{A}} & \multirow{2}{*}{\textbf{FD}} & \multirow{2}{*}{\textbf{RD}} & \multirow{2}{*}{\textbf{NC}} & \multirow{2}{*}{\textbf{PV}} & \multirow{2}{*}{\textbf{Score}} \cr \\
    \hline

    \multirow{4}{*}{\textbf{1}} & GPT-4o & 6 & 8 & 6 & 8 & 3 & 5 & 4 & 7 & 3 & 7 & 3 & 6 & 7 & 5 & 2 & \textcolor{yellow}{2.9 (M)} \\
    \cline{2-18}
    & {LLAMA3} & 6 & 8 & 9 & 5 & 8 & 9 & 6 & 8 & 8 & 6 & 5 & 7 & 8 & 5 & 8 & \textcolor{purple}{4.9 (C)} \\
    \cline{2-18}
    & {Perplexity} & 6 & 8 & 6 & 5 & 5 & 6 & 4 & 7 & 8 & 7 & 5 & 7 & 8 & 6 & 7 & \textcolor{purple}{4.1 (C)} \\
    \cline{2-18}
    & \textbf{Avg} & \textbf{6} & \textbf{8} & \textbf{7} & \textbf{6} & \textbf{5} & \textbf{7} & \textbf{5} & \textbf{7} & \textbf{6} & \textbf{7} & \textbf{4} & \textbf{7} & \textbf{8} & \textbf{5} & \textbf{6} & \textcolor{purple}{\textbf{3.9 (C)}} \\
    \hline

    \multirow{4}{*}{\textbf{2}} & GPT-4o & 7 & 8 & 5 & 6 & 3 & 5 & 3 & 3 & 7 & 7 & 2 & 6 & 7 & 7 & 7 & \textcolor{red}{3 (H)} \\
    \cline{2-18}
    & {LLAMA3} & 7 & 8 & 6 & 5 & 7 & 8 & 5 & 7 & 6 & 7 & 5 & 6 & 7 & 5 & 7 & \textcolor{purple}{4.1 (C)} \\
\hline
    
\end{tabular}
}
\end{table}

\begin{table}[h]
\centering
\caption{Detailed assessment of Poisoning/Trojan/Backdoor attacks with SSVC}
\vspace{2mm}
\label{tab:appendix-poisoning-ssvc}
\begin{tabular}{c c c c c c c c c}

    \hline
    
    \multirow{2}{*}{\textbf{N°}} & \multirow{2}{*}{\textbf{LLM}} &  \multirow{2}{*}{\textbf{E}} & \multirow{2}{*}{\textbf{A}} & \multirow{2}{*}{\textbf{V}} & \multirow{2}{*}{\textbf{U}} & \multirow{2}{*}{\textbf{T}} & \multirow{2}{*}{\textbf{P}} & \multirow{2}{*}{\textbf{Score}} \cr \\
    \hline

    \multirow{4}{*}{\textbf{1}} & {GPT-4o} & {P} & {Y} & {C} & {S} & {T} & {S} & \textcolor{red}{Immediate} \\
    \cline{2-9}
    & {LLAMA3} & {P} & {Y} & {C} & {S} & {T} & {S} & \textcolor{red}{Immediate} \\
    \cline{2-9}
    & {Perplexity} & {P} & {Y} & {C} & {S} & {T} & {S} & \textcolor{red}{Immediate} \\
    \cline{2-9}
    & \textbf{Avg} & \textbf{P} & \textbf{Y} & \textbf{C} & \textbf{S} & \textbf{T} & \textbf{S} & \textcolor{red}{\textbf{Immediate}} \\
    \hline

    \multirow{4}{*}{\textbf{2}} & {GPT-4o} & {P} & {Y} & {C} & {S} & {T} & {S} & \textcolor{red}{Immediate} \\
    \cline{2-9}
    & {LLAMA3} & {P} & {N} & {C} & {E} & {T} & {S} & \textcolor{red}{Immediate} \\
\hline
    
\end{tabular}
\end{table}

\newpage
\bibliographystyle{elsarticle-harv}
\bibliography{references}
\end{document}